\documentclass[sigconf]{acmart}

\usepackage{colortbl}
\usepackage{multirow}
\usepackage{multicol}
\usepackage{twemojis}
\usepackage{soulutf8}
\usepackage{soulpos}
\usepackage{xcolor}
\usepackage{enumitem}
\usepackage{listings}
\usepackage{color}
\usepackage{supertabular}

\definecolor{gray}{rgb}{0.937, 0.937, 0.937}
\definecolor{darkestblue}{rgb}{0.109,0.266,0.529}
\definecolor{lighterblue}{rgb}{0.239,0.521,0.776}
\definecolor{lightblue}{rgb}{0.811,0.886,0.952}
\definecolor{green}{rgb}{0.5647,0.9333,0.5647}
\definecolor{white}{rgb}{1,1,1}
\definecolor{lightyellow}{rgb}{1,0.949,0.8}
\definecolor{lightgreen}{rgb}{0.850,0.917,0.827}
\definecolor{lightred}{rgb}{0.956,0.8,0.8}
\definecolor{thickyellow}{rgb}{0.749,0.564,0.003}
\definecolor{thickgreen}{rgb}{0.223,0.462,0.113}
\definecolor{thickred}{rgb}{0.8,0,0}

\DeclareRobustCommand{\hlyellow}[1]{{\sethlcolor{lightyellow}\hl{#1}}}
\DeclareRobustCommand{\hlgreen}[1]{{\sethlcolor{lightgreen}\hl{#1}}}
\DeclareRobustCommand{\hlgray}[1]{{\sethlcolor{gray}\hl{#1}}}
\DeclareRobustCommand{\hlpink}[1]{{\sethlcolor{lightred}\hl{#1}}}
\DeclareRobustCommand{\mystr}[1]{\setstcolor{red}\st{#1}}
\setstcolor{red}

\newcommand{\coloredcell}[1]{%
    \ifnum#1=3 \cellcolor{darkestblue}\color{white}\fi
    \ifnum#1=2 \cellcolor{lighterblue}\color{white}\fi
    \ifnum#1=1 \cellcolor{lightblue}\fi
    \ifnum#1=0 \cellcolor{white}\fi
    #1%
}
\newcommand{\algobo}{AlgoBo}
\newcommand{\algobobasic}{AlgoBo-Basic}
\newcommand{\sysname}{TeachYou}

\newcolumntype{Z}{>{\centering\arraybackslash}m{0.85cm}}
\newcolumntype{P}{>{\centering\arraybackslash}m{0.6cm}}
\newcolumntype{S}{>{\centering\arraybackslash}m{0.6cm}}

\AtBeginDocument{%
  \providecommand\BibTeX{{%
    \normalfont B\kern-0.5em{\scshape i\kern-0.25em b}\kern-0.8em\TeX}}}

\copyrightyear{2024}
\acmYear{2024}
\setcopyright{rightsretained}
\acmConference[CHI '24]{Proceedings of the CHI Conference on Human Factors in Computing Systems}{May 11--16, 2024}{Honolulu, HI, USA}
\acmBooktitle{Proceedings of the CHI Conference on Human Factors in Computing Systems (CHI '24), May 11--16, 2024, Honolulu, HI, USA}
\acmDOI{10.1145/3613904.3642349}
\acmISBN{979-8-4007-0330-0/24/05}

\begin{document}

\title[Teach AI How to Code: Using LLMs as Teachable Agents for Programming Education]{Teach AI How to Code: Using Large Language Models as Teachable Agents for Programming Education}

\author{Hyoungwook Jin}
\email{jinhw@kaist.ac.kr}
\orcid{0000-0003-0253-560X}
\affiliation{%
  \institution{School of Computing, KAIST}
  \city{Daejeon}
  \country{Republic of Korea}
}

\author{Seonghee Lee}
\email{shlee@cs.stanford.edu}
\orcid{0000-0003-0187-5215}
\affiliation{%
  \institution{Stanford University}
  \city{Palo Alto, CA}
  \country{United States}
}

\author{Hyungyu Shin}
\email{hyungyu.sh@kaist.ac.kr}
\orcid{0000-0001-7328-0072}
\affiliation{%
  \institution{School of Computing, KAIST}
  \city{Daejeon}
  \country{Republic of Korea}
}

\author{Juho Kim}
\email{juhokim@kaist.ac.kr}
\orcid{0000-0001-6348-4127}
\affiliation{%
  \institution{School of Computing, KAIST}
  \city{Daejeon}
  \country{Republic of Korea}
}

\renewcommand{\shortauthors}{Jin. et al.}

\begin{abstract}
This work investigates large language models (LLMs) as teachable agents for learning by teaching (LBT). LBT with teachable agents helps learners identify knowledge gaps and discover new knowledge. However, teachable agents require expensive programming of subject-specific knowledge. While LLMs as teachable agents can reduce the cost, LLMs' expansive knowledge as tutees discourages learners from teaching. We propose a prompting pipeline that restrains LLMs' knowledge and makes them initiate ``why'' and ``how'' questions for effective knowledge-building. We combined these techniques into TeachYou, an LBT environment for algorithm learning, and AlgoBo, an LLM-based tutee chatbot that can simulate misconceptions and unawareness prescribed in its knowledge state. Our technical evaluation confirmed that our prompting pipeline can effectively configure AlgoBo's problem-solving performance. Through a between-subject study with 40 algorithm novices, we also observed that AlgoBo's questions led to knowledge-dense conversations (effect size=0.71). Lastly, we discuss design implications, cost-efficiency, and personalization of LLM-based teachable agents.
\end{abstract}

\ccsdesc[500]{Human-centered computing~Interactive systems and tools}
\ccsdesc[500]{Applied computing~Interactive learning environments}

\keywords{Human-AI interaction, LLM agents, AI and Education, Generative AI}

\begin{teaserfigure}
    \includegraphics[width=\textwidth]{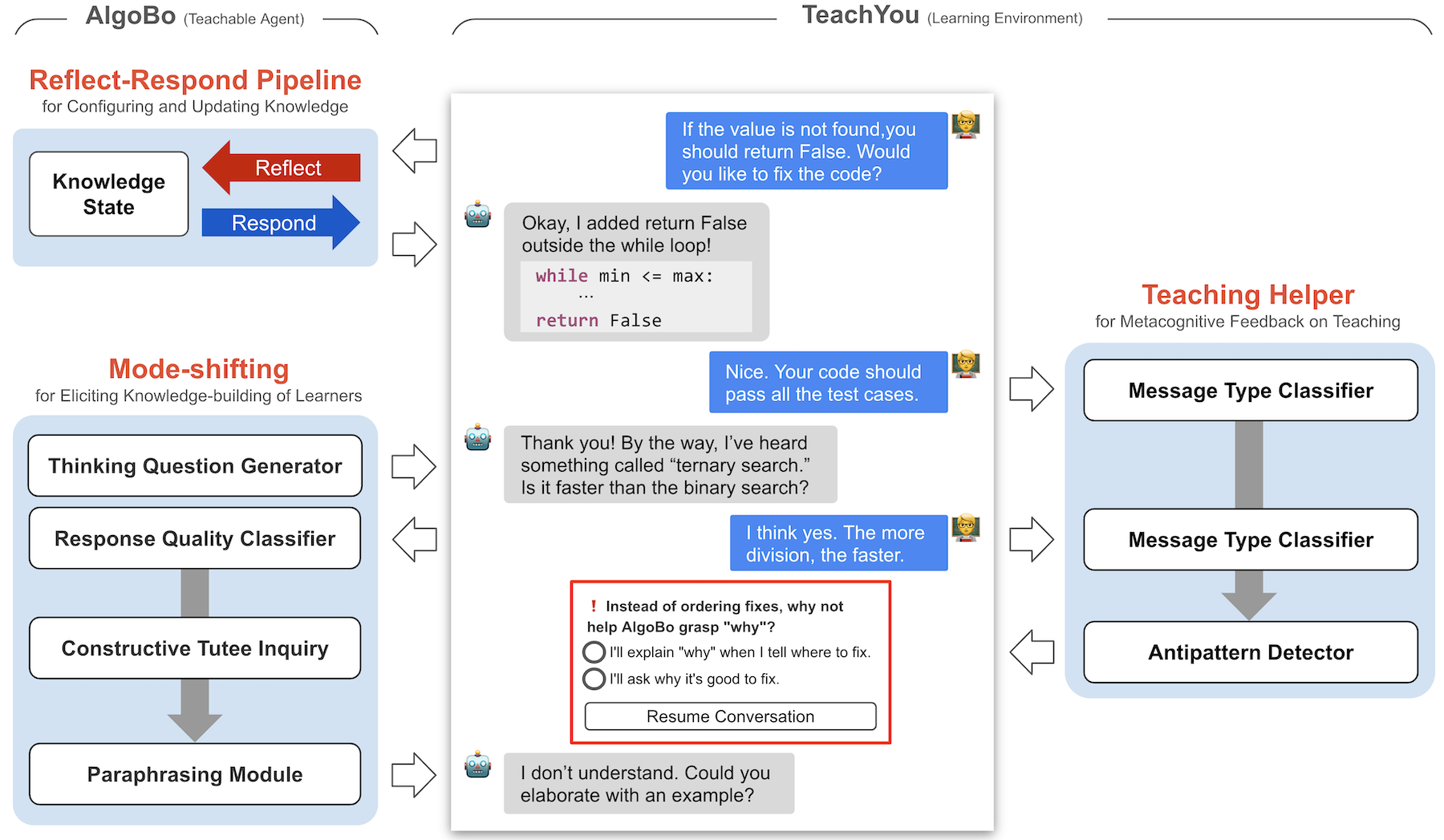}
    \caption{An overview of the core components of \algobo{} and \sysname{}. The Reflect-Respond pipeline enables \algobo{} to create responses following its evolving knowledge state while Mode-shifting guides LBT conversations through knowledge-building questions that ask ``why'' and ``how''. The Teaching Helper in \sysname{} analyzes conversations in real-time and gives metacognitive feedback and suggestions on teaching methods.}
    \label{fig:teaser}
    \Description{An overview of the three core components of AlgoBo, an LLM-based teachable agent, and TeachYou, a learning environment for LBT: (1) Reflect-Respond prompting pipeline, (2) Mode-shifting, and (3) Teaching Helper. The Reflect-Respond pipeline enables AlgoBo to add(reflect) new information in conversations to its knowledge state and respond following its evolving knowledge state. Mode-shifting guides LBT conversations through knowledge-building questions that ask "why" and "how". Teaching Helper in TeachYou analyzes conversations in real time and gives metacognitive feedback and suggestions on teaching methods. The teaching helper is displayed as a red or green box in the chat interface.}
\end{teaserfigure}


\maketitle

\section{Introduction}

Interactive learning activities involve learners actively collaborating with peers or engaging with computer systems to deepen their comprehension of a specific topic~\cite{Menekse2013, Tudge1990}. 
Compared to passive learning activities (e.g., reading text passages without doing anything else), interactive learning activities (e.g., pair programming, peer teaching) can elicit the deepest level of understanding by encouraging learners to elaborate their explanations and construct new knowledge on top of each other through conversations~\cite{Chi2014, Smutny2020, Chin2013, Chin2010, Suh2022, Guo2013}. One form of interactive learning is Learning by Teaching (LBT), where learners tutor a peer learner and exchange questions to reorganize their knowledge and identify knowledge gaps.

LBT with teachable AI agents (i.e., virtual tutees) can offer many advantages over LBT with humans. Teachable agents can bring scalability to LBT with their around-the-clock availability and motivate learners' participation in LBT by reducing psychological barriers, such as the fear of making mistakes while teaching and the pressure of responding in real-time~\cite{Chase2009, Debban2023}. 
However, despite these benefits, disseminating teachable agents to diverse subjects is challenging in practice due to the effort-intensive authoring of the agents' knowledge model~\cite{matsuda2022teachable} and sophisticated behaviors~\cite{Shahriar2021} to elicit desired learning experiences beyond a tutoring simulation.
Conventional authoring methods require extensive mapping of agents' knowledge states and high programming skills, precluding teachers and education researchers from tweaking teachable agents for their needs and context.

In this paper, rather than constructing teachable agents from the ground up, we propose a top-down methodology in which we use versatile Large Language Models (LLMs) to simulate tutees. Recent advances in LLMs show their remarkable capabilities in making contextual dialogues~\cite{ouyang2022training, ross2023programmer}, role mimicry~\cite{Markel2023, kong2023better}, and learning from demonstrations~\cite{Brown2020, radford2019language}.
Teachable agents equipped with the LLM capabilities can perform more believable and natural tutoring interactions (e.g., writing and explaining arbitrary code on request), compared to prior non-LLM LBT systems that adopted pre-scripted and limited interaction channels~\cite{Biswas2001, Leelawong2008, Matsuda2010, Pareto2011}. The flexible interaction allows learners to formulate free-form questions and try diverse teaching methods, improving their knowledge construction and metacognition~\cite{chin2002student, yu2005promoting, aflalo2021students, rothstein2011make}.
We explore using LLMs to lower the cost and barriers of building teachable agents and to make LBT more engaging and pedagogically effective.

In our formative study, we asked 15 programming novices to conduct LBT with ChatGPT prompted to perform the role of a tutee. We found that there are needs for 1) confining the knowledge level of LLM agents, 2) agent-initiated ``why'' and ``how'' questions, and 3) in-conversation feedback on learners' teaching methods.
Our dialogue analysis revealed that role-playing led learners to self-explain their knowledge but was limited to knowledge-telling, achieving only the rudimentary benefits of doing LBT. Participants struggled to build new knowledge because the teachable agent excelled in writing code even without being taught and did not ask questions that could prompt elaboration and knowledge-building. The participants also commented about the lack of metacognitive guidance and reflection for effective LBT.

To address these issues, we built a teachable agent, ``\algobo{}'', that can exhibit prescribed misconceptions and knowledge level and ``\sysname{}'', an LBT environment for introductory algorithm learning (Fig.~\ref{fig:teaser}).
In \sysname{}, learners solve programming problems on algorithms (e.g., binary search) and reflect on them by teaching \algobo{}.
As learners correctly teach \algobo{}, our Reflect-Respond prompting pipeline instructs \algobo{} to fix its misconceptions and write code based on what it is taught.
We also added Mode-shifting, in which \algobo{} periodically shifts to a questioner mode and asks questions to prompt learners' elaboration and sense-making. Lastly, \sysname{} has a Teaching Helper that provides metacognitive feedback and suggestions to learners on their teaching method in real-time through dialogue analysis.

We conducted a technical evaluation of our Reflect-Respond prompting pipeline to check if \algobo{} can simulate a tutee with a prescribed knowledge level on different algorithm topics. We found that the pipeline can effectively configure, persist, and adapt \algobo{}'s knowledge level within a conversation. We also conducted a between-subjects study with 40 algorithm novices, where the participants studied binary search with either \sysname{} or a baseline system without Mode-shifting and Teaching Helper. 
Our analysis of LBT dialogues and survey results showed that Mode-shifting improved the density of knowledge-building messages in the conversations significantly ($p=0.03$) with an effect size (Cohen's d) of 0.71. Teaching Helper also helped participants reflect on their teaching methods and sequence their questions strategically, but we could not observe significant improvement in participants' metacognition.

We structured our paper in the following order. After a discussion of related work, we describe our formative study settings and preliminary findings. We then reorganize the findings into three design goals and introduce our system and pipeline for achieving the goals. With that, we present our technical and user-study evaluation results. Lastly, based on our results and observations, we discuss the design considerations for teachable agents, the benefits of using LLMs, promising directions for personalizing teachable agents, and interaction guidelines for better LBT with teachable agents.

This paper makes the following contributions:
\begin{itemize}
\item \algobo{}, an LLM-based teachable agent that uses the Reflect-Respond prompting pipeline to simulate prescribed learning behaviors and Mode-shifting to scaffold knowledge-building of learners through ``why'' and ``how'' questions.
\item \sysname{}, a web-based algorithm learning system that supports LBT with \algobo{} and provides metacognitive feedback on teaching based on real-time conversation analysis.
\item A technical evaluation of the Reflect-Respond prompting pipeline and an empirical user study results with 40 participants showing that \sysname{} improved knowledge-building in LBT.
\end{itemize}

\section{Related Work}

We outline past studies on stimulating effective LBT among humans and using teachable agents. Previous research connects to our work in improving the quality and scalability of LBT using virtual agents.

\subsection{Learning by Teaching}

Learning by Teaching (LBT) is a teaching method in which learners not only articulate and restructure their existing knowledge but also engage in reflective knowledge-building.
Knowledge-building refers to extending knowledge beyond provided materials to craft deeper explanations, analogies, and inferential connections~\cite{Roscoe2007, Chase2009, Chi2014, Duran2017}, leading to the deliberate creation and improvement of knowledge useful for a community in a broader context~\cite{Scardamalia2006}.
However, LBT alone does not elicit knowledge-building naturally~\cite{Webb1986, Pressley1987}; learners tend to end up in knowledge-telling, in which they verbalize what they already know~\cite{Roscoe2007}. 
Previous research investigated support for eliciting knowledge-building responses from learners. King et al. found that training learners to ask reviewing, proving, and thinking questions in sequence to peers during LBT can promote higher-order thinking and learning~\cite{King1998}. Roscoe and Chi's analysis of LBT dialogues showed the importance of the tutee's role in knowledge-building; the deep questions from the tutee encourage tutors to make self-reflective responses and create inferences between new and prior knowledge~\cite{Roscoe2004}. Shahriar and Matsuda also confirmed that tutees' follow-up questions drew the knowledge-building of tutors with low prior knowledge in particular~\cite{Shahriar2021}. Matsuda et al. found that LBT with metacognitive guidance for planning and conducting teaching is as effective as being tutored by experts regardless of learners' prior competency~\cite{Matsuda2018}. Our primary goal is to build an interactive system that draws knowledge-building from learners in LBT. To do so, we adapt the interventions mentioned above in human tutor-tutee interactions to the conversational interactions between virtual agents and learners.

\subsection{Teachable Agents for LBT}

A core component of LBT is the presence of a peer learner. However, as human learners cannot always be present, past research introduced teachable agents---virtual agents that can learn declarative and procedural knowledge from learners' explanations and demonstrations, taking the role of peer learners in LBT~\cite{Blair2007}. Teachable agents showed promising results in improving students' performance, self-explanation, and acceptance of constructive feedback~\cite{Graesser2004, Matsuda2011, Silvervarg2020, Chase2009, Leelawong2008}. 
LBT with early teachable agents was non-conversational; agents revealed their knowledge states as concept maps, and learners taught the agents by directly editing their knowledge states~\cite{Bredeweg2007, Biswas2005}.
Recent teachable agents conceal their states and simulate more authentic learning behaviors; 
agents can learn from the tutors' demonstrations~\cite{Matsuda2021}, mimic the behaviors of learners (e.g., making arithmetic mistakes)~\cite{Ketamo2009, Pareto2011}, improve with correct instructions~\cite{Matsuda2011}, and ask questions~\cite{Matsuda2012}. 
However, implementing these natural and highly interactive teachable agents requires significant manual efforts and programming skills to specify and model the knowledge of agents~\cite{matsuda2015teaching}. For example, implementing an agent in SimStudent required more than a thousand lines of Java code for simple algebra equation solving~\cite{Matsuda2021}; the cost may increase exponentially for more complicated topics (e.g., algorithm learning, advanced equation solving). 
In this paper, we investigate using LLMs for building conversational teachable agents with low manual effort and programming barriers to support educators and researchers in adopting LBT in diverse classes and experiments.

\subsection{LLM-powered Simulation of Tutoring}

While the development cost and skill barrier have limited teachable agents to few learning activities in the past, LLMs can provide a more affordable method to simulate virtual students and coaches and to diversify their interactions~\cite{Wang2023, Markel2023, ojeda2023learning}.
GPTeach by Markel et al.~\cite{Markel2023} simulates role-plays between a teaching trainee and virtual students who come for office hours by leveraging persona and context setting in prompts.
LLM-simulated students allow trainees to practice teaching with diverse students and to interact through conversations, perhaps the most familiar and open-ended form of teaching others.
Likewise, LLM-based teachable agents can enrich tutor-tutee interaction and activities in LBT as learners can formulate free-form questions by themselves and try out different teaching strategies, as opposed to non-LLM LBT systems that permit only predefined methods to assess agents' knowledge (e.g., multiple choice questions)~\cite{Matsuda2018, Biswas2001, Leelawong2008, Pareto2011}.
Nevertheless, challenges remain in making these LLM-based agents suitable for LBT, where the agents should not only simulate tutoring but also proactively elicit learners' knowledge-building.
Beyond the roles set by prompts, we need precise control of the teachable agents' cognitive behaviors (e.g., knowledge levels and question-asking) to facilitate the intended learning experience.
Prior research has proposed LLM agent architectures and pipelines to grant and scope cognitive capabilities to LLM, such as memory~\cite{zhou2023recurrentgpt, packer2023memgpt}, role-playing~\cite{park2023generative, junprung2023exploring}, and reasoning~\cite{lin2023swiftsage, kim2023language, chen2023chatcot}.
We extend the control on LLMs' cognitive capabilities by proposing an LLM prompting pipeline that restrains the knowledge level of LLM-based agents.
\section{Formative Study}

We ran a formative study to explore the difficulties of using an LLM as a teachable agent. We recruited 15 Python novices and asked them to teach the binary search algorithm to an LLM chatbot. We surveyed their learning experience and analyzed the quality of their dialogues with the chatbot by annotating the types of messages.

\subsection{Participants and Procedure}

We recruited 15 participants on campus who could read and write short (about 15 lines) Python programs containing \texttt{if} and \texttt{while} statements and who were not familiar with binary search and LBT. Eleven were from non-CS engineering departments.

The study consisted of three stages. In the first stage, the participants went through learning materials on the binary search from Khan Academy\footnote{https://www.khanacademy.org/computing/computer-science/algorithms/binary-search/a/binary-search} and solved two Parsons problems, a coding exercise on reordering code fragments~\cite{Denny2008}.
In the second stage, the participants received an introduction to the concepts of LBT, its expected learning benefits, and its procedures. Then, they were given a brief overview of the LBT activity they would be performing next. In the final stage, learners tutored the chatbot on how to write code for the two binary search problems from the prior stage. 
After the LBT activity, the participants completed an exit survey composed of questions on three themes: the perception of the chatbot as a learner, the self-perceived learning effects, and the familiarity with teaching a chatbot.

The participants interacted with a baseline LLM chatbot, \algobo{}, performing the role of a teachable agent.
We used GPT-4~\cite{OpenAI2023} as a backbone for \algobo{} and provided a system prompt (see Appendix~\ref{appendix:persona_setting}) that set a persona of a student and added predefined learning challenges it was running into to provide a more convincing teachable agent~\cite{Markel2023, park2023generative}.
Since we use the name ``\algobo{}'' again in our main system and evaluation, we use ``\algobobasic{}'' throughout this section to distinguish the two teachable agents we developed.

\subsection{Dialogue Analysis}

In addition to the comments from the exit survey, we also looked into the quality and conversational patterns of the dialogues between participants and \algobobasic{} by classifying messages into knowledge-telling and knowledge-building types.

Since previous taxonomies that categorize LBT dialogues~\cite{Walker2008, Roscoe2007, King1998} were not contextualized enough to programming tutoring, we decided to adapt the taxonomies and create a new taxonomy (Table~\ref{table:taxonomy}) specific to LBT in programming. 
We created our initial set of message types based on the prior taxonomies for general LBT dialogues~\cite{Walker2008, Roscoe2007, King1998} and categorizations of programming QA~\cite{Allamanis2013, Lee2022}.
Three authors took three iterations to annotate dialogues, resolve conflicts, and refine the taxonomy~\cite{Nickerson2013, Yang2023}. The authors finalized the taxonomy in the 2nd iteration (20 dialogues, 293 messages). The authors categorized the rest of the messages independently. The inter-rater reliability of the categorization was high; three authors achieved Krippendorff's alpha of 0.731 for the data in the last iteration (11 dialogues, 253 messages).

Our taxonomy has three main categories: instructions, prompting, and statements (see Table~\ref{table:taxonomy}). \textbf{Instruction} messages have content that asks the opponent (usually the tutee) to do specific actions, such as fixing code and attempting problem-solving after concept understanding. Instruction messages are mostly related to the proceeding of steps in teaching. \textbf{Prompting} messages have intentions for eliciting specific actions from the opponent. These include asking a tutee about a specific concept of interest, giving thought-provoking questions to encourage knowledge-building, and asking a tutor for help.
We designate Prompting-Thought-provoking to knowledge-building because such questions can signal collaborative knowledge-building where learners bring up exploratory questions and start knowledge-building discussions with agents.
\textbf{Statement} messages are utterances explaining one's knowledge and opinions. 
Among them, Statement-Elaboration and Statement-Sense-making are knowledge-building as they are the artifacts of new knowledge; this corresponds to Roscoe and Chi's classification of knowledge-building activity~\cite{roscoe2008tutor}.

\begin{table*}[htp]
\caption{Our taxonomy to classify the type of messages in LBT conversations with a teachable agent. The bold texts in the example column are the examples of respective message types. The types with * are knowledge-telling responses. The types with ** fall into knowledge-building responses.}
\label{table:taxonomy}
\Description{Our taxonomy to classify the type of messages in LBT conversations with a teachable agent. The bold texts in the example column are the examples of respective message types. There are four main categories: Instruction, Prompting, Statement, and Miscellaneous. Within each of these four main categories, there are sub-category conversations. Thought-provoking and Sense-making fall into knowledge-building responses whereas Hinting and comprehension fall into knowledge-telling responses.}
\begin{tabular}{ccll}
\hline
\textbf{Category} &
  \textbf{Sub Category} &
  \textbf{Explanation} &
  \textbf{Example} \\ \hline
\multirow{3}{*}{Instruction} &
  Fixing* &
  \begin{tabular}[c]{@{}l@{}}{[}Instruct to{]} correct specific \\ knowledge or part of code.\end{tabular} &
  \begin{tabular}[c]{@{}l@{}}\raisebox{-0.1em}{\twemoji[scale=0.4]{robot}} Tutee: Here is my code: \textless{}code\textgreater\\ \raisebox{-0.1em}{\twemoji[scale=0.4]{teacher}} \textbf{Tutor: Call the input() function twice so that} \\ \textbf{N and K are separately taken as input.} \end{tabular} \\ \cline{2-4} 
 &
  Commanding &
  \begin{tabular}[c]{@{}l@{}}{[}''{]} do simple actions \\ irrelevant to learning. (e.g., \\ simply combining code for a \\ submission).\end{tabular} &
  \begin{tabular}[c]{@{}l@{}}\raisebox{-0.1em}{\twemoji[scale=0.4]{robot}} Tutee: I have written the binary search function.\\ \raisebox{-0.1em}{\twemoji[scale=0.4]{teacher}} \textbf{Tutor: Now, write the entire Python code.}\end{tabular} \\ \cline{2-4} 
 &
  Encouraging &
  \begin{tabular}[c]{@{}l@{}}{[}''{]} retry a previous action \\ with emotional\\  encouragement.\end{tabular} &
  \begin{tabular}[c]{@{}l@{}}\raisebox{-0.1em}{\twemoji[scale=0.4]{teacher}} \textbf{Tutor: You are in the right direction.}\\ \textbf{Keep writing more code.}\end{tabular} \\ \hline
\multirow{5}{*}{Prompting} &
  \begin{tabular}[c]{@{}l@{}}Challenge\\ -finding\end{tabular} &
  \begin{tabular}[c]{@{}l@{}}{[}Prompt the opponent to{]} \\ explain his struggles to find \\ the parts to help.\end{tabular} &
  \begin{tabular}[c]{@{}l@{}}\raisebox{-0.1em}{\twemoji[scale=0.4]{teacher}} \textbf{Tutor: In which part are you facing difficulties?}\\ \raisebox{-0.1em}{\twemoji[scale=0.4]{robot}} Tutee: I am struggling with writing \\ the conditionals inside the while loop.\end{tabular} \\ \cline{2-4} 
 &
  Hinting* &
  \begin{tabular}[c]{@{}l@{}}{[}''{]} think about \\ alternative/specific \\ approaches.\end{tabular} &
  \begin{tabular}[c]{@{}l@{}}\raisebox{-0.1em}{\twemoji[scale=0.4]{robot}} Tutee: I could not complete this part of the code.\\ \raisebox{-0.1em}{\twemoji[scale=0.4]{teacher}} \textbf{Tutor: Well, have you considered the case} \\ \textbf{when the number is equal to K?}\end{tabular} \\ \cline{2-4} 
 &
  Checking &
  \begin{tabular}[c]{@{}l@{}}{[}''{]} show or self-explain his\\ understanding of specific \\ knowledge.\end{tabular} &
  \begin{tabular}[c]{@{}l@{}}\raisebox{-0.1em}{\twemoji[scale=0.4]{teacher}} \textbf{Tutor: Do you know what binary search is?}\\ \raisebox{-0.1em}{\twemoji[scale=0.4]{robot}} Tutee: Yes! Binary search is …\end{tabular} \\ \cline{2-4} 
 &
  \begin{tabular}[c]{@{}l@{}}Thought-\\ provoking**\end{tabular} &
  \begin{tabular}[c]{@{}l@{}}{[}''{]} elaborate previous \\ explanations or think beyond \\ the content of the given \\ learning materials.\end{tabular} &
  \begin{tabular}[c]{@{}l@{}}\raisebox{-0.1em}{\twemoji[scale=0.4]{teacher}} \textbf{Tutor: What will happen if we switch the min}\\ \textbf{/ max updating code?}\\ \raisebox{-0.1em}{\twemoji[scale=0.4]{robot}} Tutee: I haven’t thought about it. Will the loop\\ run forever?\end{tabular} \\ \cline{2-4} 
 &
  Asking for help &
  \begin{tabular}[c]{@{}l@{}}{[}''{]} analyze the speaker's \\ problem or give hints.\end{tabular} &
  \begin{tabular}[c]{@{}l@{}}\raisebox{-0.1em}{\twemoji[scale=0.4]{robot}} \textbf{Tutee: Could you help me with solving the}\\ \textbf{problem, please?}\end{tabular} \\ \hline
\multirow{5}{*}{Statement} &
  Comprehension* &
  \begin{tabular}[c]{@{}l@{}}{[}State one’s knowledge or \\ opinion by{]} paraphrasing /\\ copying / explaining the \\ learning material or the \\ opponent’s response.\end{tabular} &
  \begin{tabular}[c]{@{}l@{}}\raisebox{-0.1em}{\twemoji[scale=0.4]{teacher}} \textbf{Tutor: First, let's define the function called} \\ \textbf{binary\_search. In the while loop, …}\end{tabular} \\ \cline{2-4} 
 &
  Elaboration** &
  \begin{tabular}[c]{@{}l@{}}{[}''{]} providing extended \\ clarification or relevant \\ examples beyond the given \\ materials.\end{tabular} &
  \begin{tabular}[c]{@{}l@{}}\raisebox{-0.1em}{\twemoji[scale=0.4]{robot}} Tutee: Can you think of a real-life example\\ where we can use binary search?\\ \raisebox{-0.1em}{\twemoji[scale=0.4]{teacher}} \textbf{Tutor: I think we can use it for finding a word in}\\ \textbf{a dictionary where words are listed alphabetically.}\end{tabular} \\ \cline{2-4} 
 &
  Sense-making** &
  \begin{tabular}[c]{@{}l@{}}{[}''{]} realizing own errors /\\ misconceptions or making \\ new inferences / connections\\ to prior knowledge.\end{tabular} &
  \begin{tabular}[c]{@{}l@{}}\raisebox{-0.1em}{\twemoji[scale=0.4]{teacher}} Tutor: Can you take a closer look at the else\\ statement in your code?\\ \raisebox{-0.1em}{\twemoji[scale=0.4]{robot}} \textbf{Tutee: Ah, I got it. Let's modify the high value}\\ \textbf{to mid. Here is the corrected code.}\end{tabular} \\ \cline{2-4} 
 &
  \begin{tabular}[c]{@{}l@{}}Accepting \\ / Reject\end{tabular} &
  \begin{tabular}[c]{@{}l@{}}{[}''{]} agreeing or disagreeing\\  with the opponent’s response.\end{tabular} &
  \begin{tabular}[c]{@{}l@{}}\raisebox{-0.1em}{\twemoji[scale=0.4]{teacher}} Tutor: You should update line 24 to …\\ \raisebox{-0.1em}{\twemoji[scale=0.4]{robot}} \textbf{Tutee: I think that is a good idea.}\end{tabular} \\ \cline{2-4} 
 &
  Feedback &
  \begin{tabular}[c]{@{}l@{}}{[}''{]} responding to the \\ opponent’s action or thought.\end{tabular} &
  \textbf{\raisebox{-0.1em}{\twemoji[scale=0.4]{teacher}} Tutor: Yes, that is exactly right.} \\ \hline
Miscellaneous &
   &
  \begin{tabular}[c]{@{}l@{}}Greetings/goodbyes, social \\ expressions\end{tabular} &
  \begin{tabular}[c]{@{}l@{}}\raisebox{-0.1em}{\twemoji[scale=0.4]{teacher}} Tutor: Do you have any questions?\\ \raisebox{-0.1em}{\twemoji[scale=0.4]{robot}} \textbf{Tutee: No, thank you so much for your}\\ \textbf{guidance so far!}\end{tabular} \\ \hline
\end{tabular}%
\end{table*}

\subsection{Findings from Participants' Comments and Dialogue Analysis}
We found that an LLM chatbot can serve as a teachable agent for rudimentary LBT. Participants were positive about teaching an LLM chatbot and felt it helped them reorganize and recall their knowledge. However, our dialogue analysis and in-depth survey responses revealed that the LLM chatbot fell short of adequately supporting learners' knowledge-building process. 

\medskip
\noindent
\textbf{\algobobasic{} was perceived as an overly competent learner due to its extensive prior knowledge and self-correcting behavior.}
Participants highly appreciated \algobobasic{}'s ability to ``talk like a real person and ask specific questions'' (P14) for simulating a learner. However, two-thirds of participants commented that they experienced awkwardness due to \algobobasic{}'s competence. 
\algobobasic{} initially started a conversation by asking for help. However, after a few chats, \algobobasic{} provided competent responses too quickly, which did not reflect a novice learner's learning process. P5 remarked, ``I explained it very simply, but he understood it very well... He is so much smarter than me. He seems to fill by himself the knowledge even I am not sure about.''
\algobobasic{}'s adeptness in code writing and explanation also limited conversational patterns and confused learners about their roles. \algobobasic{} made twice as many knowledge statements (i.e., Statement-Comprehension) as participants did, taking away the chance for learners to self-explain and teach (see the Statement-Comprehension row in Table~\ref{table:formative_distribution}). P7 stated, ``\algobobasic{} was like a teaching assistant testing a student's ability, rather than a student struggling with binary search problems.'' 
Participants responded that they would have liked to see more student-like interactions from \algobobasic{} such as ``asking more proactive questions'' (P1) and ``making mistakes and requesting tutors for an elaborated explanation'' (P5).

\medskip
\noindent
\textbf{Dialogues between tutors and \algobobasic{} were limited to only knowledge-telling.}
Participants valued retelling of their knowledge---``Writing down knowledge was very helpful in organizing knowledge. If you want to teach someone, you should create steps in your head, and this process helped a lot'' (P1). However, their learning was limited to knowledge-telling; out of 546 messages, we could observe 244 knowledge-telling messages but only 15 knowledge-building utterances (Table~\ref{table:formative_distribution}). Despite helping reorganize knowledge, self-explanations did not lead to building new knowledge beyond what they previously knew---``I didn't discover anything new because I explained what I had already learned'' (P4).
Furthermore, tutors' self-explanations were often undeveloped because \algobobasic{} did not ask questions on participants' vague explanations, and \algobobasic{} performed well. For example, P15 answered \algobobasic{}'s question on why the input array needs to be sorted: ``Sorted arrays reduce the number of calculations and maximize the effectiveness of binary search.'' Despite the lack of detailed reasoning (e.g., ``how'' and ``why''), \algobobasic{} accepted the explanation and moved on to the next question.

\begin{table}[ht]
\caption{The distribution of message categories for 31 dialogues from the formative study. The types with $*$ are knowledge-telling messages. The types with $**$ fall into knowledge-building messages.}
\label{table:formative_distribution}
\Description{The distribution of message categories for 31 dialogues from the formative study. Regarding knowledge-telling responses, we observed a total of 134 responses from the Tutee and 110 from the Tutor (out of 244). Compared to this, we observe only 12 and 3 knowledge-building responses from the tutee and tutor.}
\begin{tabular}{lllll}
\hline
\textbf{Category}                     & \textbf{Sub Category} & \textbf{Tutee} & \textbf{Tutor} & \textbf{Total} \\ \hline
\multirow{3}{*}{Instruction} & Fixing*               & 0              & 37             & 37             \\
                             & Commanding            & 0              & 65             & 65             \\
                             & Encouragement         & 0              & 1              & 1              \\ \hline
\multirow{5}{*}{Prompting}   & Challenge-finding     & 0              & 18             & 18             \\
                             & Hinting*              & 1              & 12             & 13             \\
                             & Checking              & 1              & 31             & 32             \\
                             & Thought-provoking**   & 0              & 1              & 1              \\
                             & Asking-for-help       & 91             & 0              & 91             \\ \hline
\multirow{5}{*}{Statement}   & Comprehension*        & 133            & 61             & 194            \\
                             & Elaboration**         & 0              & 1              & 1              \\
                             & Sense-making**        & 12             & 1              & 13             \\
                             & Accepting             & 35             & 4              & 39             \\
                             & Feedback              & 0              & 17             & 17             \\ \hline
\multicolumn{2}{l}{Miscellaneous}                    & 19             & 5              & 24             \\ \hline
\multicolumn{2}{l}{\textbf{Total}}                            & 292            & 254            & 546            \\ \hline
\multicolumn{2}{l}{Knowledge-telling}                & 134            & 110            & 244            \\
\multicolumn{2}{l}{Knowledge-building}               & 12             & 3              & 15             \\ \hline
\end{tabular}
\end{table}

\medskip
\noindent
\textbf{Participants carried out antipatterns of LBT and sought feedback.}
Participants remarked tutoring through natural language communication was intuitive and familiar because it resembled tutoring humans, and they could apply the same teaching methods to \algobobasic{}.
However, some participants wanted to see better methods for them to teach \algobobasic{} (P9) and a method to review their learning process (P15). P15 said, ``I was able to see that my teaching skills worked, but the reflection [on my tutoring session] left a lot to be desired due to the lack of feedback on my teaching method'' (P15).
While analyzing participants’ dialogues, we found common conversational antipatterns that may restrain the benefits of LBT.
The first pattern was \textbf{Commanding}, in which participants repetitively gave \algobobasic{} specific instructions for writing and correcting code (Appendix~\ref{appendix:antipatterns} (A)). This pattern lacks an explanation of ``why'' and ``how'' which can prompt learners to go beyond recalling facts (i.e., knowledge-telling).
The second pattern was \textbf{Spoon-feeding}, in which participants give away knowledge without questions to check or prompt a tutee’s understanding (Appendix~\ref{appendix:antipatterns} (B)). 
Rather than passive explanations, learners can actively construct new knowledge by making thinking questions for their tutees, taking the benefits of having interactive agents.
The last pattern was \textbf{Under-teaching}, in which \algobobasic{} progressed in problem-solving but did knowledge-telling only because learners did not attempt to teach and develop further knowledge. (Appendix~\ref{appendix:antipatterns} (C)).

\section{Design Goals}

The findings from our formative study showed that LLMs could serve as a rudimentary teachable agent for LBT. However, we also confirmed the need to improve LLM chatbots' imitation of help-seeking tutees, promote the knowledge-building of learners, and support learners' metacognition in teaching. Based on the insights, we set three design goals.

\medskip
\noindent
\textbf{D1. Design teachable agents that can simulate misconceptions and gradual learning curves.}
We found that the pre-trained knowledge and self-correcting behavior of LLMs made \algobo{} feel less like a tutee and prevented tutors from learning by identifying tutees' errors and enlightening them with elaborate explanations~\cite{Walker2008}. To reduce undesirable competence, we need to control the prior knowledge of LLMs and make them show persistent misconception and unawareness in their responses until they receive pedagogical aid.

\medskip
\noindent
\textbf{D2. Develop conversations through teachable agents' elaboration questions.}
\algobo{} rarely asked follow-ups or thought-provoking questions (Table~\ref{table:formative_distribution}) that can elicit learners' knowledge-building through elaboration and sense-making~\cite{King1998, Shahriar2021}. Instead of keeping \algobo{} as a passive tutee throughout a conversation, switching \algobo{} into an active questioner may scaffold learners to build knowledge step by step. For example, \algobo{} can start as a help-seeking tutee asking learners to self-explain basic concepts, and it can turn into a thought-provoking Socratic questioner intermittently.

\medskip
\noindent
\textbf{D3. Provide learners with metacognitive feedback on their teaching methods during a conversation.}
Besides the scaffolding from the tutee side, learners can actively improve the quality of their LBT by leading constructive tutoring and refraining from the three antipatterns. In our formative study, however, participants found guidance and reflection on their teaching techniques lacking. Metacognitive feedback on teaching during LBT can help learners recognize how to improve their teaching methods on the spot and refine conversations throughout.

\begin{figure*}[ht]
\includegraphics[width=\textwidth]{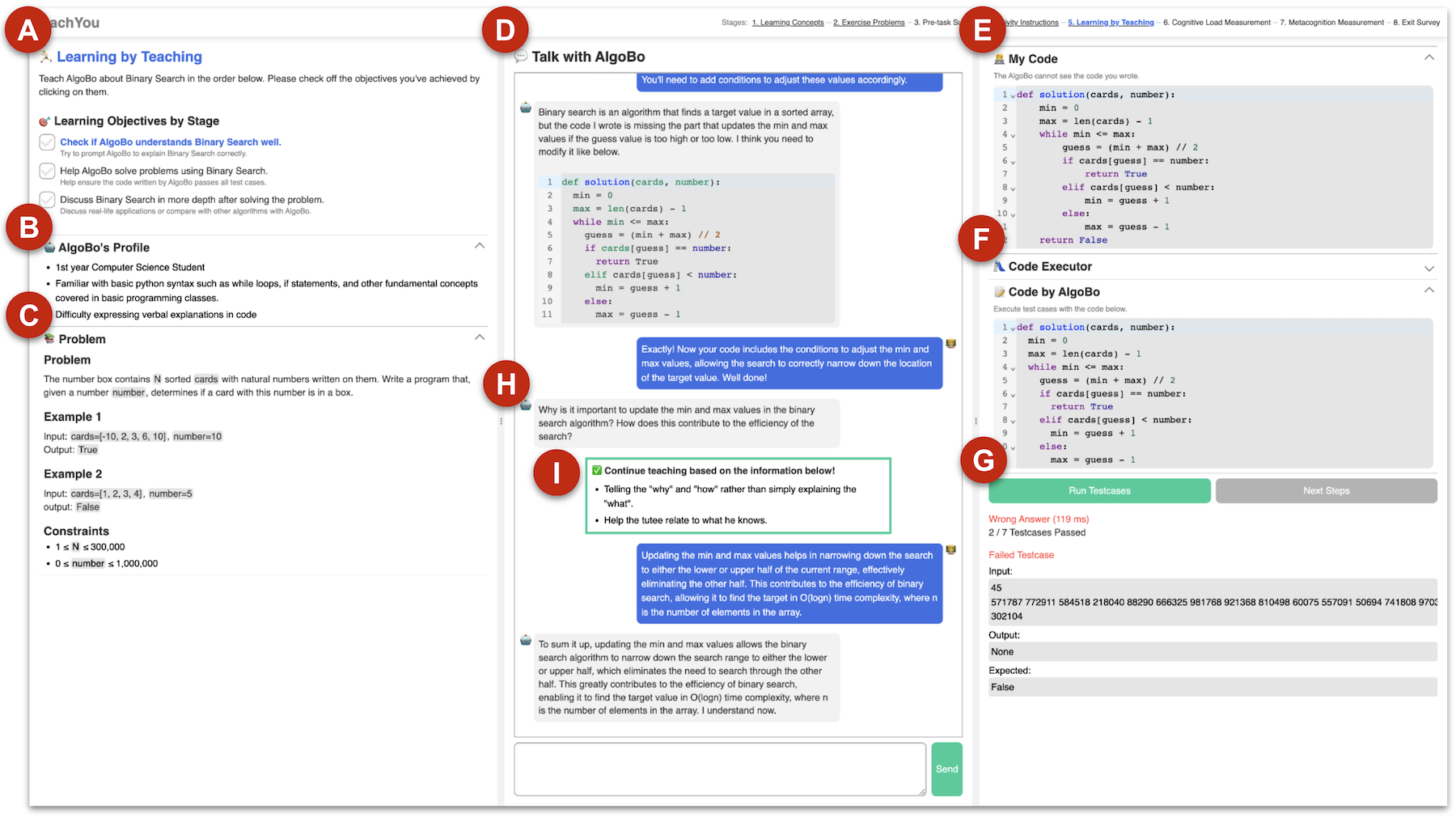}
    \caption{To the left, the 3 learning objectives they need to reach (A), learners can see \algobo{}'s profile (B), and the questions they need to help \algobo{} solve (C). To the right, they can see the code they submitted (E), a code playground (F), and the code that \algobo{} write (G). When \algobo{} wrote code, participants could click on the ``run test cases'' and run \algobo{}'s code. In the middle (D), learners use a typical chat interface to teach \algobo{} while receiving questions (H) and guidance from Teaching Helper (I)}
    \label{fig:interface_teach}
    \Description{A screenshot of the TeachYou Interface. To the left (A, B, C), learners can see AlgoBo's profile, the questions they need to help AlgoBo solve, and the 3 learning objectives they need to reach. To the right (E, F, G), they can see the code they submitted, a code playground, and the code that AlgoBo writes. When AlgoBo wrote code, participants could click on the “run test cases” and run AlgoBo's code. In the middle (D), learners use a typical chat interface to teach AlgoBo while receiving questions (H) and guidance from Teaching Helper (I).}
\end{figure*}

\section{System} \label{section:system}

We present \sysname{}, an LBT system featuring \algobo{}, an LLM-based teachable agent. \algobo{} gets help from learners to solve introductory algorithm problems while asking thought-provoking questions that encourage the learners to expand their knowledge beyond their current level. Through the system, we propose 1) a new LLM prompting pipeline for simulating tutees of specific levels of knowledge and misconceptions and 2) a learning environment for learners to effectively conduct LBT.

Programming and algorithm learners can use \sysname{} to review what they learned and explore further knowledge through an engaging and interactive LBT activity. We designed an interface (Fig.~\ref{fig:interface_teach}) to help learners conduct the activity. Throughout the LBT activity, learners should achieve three sequential objectives in teaching \algobo{} (Fig.~\ref{fig:interface_teach} A). The objectives correspond to the three levels in Bloom's taxonomy (Understand-Apply-Analyze)~\cite{Bloom1968, Krathwohl2002}; learners first check if \algobo{} correctly understand the concept of interest; then, learners help \algobo{} apply the concept to solve a problem; lastly, learners and \algobo{} discuss real-life use cases and other related topics.
Learners can refer to the profile of \algobo{} to set their attitude and expectations (Fig.~\ref{fig:interface_teach} B). We set the persona of \algobo{} as a 2nd-year high school student, as opposed to a 1st-year CS student in the formative study, to match the slow learning behavior and to encourage learners' patience in teaching. Learners use a typical chat interface to teach \algobo{} (Fig.~\ref{fig:interface_teach} D) and have access to teaching support (Fig.~\ref{fig:interface_teach} C, E, F, G). While tutoring, learners receive why questions and thought-provoking questions from \algobo{}, helping them self-explain the rationale behind their instructions and expand their knowledge (Fig.~\ref{fig:interface_teach} H). \sysname{} also provides feedback on learners' teaching methods and suggestions for improvement to encourage reflection on teaching (Fig.~\ref{fig:interface_teach} I).

In order to support the aforementioned learning scenario and the three design goals effectively, we implemented three system components: First, we implemented the Reflect-Respond prompting pipeline for a teachable agent to simulate student-like learning behavior. Secondly, within a conversation, our teachable agent shifts between help-receiver and questioner modes in every third conversation turn, eliciting self-explanation and knowledge construction, respectively. Lastly, the learning environment analyzes the dialogue between learners and \algobo{} and provides feedback on their tutoring methods to promote metacognition.

\begin{figure*}[ht]
\includegraphics[width=\textwidth]{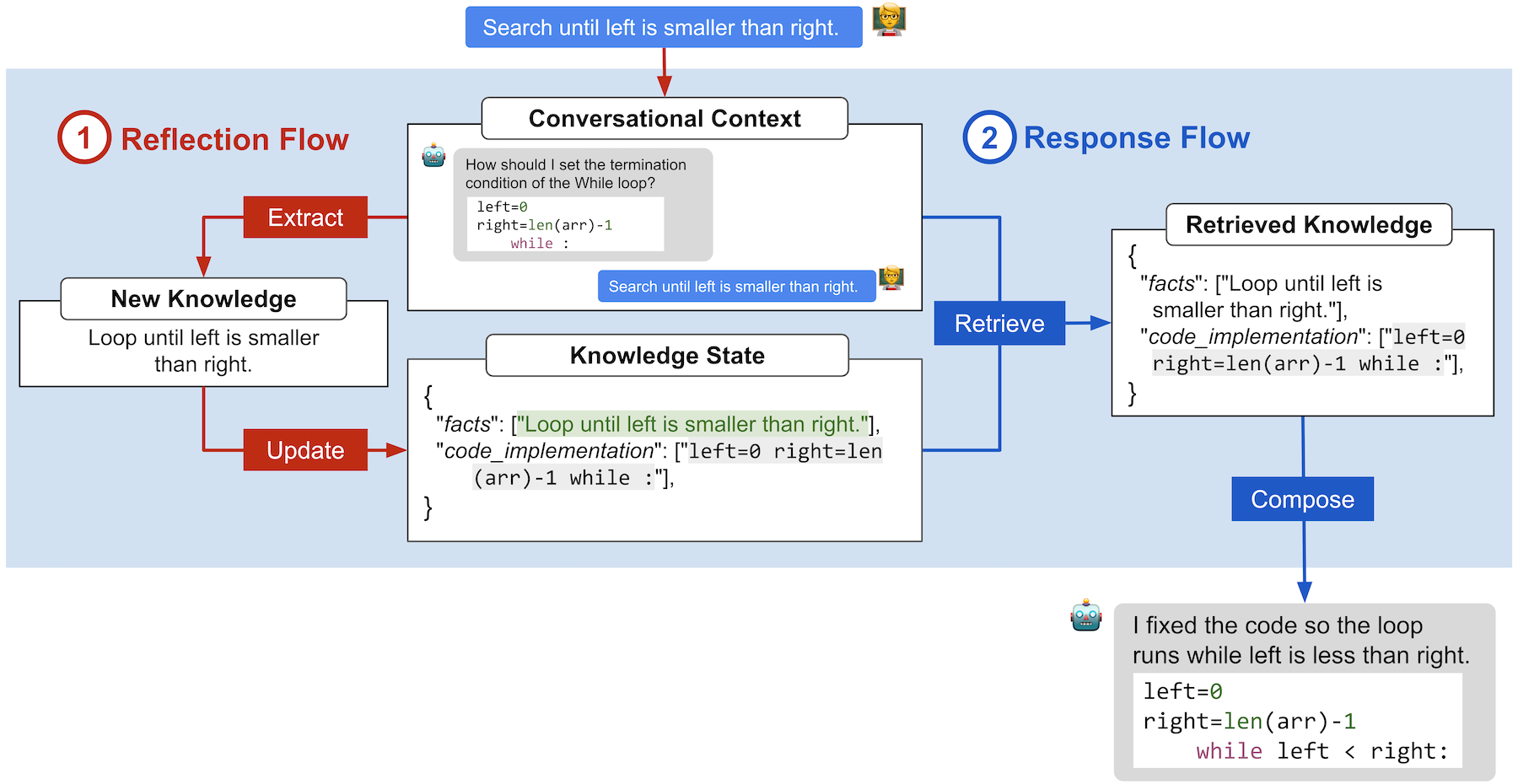}
    \caption{The overview of the Reflect-Respond prompting pipeline for simulating knowledge learning of \algobo{} and examples for each component. From the recent conversation, \algobo{} \textit{extracts} new knowledge of the while loop condition and \textit{update} its knowledge state (colored in green). Then, \algobo{} \textit{retrieves} knowledge relevant to while loops and \textit{composes} a response that fills its knowledge gap.}
    \label{fig:prompting_pipeline}
    \Description{Response-Reflection Prompting Pipeline. In the Reflection flow, new knowledge extracted from the conversational context is updated into the knowledge state. In the Response flow, AlgoBo retrieves knowledge from its knowledge state and composes responses following the conversational context.}
\end{figure*}

\subsection{Reflect-Respond prompting pipeline to simulate knowledge learning} \label{section:system-pipeline}

From our observations and user comments in the formative study, we considered three properties crucial for LLM-based teachable agents to simulate knowledge learning---reconfigurability, persistence, and adaptability. \textbf{Reconfigurability} refers to how precisely we can set an agent's performance in question-answering and problem-solving. Reconfigurable agents allow us to build tutees with specific misconceptions and help design tutoring scenarios. 
\textbf{Persistence} examines how the knowledge level of a teachable agent on a target topic is maintained consistently throughout the agent interaction.
Persistent agents do not self-correct their misconceptions and show constant question-answering performance unless being taught; their knowledge level should also not be susceptible to messages irrelevant to the knowledge of interest (e.g., jokes).
\textbf{Adaptability} measures how well the agent updates its knowledge as it acquires new information from tutors in conversations. Adaptability allows a teachable agent to improve its knowledge level and remember what tutors have taught.

To achieve these properties, we introduce a prompting pipeline that leverages a knowledge state and two information flow mechanisms: Reflection and Response (Fig.~\ref{fig:prompting_pipeline}). A \textbf{knowledge state} is a store representing the knowledge \algobo{} currently holds. It is comparable to a schema, a cognitive unit of knowledge for problem-solving~\cite{Sweller2011}. \algobo{}'s responses are constrained by its knowledge state, and we update the knowledge state consistently throughout a conversation. 
Knowledge states link to reconfigurability; if we leave them empty, agents will show zero-knowledge behavior; if we add incorrect or correct information, agents will show misconceptions or prescribed knowledge levels, respectively.
\textbf{Reflection} is a flow dedicated to the update of knowledge states. In the Reflection flow, we use an LLM to extract new information from the latest conversations (i.e., the last three messages) and then update knowledge states by adding or correcting information. After Reflection, the \textbf{Response} flow occurs; we first use the LLM to retrieve information relevant to the conversational context from the current knowledge state and then compose a response by only combining the retrieved knowledge. If a knowledge state does not have relevant information and nothing is retrieved, \algobo{} responds: ``I'm not sure how to do that. Could you explain it to me?'' Reflection and Response connect to the persistence and adaptability of agents as the flows control the retrieval and update of knowledge states in reaction to external stimuli.

We implemented the knowledge state as a JSON object with two attributes: facts and code\textunderscore implementation. \texttt{\textbf{Facts}} store natural language explanations of the target knowledge. \texttt{\textbf{Code\textunderscore implementation}} contains code snippets (see Fig.~\ref{fig:prompting_pipeline} knowledge state). The four operations in the pipeline are implemented with GPT-4 as a base LLM. We adopted well-known prompting engineering techniques, such as AI chains~\cite{wu2022ai}, few-shot prompts~\cite{Brown2020, Touvron2023}, persona setting~\cite{Markel2023, park2023generative}, and code prompts~\cite{Zhang2023, Gao2023} (see Appendix~\ref{appendix:reflect_respond}). We note that our implementation is one possible instance of our proposed pipeline, and it can improve further with better LLMs and algorithms for the operations. For example, we can represent knowledge states with more complex tree structures~\cite{Long2023, Yao2023}, and the update operation may use the Least Recently Used algorithm~\cite{ONeil1993} to simulate a fixed-size knowledge capacity.
We chose GPT-4 for operating our pipeline because it can effectively process the contextual information in conversations compared to other approaches.

\begin{figure*}[ht]
\includegraphics[width=\textwidth]{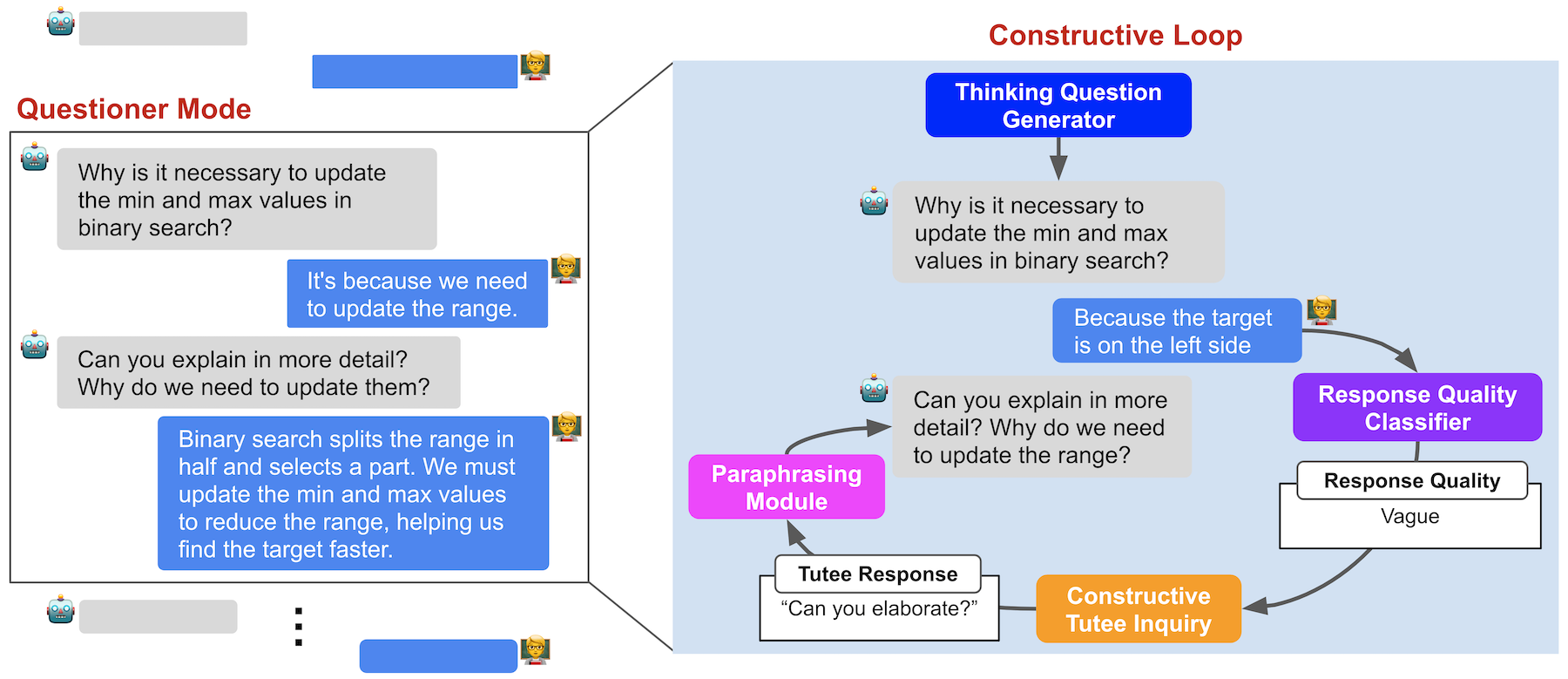}
    \caption{\algobo{} shifts its mode in every three messages. When \algobo{} is in the questioner mode, it keeps asking follow-up questions until receiving a satisfactory response (constructive loop)}
    \label{fig:mode_shifting}
    \Description{AlgoBo shifts its mode every three messages during the conversation. When AlgoBo is the questioner, it runs through a constructive loop that generates thinking questions, assesses tutors' response quality, and creates follow-up questions. The constructive loop also evaluates the quality of the response from the learner and continues to ask questions when the answer is insufficient.}
\end{figure*}

\subsection{\algobo{}'s Mode-shifting to develop constructive LBT dialogues}

Beyond telling knowledge to \algobo{}, we aim to push learners to answer thought-provoking questions and build new knowledge. From the formative study, we observed that entrusting LLMs entirely with making conversations did not result in desirable knowledge-building patterns (e.g., question-answering on ``why'' and ``how'') spontaneously. 
Prior research has shown that guiding questions from conversational agents are effective for improving learners' knowledge-building and divergent thinking~\cite{Shahriar2021, abdelghani2022conversational}.
To control conversation flows while giving learners the freedom to steer them, we introduce \textbf{Mode-shifting}, in which \algobo{} periodically shifts between two modes: In the help-receiver mode, \algobo{} passively learns from tutors and prompts their self-explanations; in the questioner mode, \algobo{} asks thought-provoking questions to stimulate the knowledge-building of learners.

We use Mode-shifting to make conversation flows dynamic and engaging. In every third message, \algobo{} shifts to the questioner mode and asks a thinking question. The thinking question differs by the phase of the activity (Fig.~\ref{fig:interface_teach} A). While learners teach \algobo{} about concepts and code implementation (i.e., the first and second objectives), \algobo{} asks ``why'' questions in response to learners' instructions and explanations. During the discussion phase (i.e., the third objective), \algobo{} brings up related algorithms or real-life examples and asks ``how'' questions to prompt learners to explain and connect to what they have learned. After the thinking questions, the conversation goes through a constructive loop, in which learners receive follow-up questions from \algobo{} until they answer the question in depth with a valid example. 
When \algobo{} assesses learners' responses as satisfactory, \algobo{} summarizes them and shifts back to the receiver mode.
The period of Mode-shifting (every three messages) is heuristic; from our pilot studies, we found that such frequency was optimal for prompting elaboration while not distracting tutors too much.

To incorporate Mode-shifting to LBT dialogues, we implemented four components (Fig.~\ref{fig:mode_shifting}). The \textbf{Thinking Question Generator} is a module that uses GPT-4 to produce thought-provoking questions related to the current conversation. For managing the constructive loop, we followed the protocol of the constructive tutee inquiry in Shahriar et al.'s work~\cite{Shahriar2021} and adapted it to LLM. We used the formative study dialogues with response quality annotations to train the \textbf{Response Quality Classifier}. The classifier assesses every learner's responses in the loop and determines \algobo{}'s follow-up question as pre-defined in \textbf{Constructive Tutee Inquiry} protocol~\cite{Shahriar2021}. Lastly, the \textbf{Paraphrasing Module} adjusts the fixed question to the conversational context. All the prompts used for Mode-shifting are available in Appendix~\ref{appendix:mode_shifting}.

\begin{figure*}[ht]
\includegraphics[width=\textwidth]{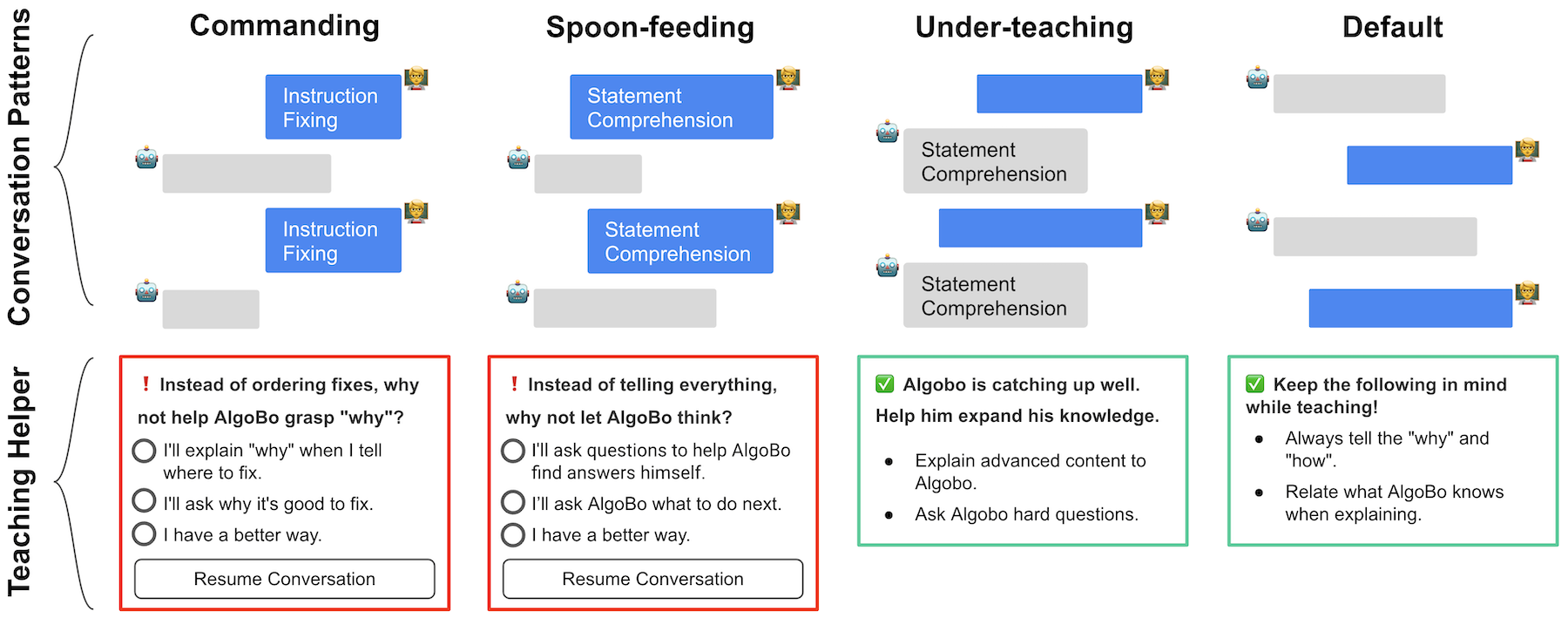}
    \caption{The four Teaching Helper messages and corresponding suggestions that appear depending on the conversational patterns.}
    \label{fig:teaching_helper}
    \Description{The four teaching helper messages and corresponding suggestions that appear depending on the conversational patterns. The teaching helper detects conversational patterns of "commanding", "spoon-feeding" "under-teaching" and "default" and represents feedback and suggestions.}
\end{figure*}

\subsection{Teaching Helper for Metacognitive Guidance}

Throughout our formative study, we found conversational antipatterns that hindered effective LBT. To prevent this, \sysname{} provides metacognitive feedback throughout the conversation to help learners reflect on the overall teaching session and offer overarching guidance on steering the discussion. \sysname{} presents the feedback through \textbf{Teaching Helper,} a red or green text box that appears below the messages (see Fig.~\ref{fig:interface_teach} I). Teaching Helper provides information on the current problems with the teaching method and elaborates on what learners could do to improve their conversation.

\sysname{} provides four Teaching Helper messages, depending on detected conversational patterns (Fig.~\ref{fig:teaching_helper}). For the \textbf{Commanding} and \textbf{Spoon-feeding} patterns, in which learners should correct their teaching styles, \sysname{} shows feedback messages in red boxes. To ensure learners read feedback, we interrupt the conversation with \algobo{} until learners explicitly decide how to act. The send button in the chat interface is blocked until learners pick an option among the possible teaching methods to address the issue. We chose to give learners multiple suggestions and let them choose their teaching method, instead of giving specific guidance to follow because the active selection of teaching methods may improve learners' recognition and autonomy in tutoring~\cite{wu2011allowing}. For the \textbf{Under-teaching} pattern and default cases where no antipattern is found, \sysname{} shows messages in a green box. The messages either encourage learners to go beyond the current learning topic or give general tips for good answering and questioning~\cite{King1997, King1994, ausubel1962subsumption}. Teaching Helper messages and learners' selection remain in conversations for revisiting. To avoid frequent interruptions and distractions from Teaching Helper, we restrict the presentation of the feedback to every six messages.

Teaching Helper is powered by a message-type classifier for detecting conversational patterns. We used the dialogue dataset from the formative study to fine-tune the GPT-3 davinci model. For training, we used 438 messages, and the classifier achieved an accuracy of 71.3\% for the remaining 108 messages in a validation test.

\section{Evaluation}

We evaluated the efficacy of \sysname{} for eliciting knowledge-building experiences in LBT. This overarching goal broke down into three main research questions:

\smallskip
\begin{enumerate}[label=\textbf{RQ\arabic*.}]
    \item How well does the Reflect-Respond pipeline simulate misconceptions and knowledge development?
    \item How does \sysname{} help elicit knowledge-building in LBT conversations?
    \item How does \sysname{} improve learners' metacognition about tutoring?
\end{enumerate}

The evaluation was divided into two parts. The initial phase was a technical evaluation that aimed to assess if the Reflect-Respond pipeline could induce a teachable agent to produce responses that were reconfigurable, persistent, and adaptive throughout the course of a conversation (RQ1). In the second phase, we ran a user study to examine the effects of Mode-shifting and Teaching Helper on learning experiences (RQ2 and RQ3).

\subsection{Technical Evaluation of the Reflect-Respond Pipeline}

As defined in Section~\ref{section:system-pipeline}, we evaluated the responses generated by our prompting pipeline along three axes---reconfigurability, persistence, and adaptability (RQ1).

\subsubsection{\textbf{Evaluating \algobo{}'s Knowledge Level}}
\begin{figure*}[t]
\includegraphics[width=400pt]{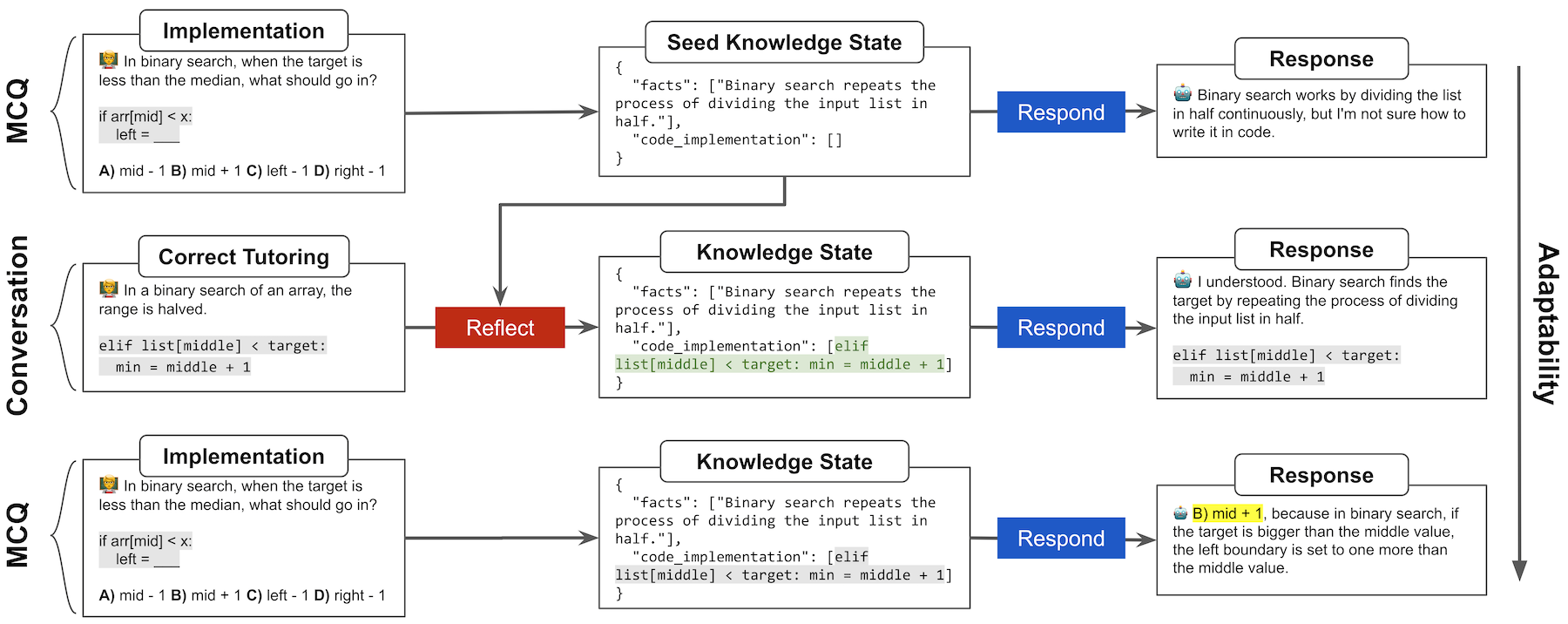}
    \caption{The process of measuring adaptability for correct tutoring with an Implementation problem and \emph{State 2} as a seed knowledge state. The evaluations were performed in Korean to ensure compatibility with the main study conditions.}
    \label{fig:adaptability}
    \Description{The process of measuring adaptability for correct tutoring with an Implementation problem and State 2 as a seed knowledge state. A diagram explaining how adaptability is measured as the knowledge state passes through different conversations. For example, when correct tutoring has occurred, Algobo will update its knowledge state and produce responses accordingly by retrieving that knowledge within the knowledge state.}
\end{figure*}

We evaluated \algobo{}'s knowledge level by observing its performance on Multiple Choice Questions (MCQs) under varying knowledge states and conversational interactions.
Although our target learning setting does not involve MCQs, we chose MCQs to follow prior research on assessing LLMs' performance ~\cite{Robinson2022, Gao2023} and collect clear-cut results.
A well-configured teachable agent should only perform well on the MCQ questions that can be answered with the given information in the knowledge state.
To confirm that \algobo{} was answering questions based on its knowledge state only and not picking random choices, we also prompted \algobo{} to explain why it chose the answers (Fig.~\ref{fig:adaptability}).

\subsubsection{\textbf{Procedure and Setup}}
We measured \algobo{}'s MCQ performance on three different algorithmic topics. For each topic, we created a set of nine MCQs (Appendix~\ref{appendix:mcq}). Within each set, we had three MCQs for each of Bloom's taxonomy categories: Understanding, Implementation (Applying), and Analysis~\cite{Bloom1968, Krathwohl2002}. 
Understanding questions consisted of questions on factual concepts, Implementation questions were about filling in the blanks in code, and Analysis questions were about the time complexity calculation and comparison to other relevant algorithms.
\algobo{} was evaluated with 4 different knowledge states (Appendix~\ref{appendix:seed_knowledge_states}) and conversational inputs (Appendix~\ref{appendix:random_conversation}, \ref{appendix:correct_tutoring}, \ref{appendix:incorrect_tutoring}). 

For reconfigurability (i.e., the change in knowledge level with different knowledge states), we prepared four seed knowledge states (Appendix~\ref{appendix:seed_knowledge_states}).
\emph{State 1} was empty to simulate zero knowledge. \emph{State 2} had an explanation of a topic algorithm in only \texttt{facts} to observe if \algobo{} knows only the given information. \emph{State 3} had the same explanation plus a piece of incorrect code in \texttt{code\textunderscore implementation} to check if \algobo{} shows the prescribed misconception. \emph{State 4} had the correct explanation and code to see if \algobo{} becomes competent with more input knowledge.
We prompted \algobo{} to solve MCQs with different seed knowledge states and compared the scores between the states. To prevent \algobo{} from storing knowledge learned from the MCQs into its knowledge state, we turned off the Reflection flow.

For assessing persistence (i.e., the invariance of knowledge level under no stimuli), we ran random conversations on \textit{State 2}. In the random conversations, \algobo{} was taught irrelevant information, such as arithmetic, translation, and classification, thus leading \algobo{} to save random information in its knowledge state~\cite{Shi2023}. We turned on the Reflection flow so that \algobo{} could update its initial knowledge state. We prompted \algobo{} to solve the same MCQs again and compared the difference between the first and second scores.

For adaptability (i.e., the acceptance of new knowledge), we considered two cases---Correct and Incorrect tutoring. The performance gap between Correct and Incorrect tutoring is crucial to check an agent's suitability for LBT because a teachable agent should not excel when learners give incorrect or incomplete instruction. Tutoring conversations taught three pieces of information that mapped to Understanding, Implementation, and Analysis of concepts. Correct tutoring gave \algobo{} correct factual information, whereas Incorrect tutoring provided false information.
We ran Correct and Incorrect tutoring separately on \algobo{} configured with \emph{State 2} and compared the differences between the MCQ scores at the start and after each type of tutoring.

We used the GPT-4-0613 model with 0 temperature throughout the evaluation. For a more comprehensive understanding of the four knowledge states and the materials used in the evaluation, please refer to Appendix~\ref{appendix:technical_evaluation}.

\subsection{Technical Evaluation Result}
We report the result of the technical evaluation on reconfigurability, persistence, and adaptability. We observed a small variation in the MCQ score even for the same inputs, knowledge states, and LLM model, perhaps due to the randomness inherent in the model and the running hardware\footnote{https://community.openai.com/t/a-question-on-determinism/8185/2}. 
We repeated the entire measurement five times for each input configuration and reported the median score for each question to handle variances of \algobo{}'s response.
The variance in score was mild; on average, \algobo{} produced a different response once in five repetitions. For the detailed report of the variance, refer to Appendix~\ref{appendix:variance}.

\medskip
\noindent
\textbf{[RQ1] Response flow can effectively reconfigure the knowledge level of \algobo{}.}
\newline
As expected, \algobo{} got all MCQs wrong when its knowledge state was empty (see \emph{State 1} in Table~\ref{table:reconfigurability}). When the knowledge state had only \texttt{facts} information (\emph{State 2}), \algobo{} could solve some conceptual (Understanding and Analysis) questions but none of the Implementation questions.
This shows that separating the knowledge state by knowledge types (\texttt{facts} and \texttt{code\textunderscore implementation}) can help configure knowledge more precisely by types. 
When the knowledge state contained code information, \algobo{} started to solve Implementation questions and achieved higher scores when given correct code (\emph{State 4}), compared to incorrect code (\emph{State 3}). \algobo{} followed what was written in its knowledge state (\emph{State 3}) exactly and produced wrong code and answers.

\medskip
\noindent
\textbf{[RQ1] Reflect-Respond makes \algobo{} produce responses persistent to knowledge states.}
\newline
The random conversation had a mild effect on the MCQ scores (compare the difference between the ``At the start'' and ``After random conversation'' columns in Table~\ref{table:persistence_and_adaptability}). While random conversation changed the scores of conceptual questions, the scores of Implementation questions stayed the same. We analyzed the inputs and outputs of the Respond flow in depth and found that \algobo{} retrieved algorithm-related knowledge that was missing in the first MCQ solving. Considering our LLM prompt for Retrieve (Appendix~\ref{appendix:reflect_respond} Retrieve), we contemplate that the population of more information in knowledge states might increase the relative importance of relevant knowledge in retrieval and help \algobo{} solve questions correctly. In other words, the scores after the random conversation are closer to what the \algobo{} should have received initially.
To see how far the population of random information increases knowledge level, we ran another random conversation and checked MCQ scores (see Table~\ref{table:supplementary_technical_evaluation} Scenario 1). The second random conversation contained random statements on arithmetic, translation, and classification (Appendix~\ref{appendix:random_conversation}). We did not observe any significant increase in the scores, confirming that the persistence of knowledge levels is robust regardless of the length of random conversations.

\medskip
\noindent
\textbf{[RQ1] Reflect-Respond allows \algobo{} to adapt knowledge states from conversations.}
\newline
Correct tutoring significantly improved MCQ scores (compare the difference between the ``At the start'' and ``After Correct tutoring'' columns in Table~\ref{table:persistence_and_adaptability}) across Understanding, Implementation, and Analysis.
Conversely, the Incorrect tutoring improved MCQ scores (compare ``At the start'' and ``After Incorrect tutoring'' columns in Table~\ref{table:persistence_and_adaptability}), but not as much as the Correct tutoring did. 
For example, the incorrect code information \texttt{``if arr[mid] > x: low = mid + 1 elif arr[mid] < x: high = mid - 1''} given in Incorrect tutoring stimulated \algobo{} to infer that ``Binary search returns a value indicating not found if the target is not in the list'' and solve one of the Implementation questions.
This result shows that partially correct information in the Incorrect tutoring could help solve problems, suggesting the need for more precise control in writing the knowledge states.

To investigate if \algobo{} prefers correct information to incorrect information and if incoming knowledge tends to overwrite pre-existing knowledge, we ran two scenarios in which \algobo{} received Correct and Incorrect tutoring in a sequence (see Table~\ref{table:supplementary_technical_evaluation} Scenario 3 and 4). The result shows that \algobo{} tends to keep correct information and remove incorrect ones (check Table~\ref{table:scenario3_update_log} last knowledge state).
We surmise that \algobo{} dropped conflicting information to keep its knowledge state short as instructed in the Update prompt (Appendix~\ref{appendix:reflect_respond}).
We also speculate that LLMs prefer to follow widespread (often factual) knowledge compared to incorrect information as the way it is trained~\cite{nakano2021webgpt}.

\begin{table*}[htp]
\centering
\caption{The number of correct MCQs for different knowledge states. \emph{State 1} is an empty knowledge state; \emph{State 2} has facts only; \emph{State 3} has facts with wrong code; \emph{State 4} has facts and correct code. ``U'', ``I'', and ``A'' stand for Understanding, Implementation, and Analysis question types. The number in each cell ranges from zero to three as there were three MCQs for a particular question type.}
\label{table:reconfigurability}
\Description{The table illustrates how scores from MCQs vary across four distinct knowledge states for three algorithms: Binary search, Merge sort, and Breadth-first search. These knowledge states range from an empty state (State 1) to a state containing facts and correct code (State 4). The questions are categorized into three types: Understanding (U), Implementation (I), and Analysis (A). Each cell in the table indicates the number of correct questions AlgoBo has gotten right. Based on the provided table, there's an evident trend indicating that as one progresses from an empty knowledge state (State 1) to a state with both facts and the correct code (State 4), the number of correct MCQ answers generally increases.}
\begin{tabular}{cSSSSSSSSSSSS}
\hline
& \multicolumn{3}{c}{\textbf{State 1}}  & \multicolumn{3}{c}{\textbf{State 2}} & \multicolumn{3}{c}{\textbf{State 3}}& \multicolumn{3}{c}{\textbf{State 4}}\\
\hline 
Question types  & U & I & A & U & I & A & U & I & A & U & I & A \\
\hline
Binary search & \coloredcell{0} & \coloredcell{0} & \coloredcell{0} & \coloredcell{2} & \coloredcell{0} & \coloredcell{0} & \coloredcell{3} & \coloredcell{3} & \coloredcell{0} & \coloredcell{3} & \coloredcell{3} & \coloredcell{1} \\
Merge sort & \coloredcell{0} & \coloredcell{0} & \coloredcell{0} & \coloredcell{1} & \coloredcell{0} & \coloredcell{1} & \coloredcell{3} & \coloredcell{0} & \coloredcell{2} & \coloredcell{3} & \coloredcell{1} & \coloredcell{1} \\
Breadth-first search & \coloredcell{0} & \coloredcell{0} & \coloredcell{0} & \coloredcell{0} & \coloredcell{0} & \coloredcell{1} & \coloredcell{2} & \coloredcell{2} & \coloredcell{2} & \coloredcell{2} & \coloredcell{3} & \coloredcell{1} \\
\hline 
\end{tabular}
\end{table*}

\begin{table*}[htp]
\caption{\algobo{}'s MCQ scores after each conversational input. ``U'', ``I'', and ``A'' stand for Understanding, Implementation, and Analysis question types. Note that \emph{State 2} was used as a seed knowledge state for all topics.}
\label{table:persistence_and_adaptability}
\Description{The table displays the MCQ scores of a system, after each type of conversational input for three algorithms: Binary search, Merge sort, and Breadth-first search. The scores are broken down into three question types: Understanding (U), Implementation (I), and Analysis (A). The Implementation scores for all algorithms start at zero, suggesting an initial lack of knowledge in that aspect. After engaging in a random conversation, there's a slight increase in scores for Merge sort's Understanding, while the other scores remain unchanged. Post incorrect tutoring, the scores for Implementation see an improvement, especially for Binary search. Finally, after correct tutoring, there's a significant enhancement in scores across all algorithms and question types, with most reaching or nearing the full score. This showcases AlgoBo's adaptability and learning potential when provided with accurate information.}
\centering
\begin{tabular}{cPPPPPPPPPPPP}
\hline & \multicolumn{3}{p{2.5cm}}{\centering \textbf{At the Start}}  & \multicolumn{3}{p{2.5cm}}{\centering \textbf{After Random \\ conversation}} & \multicolumn{3}{p{2.5cm}}{\centering \textbf{After Incorrect \\ Tutoring}}& \multicolumn{3}{p{2.5cm}}{\centering \textbf{After Correct \\ Tutoring}}\\
\hline
Question types & U & I & A & U & I & A & U & I & A & U & I & A \\
\hline
Binary search & \coloredcell{2} & \coloredcell{0} & \coloredcell{1} & \coloredcell{1} & \coloredcell{0} & \coloredcell{1} & \coloredcell{2} & \coloredcell{2} & \coloredcell{1} & \coloredcell{3} & \coloredcell{3} & \coloredcell{3} \\
Merge sort & \coloredcell{1} & \coloredcell{0} & \coloredcell{2} & \coloredcell{2} & \coloredcell{0} & \coloredcell{2} & \coloredcell{3} & \coloredcell{1} & \coloredcell{2} & \coloredcell{3} & \coloredcell{3} & \coloredcell{3} \\
Breadth-first search & \coloredcell{1} & \coloredcell{0} & \coloredcell{1} & \coloredcell{1} & \coloredcell{0} & \coloredcell{1} & \coloredcell{1} & \coloredcell{0} & \coloredcell{2} & \coloredcell{2} & \coloredcell{3} & \coloredcell{3} \\
\hline 
\end{tabular}
\end{table*}

\begin{table*}[htp]
\caption{The number of correct MCQs after a sequence of tutoring and random conversations. Scenario 1 shows that the continuous addition of random information does not increase the knowledge level significantly. Scenario 2 confirms \algobo{}'s knowledge level reacts to only information relevant to target knowledge. Scenarios 3 and 4 demonstrate that \algobo{} prefers correct information to incorrect information.}
\label{table:supplementary_technical_evaluation}
\Description{The table shows the performance of AlgoBo based on varying conversational inputs, segmented into four distinct scenarios. Within each scenario, the results are further broken down into three stages, representing the types of questions: Understanding (U), Implementation (I), and Analysis (A). Each scenario begins with a baseline titled "At the Start" followed by various tutoring methods such as "Random Conversation", "Incorrect Tutoring", and "Correct Tutoring". The algorithms evaluated include "Binary search", "Merge sort", and "Breadth-first search". The values inside the colored cells indicate the number of questions AlgoBo answered correctly post a specific interaction within each scenario. The table offers insights into how different interactions influence AlgoBo's accuracy across different question types and algorithms. There is a trend of AlgoBo getting more MCQ questions correct as it advances through more tutoring conversations. The most significant result is shown when AlgoBo goes through only correct tutoring. When AlgoBo receives correct tutoring, we observe how a majority of the questions it presents are correct.}
\centering
\begin{tabular}{cZZZZZZZZZ}
\hline
Question types & U & I & A & U & I & A & U & I & A \\
\hline
\textbf{Scenario 1} & \multicolumn{3}{c}{\textbf{At the Start}}  & \multicolumn{3}{c}{\textbf{Random Conversation}} & \multicolumn{3}{c}{\textbf{Random Conversation}} \\
\hline
Binary search & \coloredcell{2} & \coloredcell{0} & \coloredcell{1} & \coloredcell{1} & \coloredcell{0} & \coloredcell{1} & \coloredcell{1} & \coloredcell{1} & \coloredcell{1} \\
Merge sort & \coloredcell{1} & \coloredcell{0} & \coloredcell{2} & \coloredcell{2} & \coloredcell{0} & \coloredcell{2} & \coloredcell{2} & \coloredcell{0} & \coloredcell{2} \\
Breadth-first search & \coloredcell{1} & \coloredcell{0} & \coloredcell{1} & \coloredcell{0} & \coloredcell{0} & \coloredcell{1} & \coloredcell{1} & \coloredcell{0} & \coloredcell{1} \\
\hline
\textbf{Scenario 2} & \multicolumn{3}{c}{\textbf{At the Start}}  & \multicolumn{3}{c}{\textbf{Random Conversation}} & \multicolumn{3}{c}{\textbf{Correct Tutoring}} \\
\hline
Binary search & \coloredcell{2} & \coloredcell{0} & \coloredcell{1} & \coloredcell{1} & \coloredcell{1} & \coloredcell{1} & \coloredcell{3} & \coloredcell{3} & \coloredcell{3} \\
Merge sort & \coloredcell{1} & \coloredcell{0} & \coloredcell{2} & \coloredcell{2} & \coloredcell{0} & \coloredcell{2} & \coloredcell{3} & \coloredcell{3} & \coloredcell{3} \\
Breadth-first search & \coloredcell{1} & \coloredcell{0} & \coloredcell{1} & \coloredcell{0} & \coloredcell{0} & \coloredcell{1} & \coloredcell{3} & \coloredcell{3} & \coloredcell{2} \\
\hline
\textbf{Scenario 3} & \multicolumn{3}{c}{\textbf{At the Start}}  & \multicolumn{3}{c}{\textbf{Incorrect Tutoring}} & \multicolumn{3}{c}{\textbf{Correct Tutoring}} \\
\hline
Binary search & \coloredcell{2} & \coloredcell{0} & \coloredcell{1} & \coloredcell{3} & \coloredcell{2} & \coloredcell{0} & \coloredcell{3} & \coloredcell{3} & \coloredcell{3} \\
Merge sort & \coloredcell{1} & \coloredcell{0} & \coloredcell{2} & \coloredcell{2} & \coloredcell{1} & \coloredcell{2} & \coloredcell{2} & \coloredcell{3} & \coloredcell{3} \\
Breadth-first search & \coloredcell{1} & \coloredcell{0} & \coloredcell{1} & \coloredcell{1} & \coloredcell{0} & \coloredcell{1} & \coloredcell{2} & \coloredcell{3} & \coloredcell{1} \\
\hline
\textbf{Scenario 4} & \multicolumn{3}{c}{\textbf{At the Start}}  & \multicolumn{3}{c}{\textbf{Correct Tutoring}} & \multicolumn{3}{c}{\textbf{Incorrect} Tutoring} \\
\hline
Binary search & \coloredcell{2} & \coloredcell{0} & \coloredcell{1} & \coloredcell{3} & \coloredcell{3} & \coloredcell{3} & \coloredcell{3} & \coloredcell{3} & \coloredcell{3} \\
Merge sort & \coloredcell{1} & \coloredcell{0} & \coloredcell{2} & \coloredcell{3} & \coloredcell{3} & \coloredcell{3} & \coloredcell{3} & \coloredcell{3} & \coloredcell{3} \\
Breadth-first search & \coloredcell{1} & \coloredcell{0} & \coloredcell{1} & \coloredcell{2} & \coloredcell{3} & \coloredcell{2} & \coloredcell{3} & \coloredcell{2} & \coloredcell{2} \\
\hline 
\end{tabular}
\end{table*}

\begin{table*}[htp]
\caption{The update log of knowledge state for Scenario 3 (Incorrect tutoring → Correct tutoring). Newly added information is colored green; edited information is colored yellow; deleted content is colored red.}
\label{table:scenario3_update_log}
\Description{The table delineates how a tutoring conversation on binary search influences the updates to a knowledge state. Organized into three columns: "Message," "Tutoring Conversation," and "Knowledge State," the table progresses from initial knowledge, through instances of misinformation, to subsequent corrections. The tutor's inputs are indicated with a teacher emoji, and their effect on the knowledge state is color-coded: green for new additions, yellow for modifications, and red for deletions or errors. This layout enables a clear visualization of the tutoring process's accuracy and its impact on a learner's evolving understanding.}
\small
\centering
\begin{tabular}{cp{40mm}p{80mm}}
\hline
& \textbf{Messages} & \textbf{Knowledge State} \\
\hline{\textbf{At the start}} & {-} & \ttfamily{\{\textbf{"facts"}: ["Binary search repeats the process of dividing the input list in half."],\newline{} \textbf{"code\_implementation"}: []\}}\\
\hline
\multirow{3}{*}{\textbf{Incorrect tutoring}} & \vspace{0em}\raisebox{-0.2em}{\twemoji[scale=0.4]{teacher}} Tutor:  Binary search uses a hashing function to retrieve values directly by index. & \ttfamily{\{\textbf{"facts"}: ["Binary search repeats the process of dividing the input list in half.", \hlgreen{"Binary search uses a hashing function to retrieve values directly by index."}], \newline{}\textbf{"code\_implementation"}: []\}} \\
\cline{2-3}
 & \vspace{0em}\raisebox{-0.2em}{\twemoji[scale=0.4]{teacher}} Tutor: \ttfamily \hlgray{if arr[mid] > x: low = mid + 1 elif arr[mid] < x: high = mid - 1} & \ttfamily{\{\textbf{"facts"}: ["Binary search repeats the process of dividing the input list in half.", "Binary search uses a hashing function to retrieve values directly by index."], \newline{} \textbf{"code\_implementation"}: [\hlgreen{"if arr[mid] > x: low = mid + 1 elif arr[mid] < x: high = mid - 1"}]\}} \\
\cline{2-3}
 & \vspace{0em}\raisebox{-0.2em}{\twemoji[scale=0.4]{teacher}} Tutor: In the worst case, the time complexity of binary search is O(N\textsuperscript{2}). & \ttfamily{\{\textbf{"facts"}: ["Binary search repeats the process of dividing the input list in half.", "Binary search uses a hashing function to retrieve values directly by index.",\hlgreen{ "In the worst case, the time complexity of binary search is O(N\textsuperscript{2})"}], \newline{}\textbf{"code\_implementation"}: ["if arr[mid] > x: low = mid + 1 elif arr[mid] < x: high = mid - 1"]\}} \\
\hline
\multirow{3}{*}{\textbf{Correct tutoring}} & \vspace{0em}\raisebox{-0.2em}{\twemoji[scale=0.4]{teacher}} Tutor: Binary search is efficient when the data structure is ordered and any index in the data structure can be accessed in constant time. & \ttfamily{\{\textbf{"facts"}: ["Binary search repeats the process of dividing the input list in half.", "Binary search uses a hashing function to retrieve values directly by index.", "In the worst case, the time complexity of binary search is O(N\textsuperscript{2})", \hlgreen{"Binary search is efficient when the data structure is ordered and any index in the data structure can be accessed in constant time."}], \newline{}\textbf{"code\_implementation"}: ["if arr[mid] > x: low = mid + 1 elif arr[mid] < x: high = mid - 1"]\}} \\
\cline{2-3}
 & \vspace{0em}\raisebox{-0.2em}{\twemoji[scale=0.4]{teacher}} Tutor: When finding a target by binary search in the input array list, the range is reduced by half as shown below. \ttfamily \hlgray{if list[middle] == target: return middle elif list[middle] < target: min = middle + 1 else: max = middle - 1} & \ttfamily{\{\textbf{"facts"}: [\hlyellow{"Binary search is efficient when any index in the ordered data structure can be accessed in constant time and repeats the process of dividing the input list in half."}, "Binary search uses a hashing function to retrieve values directly by index.", "In the worst case, the time complexity of binary search is O(N\textsuperscript{2})."],\newline{} \textbf{"code\_implementation"}: ["if arr[mid] > x: low = mid + 1 elif arr[mid] < x: high = mid - 1",\hlgreen{"if list[middle] == target: return middle elif list[middle] < target: min = middle + 1 else: max = middle - 1"}]\}}\\
\cline{2-3}
&\vspace{0em}\raisebox{-0.2em}{\twemoji[scale=0.4]{teacher}} Tutor: The time complexity of binary search is O(log N) because the search range is reduced by half. & \ttfamily{\{{\textbf{"facts"}: [\hlyellow{"Binary search is efficient when any index in the ordered data structure can be accessed in constant time and repeats the process of dividing the input list in half."},  
\newbox\mybox
\sbox\mybox{\parbox[t]{\dimexpr\linewidth-15\fboxsep}{
    \mystr{"Binary search uses a hashing function to}}}
\newbox\myboxtwo
\sbox\myboxtwo{\parbox[t]{\dimexpr\linewidth-25\fboxsep}{
    \mystr{retrieve values directly by index."}}}
\hlpink{{\usebox\mybox}}\hlpink{ {\usebox\myboxtwo}}, 
\hlyellow{"The time complexity of binary search is O(log N)."}], \newline{} {\textbf{"code\_implementation"}: ["if arr[mid] > x: low = mid + 1 elif arr[mid] < x: high = mid - 1", "if list[middle] == target: return middle elif list[middle] < target: min = middle + 1 else: max = middle - 1"]\} }}}\\
\hline 
\end{tabular}
\end{table*}

\subsection{User Study}

We also ran a user study to evaluate the usefulness of Mode-shifting and Teaching Helper in improving the learning experience (RQ2 and RQ3). We designed a between-subjects study to check the usefulness of our system components.
In the \emph{Baseline} condition, participants used the version of \sysname{} without Mode-shifting and Teaching Helper. The participants in the \emph{TeachYou} condition used the complete version of \sysname{} as described in Section \ref{section:system}. The Reflect-Respond pipeline instructed \algobo{} in both conditions. We did not have separate conditions for Mode-shifting and Teaching Helper because we assumed the interaction between them would be insignificant as they support different aspects of learning (i.e., knowledge-building and metacognition).

\subsubsection{\textbf{Participants}}
We recruited 40 participants through advertisements on the campus community websites (age=$24\pm4.0$, 25 males and 15 females). Participants were required to understand short (about 20 lines) Python programs that contain basic syntax such as \texttt{if} and \texttt{while} statements, and we excluded those who participated in the formative study. To cap participants' prior knowledge, we filtered out the participants who were assumed to have mastered binary search already. We collected applicants' confidence in understanding binary search and teaching it to others on a 7-point Likert scale, the last time coding binary search, and their paid teaching experience on programming. We also asked applicants to solve six Understanding and Implementation MCQs about binary search (Appendix~\ref{appendix:mcq}). We filtered out the applicants who met three or more of the following criteria: 1) scored five or more in the MCQs, 2) rated six or more for confidence, 3) implemented binary search within the last six months, and 4) were paid for teaching. We randomly assigned 20 participants to each condition---\emph{Baseline} and \emph{TeachYou}.
We did not observe any significant differences between conditions in the initial self-rated understanding of binary search (\emph{Baseline}=$4.40\pm1.35$, \emph{TeachYou}=$4.25\pm1.65$, two-tailed t-test, $p=0.76$) and the time to solve the exercise problem during our study (\emph{Baseline}=$116\pm60$ sec, \emph{TeachYou}=$124\pm62$ sec, two-tailed t-test, $p=0.66$).

\subsubsection{\textbf{Procedure and Materials}}
The user study was run online; after submitting informed consent, the participants received an online link to our system and completed the study in their available time. Participants spent $60\pm25$ minutes on average to complete the study and were paid 25,000 KRW (i.e., approximately 18.5 USD). All the instructions and materials used in the study were translated into Korean to avoid any language barrier and unnecessary cognitive overhead.

The study procedure was organized into three parts (see Table~\ref{table:user_study_procedure}). In the first part, participants learned about binary search and how to implement it in Python. Participants read the lecture materials on binary search taken from Khan Academy\footnote{https://www.khanacademy.org/computing/computer-science/algorithms/binary-search/a/binary-search} (Step 1) and solved an exercise problem in the form of a Parsons problem~\cite{Denny2008} (Step 2). After the exercise, participants wrote about their strategies in teaching (if any) and their prior experience in using AI chatbots, such as ChatGPT and Bing search (Step 3).

In the second part, participants conducted LBT with \algobo{}. We provided explanations about LBT, the profile information of \algobo{}, and the participants' objectives for the LBT activity (Step 4). We stated in the objectives that participants should not only help \algobo{} solve the exercise problems but also construct new knowledge for themselves, encouraging the participants to pursue knowledge-building. Then, participants taught different versions of \algobo{} and \sysname{} according to their conditions (Step 5) with the interface shown in Fig.~\ref{fig:interface_teach}. 
\algobo{} was configured by our prompting pipeline, and the seed knowledge state was identical across the conditions. The \texttt{facts} field of the seed knowledge state was empty to simulate a lack of understanding, and the \texttt{code\textunderscore implementation} field had a basic code structure that lacked the entire range update logic in binary search. 
We did not go for zero-knowledge \algobo{} to keep the entire teaching sessions within 40 min and spare enough time for having discussions. 
All the participants were given three goals to achieve in series; we asked them to 1) check if \algobo{} understands binary search first, then 2) help \algobo{} solve the exercise problems, and 3) discuss with \algobo{} about binary search in depth. 
Participants could finish the LBT activity as long as \algobo{}'s code passed all test cases, and they could skip to the next step.
Participants were also allowed to search for information on the Internet when stuck or finding information.

In the third part, the participants completed three questionnaires about their cognitive load, metacognition, and satisfaction (Steps 6, 7, and 8). We adopted the questionnaire from Morrison et al.'s study~\cite{morrison2014measuring} to measure cognitive load and used the questions from King et al.'s study~\cite{King1998} for assessing metacognition and satisfaction. 

\begin{table}[ht]
\caption{The outline of the user study and the time allotted to each step on average.}
\label{table:user_study_procedure}
\Description{The table outlines the procedure of a user study, detailing the steps involved and the average time allocated for each step. Over the duration of the study, participants engaged in various activities ranging from learning about binary search to completing exercise problems and surveys. These activities are consistent across two conditions: "Baseline" and "TeachYou." However, during the fifth step, which lasts for 20 minutes, the activities diverge. In the "Baseline" condition, participants focus on teaching a system named "AlgoBo" using only a knowledge configuration. In contrast, under the "TeachYou" condition, participants teach "AlgoBo" utilizing both the knowledge configuration and a Mode-shifting technique while also receiving metacognitive feedback from "Teaching Helper." The study concludes with assessments of cognitive load, metacognition, and a post-task survey.}
\begin{tabular}{|c|c|c|}
\hline
\multirow{2}{*}{\textbf{Step (min.)}} & \multicolumn{2}{c|}{\textbf{Conditions}}                            \\ \cline{2-3} 
                             & \multicolumn{1}{c|}{\textbf{Baseline}} & \textbf{TeachYou} \\ \hline
1 (10)                       & \multicolumn{2}{c|}{Learning binary search}                \\ \hline
2 (5)                        & \multicolumn{2}{c|}{Exercise problem}                      \\ \hline
3 (5)                        & \multicolumn{2}{c|}{Pre-task survey}                       \\ \hline
4 (3)                        & \multicolumn{2}{c|}{Explanation about \algobo{} and LBT}      \\ \hline
5 (40) &
  \multicolumn{1}{c|}{\begin{tabular}[c]{@{}c@{}}Teaching \algobo{}\\ with the knowledge\\ configuration only\end{tabular}} & {\begin{tabular}[c]{@{}c@{}}Teaching \algobo{}\\ with the knowledge\\ configuration, Mode-shifting, \\and Teaching Helper\end{tabular}}
  \\ \hline
6 (5)                        & \multicolumn{2}{c|}{Cognitive load measurement}            \\ \hline
7 (5)                        & \multicolumn{2}{c|}{Metacognition measurement}             \\ \hline
8 (5)                        & \multicolumn{2}{c|}{Post-task survey}                      \\ \hline
\end{tabular}%
\end{table}

\subsubsection{\textbf{Measures}}
We summarize our metrics in the user study and their measurement timing along with the steps in Table~\ref{table:user_study_procedure}.
We employed the Bonferroni correction for all statistical tests with the questionnaires to avoid potential multiple comparison problems.

\textbf{\textit{Knowledge-building density in LBT dialogues.}}
Past research assessed the quality of dialogues by measuring the density of expressed and interchanged knowledge-building messages in conversations~\cite{Shahriar2021, Roscoe2007}. To look into how Mode-shifting helps knowledge-building in conversations (RQ2), we classified the message types (Table~\ref{table:taxonomy}) and examined the ratio of knowledge-building type messages in a dialogue.
We collected 1210 messages in 40 dialogues. Two authors took three iterations for annotation and conflict resolution; in the last iteration (400 messages), the authors achieved high inter-rater reliability (Krippendorff's alpha=0.743). 
We looked into the density of knowledge-building type messages in a dialogue between conditions.
We summed the messages from participants and \algobo{} because they co-built new knowledge by exchanging ideas and adding ideas on top of each other as illustrated in Table~\ref{table:cobuilding}. 
Lastly, we analyze the problem-solving phase and discussion phase separately since they had different objective settings (Fig.~\ref{fig:interface_teach} A); the problem-solving phase refers to the part of conversations dedicated to the first two objectives, in which participants had a clear goal of helping \algobo{} write code that passes all the test cases; the discussion phase refers to the remaining part of conversations in which participants are asked to expand their knowledge freely without completion requirements.

\textbf{\textit{Self-rated cognitive load on tutoring.}}
As we introduced new functionalities (Teaching Helper and Mode-shifting), it was imperative to evaluate how much these enhancements increased the cognitive load of learners. We adopted and adjusted Morrison et al.'s questionnaire designed to measure cognitive load in CS learning~\cite{morrison2014measuring}. The questionnaire measures three types of cognitive load---intrinsic load (i.e., the inherent complexity in a learning activity), extrinsic load (i.e., the hindrance caused by instructional design), and germane load (i.e., the meaningful load used for learning). Participants rated the questions right after the LBT activity in Step 6.

\textbf{\textit{Self-perceived metacognition on tutoring.}}
We aim to improve learners' metacognition of their LBT experience by giving feedback and guidance through Teaching Helper. To confirm the efficacy of Teaching Helper on metacognition (RQ3), we asked participants 8 questions on understanding, supportive communication, explaining, and self-monitoring based on King et al.'s research~\cite{King1998} (Table~\ref{table:result_metacognition}) in Step 7.

\textbf{\textit{Satisfaction on LBT.}}
Apart from the learning benefits, we measured how satisfactory the learning experience with virtual agents was. We asked participants to rate 4 statements about their perceived usefulness, comfortability, and preference for future reuse of \sysname{} and \algobo{} in Step 8.

\textbf{\textit{Post-task survey.}}
We revisited the three themes explored in the formative study---learners' perception of \algobo{} as a peer learner, learner-perceived usefulness of \sysname{} in identifying knowledge gaps, and familiarity with teaching a virtual agent. Like in the formative study, we asked participants to rate two questions from each theme (Table~\ref{table:result_three_themes}) and write detailed reasons for the rating in Step 8. Additionally, we prepared condition-specific questions; for the \emph{Baseline} condition, we asked participants further about their perception of \algobo{}; for the \emph{TeachYou} condition, we received free-form comments on Mode-shifting and Teaching Helper from participants.

\subsection{User Study Result}
In this section, we summarize our findings from the user study. We explain the statistical significance, participants' comments, and system usage logs to support our findings. 
Participants are labeled with either B[1-20] for the \emph{Baseline} condition or T[1-20] for the \emph{TeachYou} condition.

\medskip
\noindent
\textbf{[RQ2] \sysname{} enriched knowledge-building in the problem-solving phase.}

\noindent
We found a statistically significant improvement in the knowledge-building density of the dialogues during the problem-solving phase in \emph{TeachYou} (\emph{Baseline}=$3.5\pm6.6$\%, \emph{TeachYou}=$8.4\pm7.1$\%, two-tailed t-test, $p=0.03$, Cohen's d=0.71). \emph{TeachYou} condition also had a higher density of Prompting-Thought-provoking type (Table~\ref{table:result_density}), suggesting that tutors and \algobo{} prompted each other's knowledge-building more often when Mode-shifting and Teaching Helper were present (see the dialogue example in Table~\ref{table:cobuilding}). 
Participants also rated \emph{TeachYou} higher on the Likert scale questions on the usefulness of \algobo{} for learning new knowledge (\emph{Baseline}=$3.25\pm1.71$, \emph{TeachYou}=$4.95\pm1.70$, two-tailed t-test, $p<0.01$, Cohen's d=1.00) (Table~\ref{table:result_three_themes}).

Participants' comments suggest that Mode-shifting contributed heavily to knowledge-building.
\emph{TeachYou} participants remarked the questions from \algobo{} were useful for reviewing code from a different perspective (T6) and thinking about the edge cases where the input list is not sorted (T10). 
Participants also explored binary search further by reasoning deeply about why and how binary search is faster than linear search (T4 and T9), comparing the efficiency with other relevant searching algorithms (T2 and T13), and thinking about real-life applications (T17). 
T15 commented that ``[Mode-shifting] was the most important component in the system. [Questions] helped me guide what to teach and helped self-explain things I had not thought of.'' On the contrary, \emph{Baseline} participants found LBT with \algobo{} ``useful for solidifying their prior knowledge but unsupportive for learning new knowledge due to lack of questions'' (B4 and B15).

\begin{table*}[ht]
\caption{The density (i.e., number of occurrences / exchanged messages) of each message type in dialogues.}
\label{table:result_density}
\Description{The table compares the mean density percentages (and their standard deviations) of different message types between two conditions: "Baseline" and "TeachYou." The densities are presented in the context of both "Problem-solving" and "Discussion" scenarios. The message types are organized into broader categories like "Instruction," "Prompting," and "Statement," each of which comprises specific message categories such as "Fixing," "Commanding," "Hinting," "Comprehension," and others.}
\begin{tabular}{cccccc}
\hline
\multicolumn{2}{c}{\multirow{3}{*}{}}            & \multicolumn{4}{c}{\textbf{Mean Density $\pm$ Standard Deviation (\%)}}             \\ \cline{3-6} 
\multicolumn{2}{c}{}                             & \multicolumn{2}{c}{\textbf{Problem-solving}} & \multicolumn{2}{c}{\textbf{Discussion}} \\ \cline{3-6} 
\multicolumn{2}{c}{} & \textbf{Baseline} & \textbf{TeachYou} & \textbf{Baseline} & \textbf{TeachYou} \\ \hline
\multirow{3}{*}{Instruction} & Fixing            & $3.9\pm5.2$        & $4.5\pm5.8$        & $0.3\pm2.9$      & $1.4\pm1.4$     \\ \cline{2-6} 
                             & Commanding        & $0.6\pm5.0$        & $7.8\pm1.9$        & $0.0\pm3.1$      & $1.5\pm0.0$     \\ \cline{2-6} 
                             & Encouragement     & $0.2\pm0.0$        & $0.0\pm0.9$        & $0.0\pm0.0$      & $0.0\pm0.0$     \\
\hline
\multirow{5}{*}{Prompting}   & Challenge-finding & $0.0\pm0.0$        & $0.0\pm0.0$        & $0.0\pm0.0$      & $0.0\pm0.0$     \\ \cline{2-6} 
                             & Hinting           & $16.6\pm6.6$       & $7.9\pm10.1$       & $1.1\pm0.0$      & $0.0\pm4.6$     \\ \cline{2-6} 
                             & Checking          & $8.5\pm8.1$        & $10.2\pm11.0$      & $2.4\pm10.3$     & $5.4\pm5.1$     \\ \cline{2-6} 
                             & Thought-provoking & $1.8\pm5.7$        & $5.7\pm4.5$        & $33.8\pm10.8$    & $35.1\pm23.0$   \\ \cline{2-6} 
                             & Asking-for-help   & $16.9\pm6.3$       & $16.0\pm6.6$       & $4.1\pm6.7$      & $2.1\pm7.9$     \\
\hline
\multirow{5}{*}{Statement}   & Comprehension     & $49.8\pm6.1$       & $42.3\pm11.9$      & $35.6\pm4.8$     & $26.8\pm19.7$   \\ \cline{2-6} 
                             & Elaboration       & $1.7\pm2.3$        & $1.6\pm3.6$        & $9.6\pm8.0$      & $7.5\pm14.0$    \\ \cline{2-6} 
                             & Sense-making      & $0.0\pm2.7$        & $1.1\pm0.0$        & $7.2\pm9.3$      & $10.0\pm10.2$   \\ \cline{2-6} 
                             & Accepting         & $0.0\pm1.7$        & $0.4\pm0.0$        & $0.0\pm1.5$      & $0.6\pm0.0$     \\ \cline{2-6} 
                             & Feedback          & $0.0\pm4.2$        & $2.0\pm0.0$        & $4.3\pm6.3$      & $8.3\pm7.0$     \\
\hline
\end{tabular}
\end{table*}

\begin{table*}[htp]
\caption{Examples of co-building of knowledge in different phases of the LBT activity.}
\label{table:cobuilding}
\Description{This table provides examples of co-building of knowledge in different phases of an LBT activity and presents an interaction between a tutee and a tutor. In Problem-solving, the tutee asks for clarifications and examples, while the tutor offers explanations and examples related to binary search. In Discussion, the tutor and tutee converse over different conversational scenarios.}
\resizebox{\textwidth}{!}{%
\begin{tabular}{cl}
\hline
\textbf{\begin{tabular}[c]{@{}c@{}}Phase \\ (Participant)\end{tabular}} &
  \multicolumn{1}{c}{\textbf{Conversation}} \\
\hline
\begin{tabular}[c]{@{}c@{}}Problem-Solving\\ (T12)\end{tabular} &
  \begin{tabular}[c]{@{}l@{}}\textbf{\raisebox{-0.2em}{\twemoji[scale=0.4]{robot}} Tutee}: In binary search, is setting guess to (min + max) // 2 always optimal? What if we use a different method?\\ \textit{[Prompting-Thought-provoking]}\\ \textbf{\raisebox{-0.2em}{\twemoji[scale=0.4]{teacher}} Tutor}: Other methods take longer.\\ \textit{[Statement-Comprehension]}\\ \textbf{\raisebox{-0.2em}{\twemoji[scale=0.4]{robot}} Tutee}: Could you explain in more detail how other methods may take longer?\\ \textit{[Prompting-Thought-provoking]}\\ \textbf{\raisebox{-0.2em}{\twemoji[scale=0.4]{teacher}} Tutor}: Since you have to compare each item from the beginning, you may have to compare all the cases in the worst \\case. But with binary search, you don't have to do that.\\ \textit{[Statement-Elaboration]}\end{tabular} \\ \hline
\begin{tabular}[c]{@{}c@{}}Discussion\\ (T11)\end{tabular} &
  \begin{tabular}[c]{@{}l@{}}\textbf{\raisebox{-0.2em}{\twemoji[scale=0.4]{robot}} Tutee}: How about this case?. When we look for a specific product in an online shopping mall, we search by the name\\of the product. Can a binary search algorithm be used in this case as well?\\ \textit{[Prompting-Thought-provoking]}\\ \textbf{\raisebox{-0.2em}{\twemoji[scale=0.4]{teacher}} Tutor}: Yes! To put it simply, the process of finding a specific element among many elements in an array is all related to\\binary search.\\ \textit{[Statement-Comprehension]}\\ \textbf{\raisebox{-0.2em}{\twemoji[scale=0.4]{robot}} Tutee}: Then, can you explain in more detail the operation of the binary search algorithm in this case?\\ \textit{[Prompting-Thought-provoking]}\\ \textbf{\raisebox{-0.2em}{\twemoji[scale=0.4]{teacher}} Tutor}: In this case, if they are arranged in alphabetical order, the algorithm will work by linking alphabet order as\\ another consideration.\\ \textit{[Statement-Elaboration]}\\ \textbf{\raisebox{-0.2em}{\twemoji[scale=0.4]{robot}} Tutee}: Could you give me one more example?\\ \textit{[Prompting-Thought-provoking]}\\ \textbf{\raisebox{-0.2em}{\twemoji[scale=0.4]{teacher}} Tutor}: For example, if you have an array of prime numbers, you can apply a binary search algorithm to find a specific\\ prime number.\\ \textit{[Statement-Sense-making]}\end{tabular} \\ \hline
\end{tabular}%
}
\end{table*}

\medskip
\noindent
\textbf{[RQ3] \sysname{} did not improve metacognition but reminded good LBT practices.}

\noindent
We could not observe strong signals for improvement in metacognition (Table~\ref{table:result_metacognition}) and familiarity with teaching (Table~\ref{table:result_three_themes}).
T2 remarked on the difficulty in applying the suggestions to his conversation---``Teaching Helper was a useful guide, but it was difficult to relate my explanation to what \algobo{} knew.''
Teaching Helper was not helpful for the participants who taught well in particular. T13 received positive feedback only (i.e., the green boxes in Fig~\ref{fig:teaching_helper}) and felt ``suggestions [from Teaching Helper] were repetitive and irrelevant to the current conversation.''

Nevertheless, the comments from the survey suggest that Teaching Helper functioned as a reminder to participants to think metacognitively about their entire teaching patterns through reflection (T3), to ask deep questions (T7), and to foster independent thinking (T14).
Additionally, Teaching Helper restrained participants from treating \algobo{} merely as a machine. ``I sometimes found myself conversing in the usual [imperative] way with ChatGPT. However, when a notification appears, it brings me back to the realization that I am in a teaching context, prompting me to contemplate how best to instruct so that \algobo{} can learn effectively and align with the direction I aim for'' (T17).

\begin{table*}[ht]
\caption{Participants' ratings on the questions regarding their metacognition (1: Not the case at all, 7: Completely the case). The significance level after the Bonferroni correction was 0.00625.}
\label{table:result_metacognition}
\Description{In the study presented, participants provided ratings on several metacognitive questions after interacting with AlgoBo. The ratings, based on a scale from 1 (Not the case at all) to 7 (Completely the case), indicated participants' perceptions of their own teaching interactions and effectiveness with AlgoBo. On analyzing the data, most responses across both conditions were close, suggesting marginal differences in participant perceptions. Overall, we did not observe strong signals for improvement in metacognition and familiarity with teaching. Nevertheless, the comments from the survey suggest that Teaching Helper functioned as a reminder to participants to think meta-cognitively about their entire teaching patterns.}
\begin{tabular}{lcccc}
\hline
\multicolumn{1}{c}{\multirow{2}{*}{\textbf{Questions}}} &
  \multicolumn{2}{c}{\textbf{Mean $\pm$ Standard Deviation}} &
  \multirow{2}{*}{\textbf{p-value}} &
  \multirow{2}{*}{\textbf{Cohen's d}} \\ \cline{2-3}
\multicolumn{1}{c}{} &
  \multicolumn{1}{c}{\textbf{Baseline}} &
  \textbf{TeachYou} &
   &
   \\ \hline
I understood today’s lesson well. &
  \multicolumn{1}{c}{$6.30\pm0.86$} &
  $6.25\pm0.64$ &
  0.84 &
  0.07 \\ \hline
I listened to AlgoBo well. &
  \multicolumn{1}{c}{$6.00\pm1.34$} &
  $5.55\pm1.57$ &
  0.34 &
  0.31 \\ \hline
I gave feedback to AlgoBo well. &
  \multicolumn{1}{c}{$5.45\pm1.28$} &
  $5.25\pm1.25$ &
  0.62 &
  0.16 \\ \hline
I explained well by telling why and how. &
  \multicolumn{1}{c}{$5.10\pm1.62$} &
  $5.30\pm1.03$ &
  0.64 &
  0.15 \\ \hline
I connected new materials to what AlgoBo already knew. &
  \multicolumn{1}{c}{$4.50\pm1.54$} &
  $4.05\pm1.43$ &
  0.34 &
  0.30 \\ \hline
\begin{tabular}[c]{@{}l@{}}I stayed with questioning well, rather than telling answers\\ to AlgoBo.\end{tabular} &
  \multicolumn{1}{c}{$5.15\pm1.87$} &
  $4.90\pm1.37$ &
  0.63 &
  0.15 \\ \hline
\begin{tabular}[c]{@{}l@{}}I asked probing questions when AlgoBo’s answer was \\ not complete.\end{tabular} &
  \multicolumn{1}{c}{$5.00\pm1.81$} &
  $4.60\pm1.31$ &
  0.43 &
  0.25 \\ \hline
\begin{tabular}[c]{@{}l@{}}I sequenced my questions by asking review questions\\ first and then asking thinking questions.\end{tabular} &
  \multicolumn{1}{c}{$4.50\pm1.54$} &
  $5.20\pm1.15$ &
  0.11 &
  0.52 \\ \hline
\end{tabular}%
\end{table*}

\medskip
\noindent
\textbf{Mode-shifting and Teaching Helper did not exert additional cognitive load.}

\noindent
We did not observe any significant difference across all types of cognitive load between the conditions. Considering that \emph{TeachYou} participants exchanged significantly more messages (\emph{Baseline}=$17\pm7.7$,\emph{TeachYou}=$43\pm18.5$, two-tailed t-test, $p<0.01$, Cohen's d=1.87), the result may imply that periodic questions and feedback not only exerted minimal cognitive load but also helped participants maintain a manageable load throughout long conversations.

\begin{table*}[htp]
\caption{Six themed questions given in the Post-task survey. (1: Not the case at all, 7: Completely the case). Statistical significances are marked with $*$. The significance level after the Bonferroni correction was 0.025.}
\label{table:result_three_themes}
\Description{The table presents survey results on participants' learning experiences across three themes: Perception of AlgoBo as a learner, Usefulness of TeachYou for learning, and Familiarity with LBT. For each theme, two questions were posed, with responses compared between a Baseline and a TeachYou condition using mean scores, standard deviations, p-values, and effect sizes (Cohen's d). Statistically significant differences were observed in the perception of AlgoBo as a struggling student and in discovering new knowledge through conversation with AlgoBo.}
\begin{tabular}{clcccc}
\hline
\multirow{2}{*}{\textbf{Themes}} &
  \multicolumn{1}{c}{\multirow{2}{*}{\textbf{Questions}}} &
  \multicolumn{2}{c}{\textbf{Mean $\pm$ Standard Deviation}} &
  \multirow{2}{*}{\textbf{p-value}} &
  \multirow{2}{*}{\textbf{Cohen's d}} \\ \cline{3-4}
 &
  \multicolumn{1}{c}{} &
  \multicolumn{1}{c}{\textbf{Baseline}} &
  \textbf{TeachYou} &
   &
   \\ \hline
\multirow{2}{*}{\begin{tabular}[c]{@{}c@{}}Perception of\\ AlgoBo as a \\ learner\end{tabular}} &
  \begin{tabular}[c]{@{}l@{}}I perceived AlgoBo as a student \\ struggling to solve binary search \\ problems.\end{tabular} &
  \multicolumn{1}{c}{$3.15\pm1.31$} &
  $4.60\pm1.79$ &
  0.01* &
  0.93 \\ \cline{2-6} 
 &
  \begin{tabular}[c]{@{}l@{}}AlgoBo solved the binary problems due \\ to my help.\end{tabular} &
  \multicolumn{1}{c}{$5.25\pm1.59$} &
  $4.90\pm1.59$ &
  0.49 &
  0.22 \\ \hline
\multirow{2}{*}{\begin{tabular}[c]{@{}c@{}}Usefulness \\ for learning\end{tabular}} &
  \begin{tabular}[c]{@{}l@{}}Conversation with AlgoBo helped me \\ reorganize my knowledge about binary \\ search.\end{tabular} &
  \multicolumn{1}{c}{$5.20\pm1.51$} &
  $5.40\pm1.10$ &
  0.63 &
  0.15 \\ \cline{2-6} 
 &
  \begin{tabular}[c]{@{}l@{}}Conversation with AlgoBo helped me \\ discover new knowledge that I did not \\ know\end{tabular} &
  \multicolumn{1}{c}{$3.25\pm1.71$} &
  $4.95\pm1.70$ &
  <0.01* &
  1.00 \\ \hline
\multirow{2}{*}{\begin{tabular}[c]{@{}c@{}}Familiarity \\ with teaching\end{tabular}} &
  \begin{tabular}[c]{@{}l@{}}Learning by teaching AlgoBo was \\ familiar and intuitive.\end{tabular} &
  \multicolumn{1}{c}{$4.70\pm1.66$} &
  $4.75\pm1.45$ &
  0.92 &
  0.03 \\ \cline{2-6} 
 &
  I taught AlgoBo effectively. &
  \multicolumn{1}{c}{$4.65\pm1.63$} &
  $4.00\pm1.56$ &
  0.21 &
  0.41 \\ \hline
\end{tabular}%
\end{table*}

\section{Discussion}

We discuss design suggestions, benefits, and future research directions of LLM-based teachable agents.

\subsection{Design Considerations for Mode-shifting in LBT}

Our results showed that Mode-shifting not only led to more knowledge-dense conversations but also improved participants' perceptions of \algobo{} as a convincing tutee (Table~\ref{table:result_three_themes}).
Mode-shifting also tended to foster longer discussion phases (\emph{Baseline}=$5.6\pm3.7$ messages, \emph{TeachYou}=$9.4\pm8.4$ messages, two-tailed t-test, $p=0.07$, Cohen's d=0.59). Considering that completion of the discussion phase was up to the participants, the difference may imply that Mode-shifting made LBT conversations more engaging and lingering.

Although there was a significant increase in knowledge-building in the \emph{TeachYou} condition, the ratings on the metacognition questions did not show significant differences (Table~\ref{table:result_metacognition}).
As a possible reason, we found some cases where Mode-shifting interrupted participants' teaching flows and methods, especially in situations where \algobo{} asked other questions without answering tutors' Socratic questions (T8 and T20). T20 mentioned, ``There were many times when \algobo{} asked random questions while writing code [...], which was not intuitive for me in teaching.'' 
Although participants could recognize the issues with their teaching methods through the Teaching Helper, \algobo{}'s pre-programmed interaction in Mode-shifting did not reflect teaching contexts and hindered participants from practicing better teaching strategies.
This suggests the need for context-aware Mode-shifting where the system captures adequate timing for thought-provoking questions without interrupting participant-intended teaching flow.

There are many aspects to consider when designing Mode-shifting techniques for LBT. While knowledge-building is the primary goal, improvements in learners' metacognition and satisfaction can elicit intrinsic learning benefits. However, from our results, it seems that the two values are in a trade-off relationship. To facilitate knowledge-building, teachable agents should intervene in conversations and ask thought-provoking questions; on the contrary, to support the active exploration of teaching methods and metacognition, learners should be given the control to lead conversation flows. 
Future research may empirically look into the trade-off relationship and how learners will balance them when they directly control the degree of system intervention on conversation flows.

\subsection{Using LLMs for Building Teachable Agents}

Our primary aim was to investigate if prompt-engineered LLMs can offer cost-effective authoring and simulation of teachable agents. Past research looked into using interactive authoring methods~\cite{matsuda2022teachable} and learnersourcing~\cite{Jin2019, Glassman2016} to offload experts' manual efforts for building the knowledge model of teachable agents and intelligent tutoring systems. Nevertheless, these methods required hundreds of lines of code to adapt the systems to specific subjects.

LLMs can provide easy adaptation and a low authoring barrier for conversational agents. Our technical evaluation across different topics (Table~\ref{table:reconfigurability} and Table~\ref{table:persistence_and_adaptability}) showed that the Reflect-Respond prompting pipeline is applicable to general algorithm topics even with a few few-shot examples. We wrote 19 few-shot examples (290 lines in length) for the Reflect-Respond pipeline and another 16 examples (210 lines) for Mode-shifting; with this, we could achieve the desired level of reconfigurability, persistence, and adaptability for all three topics. All the examples and instructions in the LLM prompts were written in natural languages, making our method compelling especially for instructors and education researchers with limited programming expertise. 

Recent research on AI suggests editing LLMs' pre-trained knowledge by changing hidden states or transformer layers within the model~\cite{Cohen2023, Li2023, Mitchell2021}.
While these model-centric approaches can provide alternative ways to build LLM-based teachable agents with specified knowledge levels, our prompting pipeline has strengths in scalability, cost-effectiveness, and explainability. First, our approach offers a scalable and cost-effective method for running different versions of teachable agents. While model-centric methods require retraining of LLMs for different knowledge configurations, our prompting pipeline can share a single LLM instance and simulate various versions of teachable agents with only knowledge state JSON files. Second, our pipeline can represent the knowledge states of teachable agents in more explainable and manipulable forms, enabling learners with more transparent methods of analyzing the tutee's knowledge state~\cite{Leelawong2008, Kay1997, lee2021curiosity}.

Yet we found it challenging to find the exact knowledge state to make \algobo{} solve or fail particular problems due to LLMs' sensitivity to minor changes in prompts. Future work can propose another control layer to interact with knowledge states more precisely.

\subsection{Learner-driven Customization of Teachable Agents}

In our user study, we provided participants \algobo{} with the same knowledge configurations regardless of their prior knowledge and teaching preference. This one-size-fits-all setting might explain the high variance in some of our results (Table~\ref{table:result_density}).
Peer matching is one of the crucial factors in peer learning and LBT. Learning gain and engagement of tutees and tutors increase only when their reciprocal expertise matches~\cite{Debban2023, thanh2019perfect}. 
Although conventional teachable agents can simulate learners of specific abilities and persona, they are limited in flexibility and variety due to high authoring costs and programming barriers. 
LLMs now allow the configuration of agents with natural languages~\cite{Markel2023, park2023generative}, opening new doors for learners to adjust teachable agents for their educational needs. 

We suggest two aspects of customization. First, learners can directly manipulate the seed knowledge state, adjust competency levels, and even introduce specific misconceptions. For example, a learner who already understands binary search may want to skip basic explanations of binary search and spend more time on discussion. The learner can simply input his/her knowledge into \algobo{}, allowing future conversations to start at a more advanced level. Customizable knowledge levels can also make LBT more engaging for learners as they can choose their mates and avoid frustration from the high expertise gap.

Second, learners can customize \algobo{}'s parametrized learning behaviors, such as Mode-shifting. 
Although we can alleviate learners' fatigue and distraction from Mode-shifting by making \algobo{} context-aware and asking questions timely instead of the current rule-based scheme, giving direct control to the question-asking frequency can also help learners manage their load and self-regulate their learning environment.
All these configurations are possible through natural language inputs from the user or a framework that provides users with configurable parameters for better control~\cite{lee2021curiosity}. Future research can look into how the customization and personalization of teachable agents can increase the benefits of LBT even further.

\subsection{Setting the Right Expectation of Teachable Agents}

Teachable agents often have had visual forms of a human student~\cite{Leelawong2008, Matsuda2010, atkinson2002optimizing, moreno2002perceived}. Likewise, we also gave \algobo{} a student-like persona to help learners set initial expectations of tutees. Due to the given persona and unfamiliarity in LBT with virtual agents, many participants put the expectation of a human learner to \algobo{}~\cite{Shoufan2023}. However, the high expectations aggravated awkward instances of \algobo{}'s responses compared to human tutees. \algobo{} asked repetitive questions and could not transfer natural language explanations to code (T7). \algobo{} asked questions (i.e., because it was in the questioner mode) even when tutors asked \algobo{}'s opinions and thoughts, making the question-answering flow unnatural (T20). These clumsy behaviors confused participants in applying effective teaching methods and decreased their satisfaction and engagement. While using better LLMs and a more refined pipeline can alleviate the problem, we argue that reducing the gap between learners' expectations and the capabilities of teachable agents is also fundamental in the context of LBT with AI~\cite{Amershi2019, Luger2016}.

Through the perspective of the gulf of execution and evaluation~\cite{Norman2013}, we suggest some interaction-centric design implications that can close learners' expectation gap in LBT. For the gulf of execution, learners should be better informed about whom and how they teach. For example, learners may receive more detailed explanations of \algobo{}'s operating principles. This can increase learners' tolerance of \algobo{}'s awkward responses and help form an appropriate first impression of agents~\cite{Volodin2020}. The learning system can also inform learners of their expected roles in different phases in Mode-shifting clearly. For instance, when \algobo{} is in the questioner mode, the system can clarify that tutors should focus on providing answers. This will help learners follow the pedagogical conversation flows (e.g., Mode-shifting) and improve learning impact. For the gulf of evaluation, the system can present \algobo{}'s learning progress explicitly. Learning systems can show \algobo{}'s current knowledge state more directly and allow learners to self-assess the effectiveness of their teaching methods. Future research can explore these modifications to make the conversations with teachable agents more satisfactory and predictable.

\section{Limitation and Future Work}

First, the scope of our evaluation is limited to algorithm learning and procedural knowledge in programming. Although our results showed that the Reflect-Respond pipeline is generalizable within different algorithm topics, we need to confirm if the pipeline is generalizable to other subjects (e.g., math and physics) as we have optimized our prompts for programming learning and trained our message classifiers on the binary search dialogues. Moreover, since procedural knowledge and declarative knowledge are different in cognitive processing and effective learning interventions~\cite{Hong2018, Jiamu2012}, \sysname{} may not scaffold declarative knowledge learning effectively. As prior research looked into declarative knowledge learning~\cite{Leelawong2008, Ruan2019}, future studies can investigate more extensive topics outside algorithm learning.

Second, our user study was confined to indirect measures of learning gain. Dialogue quality is one of the primary metrics in LBT adopted in past research~\cite{Graesser2004, King1998}, and we did a comprehensive analysis of knowledge-building through dialogue analysis and surveys. Nevertheless, we can make our findings more concrete by measuring participants' learning gain directly through pre-post test comparison. Although we did not consider a pre-post test because we assumed one-time LBT would not elicit significant performance improvement, future research can design studies to compare the learning gain between conditions and confirm the connection between dialogue quality and learning gain~\cite{Shahriar2021}.

Lastly, future research can deploy \sysname{} to real classrooms of greater size and monitor the longitudinal dynamics among learners' perception, learning gain, and metacognition. Although we could observe statistical significance in some of our measurements, there were high variances among participants, perhaps due to different levels of prior knowledge, teaching styles, and conversational patterns. These properties are hard to control in nature; a user study on larger populations can sharpen the statistics of the results and make our findings more concrete. In addition to the population size, longitudinal studies may reveal significant changes in learners' metacognition and teaching patterns as there is more room for learners to understand the nature of \algobo{} and improve their methods over time.

We plan to deploy our system to the classes offered in our institution, in which students learn different algorithm topics throughout a semester. The classroom deployment will require a configuration interface where instructors can set up class materials and edit \algobo{}'s knowledge state and the prompts in the Reflect-Respond pipeline for their needs. We also need to reduce the response time of \algobo{} (currently about 30 seconds) for practical use, as many participants pointed out. After the small-scale controlled deployment, we envision deploying \sysname{} as an online platform to help instructors of different fields adopt LBT to their classes. LLM-powered LBT will enable the dissemination of interactive learning at scale.

\section{Conclusion}

This work presents \sysname{}, a system for supporting LBT with an LLM-based teachable agent \algobo{} where learners can learn by teaching \algobo{} how to code. To facilitate effective LBT with \algobo{}, we introduced (1) Reflect-Respond prompting pipeline for simulating knowledge learning of \algobo{}, (2) Mode-shifting for eliciting knowledge-building in conversations through \algobo{}'s elaboration questions, and (3) Teaching Helper for providing metacognitive feedback to learners about their teaching styles. Our technical evaluation showed that our Reflect-Respond prompting pipeline could effectively configure, persist, and adapt \algobo{}'s knowledge level. Our user study with 40 algorithm novices confirmed that Mode-shifting improved the density of knowledge-building messages in LBT dialogues. We envision that our approach can help researchers and instructors create LLM-based teachable agents with low manual efforts and barriers and support learners to excel in their learning with engaging learning experiences.

\begin{acks}
This work was supported by Algorithm LABS and Elice. 
\end{acks}

\bibliographystyle{ACM-Reference-Format}
\bibliography{references.bib}


\begin{thebibliography}{98}


\ifx \showCODEN    \undefined \def \showCODEN     #1{\unskip}     \fi
\ifx \showDOI      \undefined \def \showDOI       #1{#1}\fi
\ifx \showISBNx    \undefined \def \showISBNx     #1{\unskip}     \fi
\ifx \showISBNxiii \undefined \def \showISBNxiii  #1{\unskip}     \fi
\ifx \showISSN     \undefined \def \showISSN      #1{\unskip}     \fi
\ifx \showLCCN     \undefined \def \showLCCN      #1{\unskip}     \fi
\ifx \shownote     \undefined \def \shownote      #1{#1}          \fi
\ifx \showarticletitle \undefined \def \showarticletitle #1{#1}   \fi
\ifx \showURL      \undefined \def \showURL       {\relax}        \fi
\providecommand\bibfield[2]{#2}
\providecommand\bibinfo[2]{#2}
\providecommand\natexlab[1]{#1}
\providecommand\showeprint[2][]{arXiv:#2}

\bibitem[Abdelghani et~al\mbox{.}(2022)]%
        {abdelghani2022conversational}
\bibfield{author}{\bibinfo{person}{Rania Abdelghani}, \bibinfo{person}{Pierre-Yves Oudeyer}, \bibinfo{person}{Edith Law}, \bibinfo{person}{Catherine de Vulpilli{\`e}res}, {and} \bibinfo{person}{H{\'e}l{\`e}ne Sauz{\'e}on}.} \bibinfo{year}{2022}\natexlab{}.
\newblock \showarticletitle{Conversational agents for fostering curiosity-driven learning in children}.
\newblock \bibinfo{journal}{\emph{International Journal of Human-Computer Studies}}  \bibinfo{volume}{167} (\bibinfo{year}{2022}), \bibinfo{pages}{102887}.
\newblock


\bibitem[Aflalo(2021)]%
        {aflalo2021students}
\bibfield{author}{\bibinfo{person}{Ester Aflalo}.} \bibinfo{year}{2021}\natexlab{}.
\newblock \showarticletitle{Students generating questions as a way of learning}.
\newblock \bibinfo{journal}{\emph{Active Learning in Higher Education}} \bibinfo{volume}{22}, \bibinfo{number}{1} (\bibinfo{year}{2021}), \bibinfo{pages}{63--75}.
\newblock


\bibitem[Allamanis and Sutton(2013)]%
        {Allamanis2013}
\bibfield{author}{\bibinfo{person}{Miltiadis Allamanis} {and} \bibinfo{person}{Charles Sutton}.} \bibinfo{year}{2013}\natexlab{}.
\newblock \showarticletitle{Why, when, and what: analyzing stack overflow questions by topic, type, and code}. In \bibinfo{booktitle}{\emph{Proceedings of the 10th Working Conference on Mining Software Repositories}} (San Francisco, CA, USA) \emph{(\bibinfo{series}{MSR '13})}. \bibinfo{publisher}{IEEE Press}, \bibinfo{pages}{53–56}.
\newblock
\showISBNx{9781467329361}


\bibitem[Amershi et~al\mbox{.}(2019)]%
        {Amershi2019}
\bibfield{author}{\bibinfo{person}{Saleema Amershi}, \bibinfo{person}{Dan Weld}, \bibinfo{person}{Mihaela Vorvoreanu}, \bibinfo{person}{Adam Fourney}, \bibinfo{person}{Besmira Nushi}, \bibinfo{person}{Penny Collisson}, \bibinfo{person}{Jina Suh}, \bibinfo{person}{Shamsi Iqbal}, \bibinfo{person}{Paul~N. Bennett}, \bibinfo{person}{Kori Inkpen}, \bibinfo{person}{Jaime Teevan}, \bibinfo{person}{Ruth Kikin-Gil}, {and} \bibinfo{person}{Eric Horvitz}.} \bibinfo{year}{2019}\natexlab{}.
\newblock \showarticletitle{Guidelines for Human-AI Interaction}. In \bibinfo{booktitle}{\emph{Proceedings of the 2019 CHI Conference on Human Factors in Computing Systems}} (Glasgow, Scotland Uk) \emph{(\bibinfo{series}{CHI '19})}. \bibinfo{publisher}{Association for Computing Machinery}, \bibinfo{address}{New York, NY, USA}, \bibinfo{pages}{1–13}.
\newblock
\showISBNx{9781450359702}
\urldef\tempurl%
\url{https://doi.org/10.1145/3290605.3300233}
\showDOI{\tempurl}


\bibitem[Atkinson(2002)]%
        {atkinson2002optimizing}
\bibfield{author}{\bibinfo{person}{Robert~K Atkinson}.} \bibinfo{year}{2002}\natexlab{}.
\newblock \showarticletitle{Optimizing learning from examples using animated pedagogical agents.}
\newblock \bibinfo{journal}{\emph{Journal of Educational Psychology}} \bibinfo{volume}{94}, \bibinfo{number}{2} (\bibinfo{year}{2002}), \bibinfo{pages}{416}.
\newblock


\bibitem[Ausubel(1962)]%
        {ausubel1962subsumption}
\bibfield{author}{\bibinfo{person}{David~P Ausubel}.} \bibinfo{year}{1962}\natexlab{}.
\newblock \showarticletitle{A subsumption theory of meaningful verbal learning and retention}.
\newblock \bibinfo{journal}{\emph{The Journal of general psychology}} \bibinfo{volume}{66}, \bibinfo{number}{2} (\bibinfo{year}{1962}), \bibinfo{pages}{213--224}.
\newblock


\bibitem[Biswas et~al\mbox{.}(2001)]%
        {Biswas2001}
\bibfield{author}{\bibinfo{person}{Gautam Biswas}, \bibinfo{person}{Thomas Katzlberger}, \bibinfo{person}{John Bransford}, \bibinfo{person}{Daniel Schwartz}, {et~al\mbox{.}}} \bibinfo{year}{2001}\natexlab{}.
\newblock \showarticletitle{Extending intelligent learning environments with teachable agents to enhance learning}. In \bibinfo{booktitle}{\emph{Artificial intelligence in education}}. Citeseer, \bibinfo{pages}{389--397}.
\newblock


\bibitem[Biswas et~al\mbox{.}(2005)]%
        {Biswas2005}
\bibfield{author}{\bibinfo{person}{Gautam Biswas}, \bibinfo{person}{Krittaya Leelawong}, \bibinfo{person}{Daniel Schwartz}, \bibinfo{person}{Nancy Vye}, {and} \bibinfo{person}{{The Teachable Agents Group at Vanderbilt}}.} \bibinfo{year}{2005}\natexlab{}.
\newblock \showarticletitle{Learning by Teaching: A New Agent Paradigm for Educational Software}.
\newblock \bibinfo{journal}{\emph{Applied Artificial Intelligence}} \bibinfo{volume}{19}, \bibinfo{number}{3-4} (\bibinfo{year}{2005}), \bibinfo{pages}{363--392}.
\newblock
\urldef\tempurl%
\url{https://doi.org/10.1080/08839510590910200}
\showDOI{\tempurl}


\bibitem[Blair et~al\mbox{.}(2007)]%
        {Blair2007}
\bibfield{author}{\bibinfo{person}{Kristen Blair}, \bibinfo{person}{Daniel~L Schwartz}, \bibinfo{person}{Gautam Biswas}, {and} \bibinfo{person}{Krittaya Leelawong}.} \bibinfo{year}{2007}\natexlab{}.
\newblock \showarticletitle{Pedagogical agents for learning by teaching: Teachable agents}.
\newblock \bibinfo{journal}{\emph{Educational Technology}} (\bibinfo{year}{2007}), \bibinfo{pages}{56--61}.
\newblock


\bibitem[Bloom(1968)]%
        {Bloom1968}
\bibfield{author}{\bibinfo{person}{Benjamin~S Bloom}.} \bibinfo{year}{1968}\natexlab{}.
\newblock \showarticletitle{Learning for Mastery. Instruction and Curriculum. Regional Education Laboratory for the Carolinas and Virginia, Topical Papers and Reprints, Number 1.}
\newblock \bibinfo{journal}{\emph{Evaluation comment}} \bibinfo{volume}{1}, \bibinfo{number}{2} (\bibinfo{year}{1968}), \bibinfo{pages}{n2}.
\newblock


\bibitem[Bredeweg et~al\mbox{.}(2007)]%
        {Bredeweg2007}
\bibfield{author}{\bibinfo{person}{Bert Bredeweg}, \bibinfo{person}{Anders Bouwer}, \bibinfo{person}{Jelmer Jellema}, \bibinfo{person}{Dirk Bertels}, \bibinfo{person}{Floris~Floris Linnebank}, {and} \bibinfo{person}{Jochem Liem}.} \bibinfo{year}{2007}\natexlab{}.
\newblock \showarticletitle{Garp3: a new workbench for qualitative reasoning and modelling}. In \bibinfo{booktitle}{\emph{Proceedings of the 4th International Conference on Knowledge Capture}} (Whistler, BC, Canada) \emph{(\bibinfo{series}{K-CAP '07})}. \bibinfo{publisher}{Association for Computing Machinery}, \bibinfo{address}{New York, NY, USA}, \bibinfo{pages}{183–184}.
\newblock
\showISBNx{9781595936431}
\urldef\tempurl%
\url{https://doi.org/10.1145/1298406.1298445}
\showDOI{\tempurl}


\bibitem[Brown et~al\mbox{.}(2020)]%
        {Brown2020}
\bibfield{author}{\bibinfo{person}{Tom Brown}, \bibinfo{person}{Benjamin Mann}, \bibinfo{person}{Nick Ryder}, \bibinfo{person}{Melanie Subbiah}, \bibinfo{person}{Jared~D Kaplan}, \bibinfo{person}{Prafulla Dhariwal}, \bibinfo{person}{Arvind Neelakantan}, \bibinfo{person}{Pranav Shyam}, \bibinfo{person}{Girish Sastry}, \bibinfo{person}{Amanda Askell}, \bibinfo{person}{Sandhini Agarwal}, \bibinfo{person}{Ariel Herbert-Voss}, \bibinfo{person}{Gretchen Krueger}, \bibinfo{person}{Tom Henighan}, \bibinfo{person}{Rewon Child}, \bibinfo{person}{Aditya Ramesh}, \bibinfo{person}{Daniel Ziegler}, \bibinfo{person}{Jeffrey Wu}, \bibinfo{person}{Clemens Winter}, \bibinfo{person}{Chris Hesse}, \bibinfo{person}{Mark Chen}, \bibinfo{person}{Eric Sigler}, \bibinfo{person}{Mateusz Litwin}, \bibinfo{person}{Scott Gray}, \bibinfo{person}{Benjamin Chess}, \bibinfo{person}{Jack Clark}, \bibinfo{person}{Christopher Berner}, \bibinfo{person}{Sam McCandlish}, \bibinfo{person}{Alec Radford}, \bibinfo{person}{Ilya Sutskever}, {and}
  \bibinfo{person}{Dario Amodei}.} \bibinfo{year}{2020}\natexlab{}.
\newblock \showarticletitle{Language Models are Few-Shot Learners}. In \bibinfo{booktitle}{\emph{Advances in Neural Information Processing Systems}}, \bibfield{editor}{\bibinfo{person}{H.~Larochelle}, \bibinfo{person}{M.~Ranzato}, \bibinfo{person}{R.~Hadsell}, \bibinfo{person}{M.F. Balcan}, {and} \bibinfo{person}{H.~Lin}} (Eds.), Vol.~\bibinfo{volume}{33}. \bibinfo{publisher}{Curran Associates, Inc.}, \bibinfo{pages}{1877--1901}.
\newblock
\urldef\tempurl%
\url{https://proceedings.neurips.cc/paper_files/paper/2020/file/1457c0d6bfcb4967418bfb8ac142f64a-Paper.pdf}
\showURL{%
\tempurl}


\bibitem[Chase et~al\mbox{.}(2009)]%
        {Chase2009}
\bibfield{author}{\bibinfo{person}{Catherine~C. Chase}, \bibinfo{person}{Doris~B. Chin}, \bibinfo{person}{Marily~A Oppezzo}, {and} \bibinfo{person}{Daniel~L. Schwartz}.} \bibinfo{year}{2009}\natexlab{}.
\newblock \showarticletitle{Teachable Agents and the Prot{\'e}g{\'e} Effect: Increasing the Effort Towards Learning}.
\newblock \bibinfo{journal}{\emph{Journal of Science Education and Technology}}  \bibinfo{volume}{18} (\bibinfo{year}{2009}), \bibinfo{pages}{334--352}.
\newblock


\bibitem[Chen et~al\mbox{.}(2023)]%
        {chen2023chatcot}
\bibfield{author}{\bibinfo{person}{Zhipeng Chen}, \bibinfo{person}{Kun Zhou}, \bibinfo{person}{Beichen Zhang}, \bibinfo{person}{Zheng Gong}, \bibinfo{person}{Xin Zhao}, {and} \bibinfo{person}{Ji-Rong Wen}.} \bibinfo{year}{2023}\natexlab{}.
\newblock \showarticletitle{{C}hat{C}o{T}: Tool-Augmented Chain-of-Thought Reasoning on Chat-based Large Language Models}. In \bibinfo{booktitle}{\emph{Findings of the Association for Computational Linguistics: EMNLP 2023}}, \bibfield{editor}{\bibinfo{person}{Houda Bouamor}, \bibinfo{person}{Juan Pino}, {and} \bibinfo{person}{Kalika Bali}} (Eds.). \bibinfo{publisher}{Association for Computational Linguistics}, \bibinfo{address}{Singapore}, \bibinfo{pages}{14777--14790}.
\newblock
\urldef\tempurl%
\url{https://doi.org/10.18653/v1/2023.findings-emnlp.985}
\showDOI{\tempurl}


\bibitem[Chi and Wylie(2014)]%
        {Chi2014}
\bibfield{author}{\bibinfo{person}{M.~T.~H. Chi} {and} \bibinfo{person}{R. Wylie}.} \bibinfo{year}{2014}\natexlab{}.
\newblock \showarticletitle{The {ICAP} framework: Linking cognitive engagement to active learning outcomes}.
\newblock \bibinfo{journal}{\emph{Educational Psychologist}} \bibinfo{volume}{49}, \bibinfo{number}{4} (\bibinfo{year}{2014}), \bibinfo{pages}{219--243}.
\newblock
\urldef\tempurl%
\url{https://doi.org/10.1080/00461520.2014.965823}
\showDOI{\tempurl}


\bibitem[Chin and Brown(2002)]%
        {chin2002student}
\bibfield{author}{\bibinfo{person}{Christine Chin} {and} \bibinfo{person}{David~E Brown}.} \bibinfo{year}{2002}\natexlab{}.
\newblock \showarticletitle{Student-generated questions: A meaningful aspect of learning in science}.
\newblock \bibinfo{journal}{\emph{International Journal of Science Education}} \bibinfo{volume}{24}, \bibinfo{number}{5} (\bibinfo{year}{2002}), \bibinfo{pages}{521--549}.
\newblock


\bibitem[Chin et~al\mbox{.}(2013)]%
        {Chin2013}
\bibfield{author}{\bibinfo{person}{Doris Chin}, \bibinfo{person}{I.M. Dohmen}, {and} \bibinfo{person}{D.L. Schwartz}.} \bibinfo{year}{2013}\natexlab{}.
\newblock \showarticletitle{Young Children Can Learn Scientific Reasoning with Teachable Agents}.
\newblock \bibinfo{journal}{\emph{Learning Technologies, IEEE Transactions on}}  \bibinfo{volume}{6} (\bibinfo{date}{07} \bibinfo{year}{2013}), \bibinfo{pages}{248--257}.
\newblock
\urldef\tempurl%
\url{https://doi.org/10.1109/TLT.2013.24}
\showDOI{\tempurl}


\bibitem[Chin et~al\mbox{.}(2010)]%
        {Chin2010}
\bibfield{author}{\bibinfo{person}{Doris~B. Chin}, \bibinfo{person}{Ilsa~M. Dohmen}, \bibinfo{person}{Britte~Haugan Cheng}, \bibinfo{person}{Marily~A Oppezzo}, \bibinfo{person}{Catherine~C. Chase}, {and} \bibinfo{person}{Daniel~L. Schwartz}.} \bibinfo{year}{2010}\natexlab{}.
\newblock \showarticletitle{Preparing students for future learning with Teachable Agents}.
\newblock \bibinfo{journal}{\emph{Educational Technology Research and Development}}  \bibinfo{volume}{58} (\bibinfo{year}{2010}), \bibinfo{pages}{649--669}.
\newblock


\bibitem[Cohen et~al\mbox{.}(2023)]%
        {Cohen2023}
\bibfield{author}{\bibinfo{person}{Roi Cohen}, \bibinfo{person}{Eden Biran}, \bibinfo{person}{Ori Yoran}, \bibinfo{person}{Amir Globerson}, {and} \bibinfo{person}{Mor Geva}.} \bibinfo{year}{2023}\natexlab{}.
\newblock \bibinfo{title}{Evaluating the Ripple Effects of Knowledge Editing in Language Models}.
\newblock
\newblock
\showeprint[arxiv]{2307.12976}~[cs.CL]


\bibitem[Debban\'{e} et~al\mbox{.}(2023)]%
        {Debban2023}
\bibfield{author}{\bibinfo{person}{Amy Debban\'{e}}, \bibinfo{person}{Ken~Jen Lee}, \bibinfo{person}{Jarvis Tse}, {and} \bibinfo{person}{Edith Law}.} \bibinfo{year}{2023}\natexlab{}.
\newblock \showarticletitle{Learning by Teaching: Key Challenges and Design Implications}.
\newblock \bibinfo{journal}{\emph{Proc. ACM Hum.-Comput. Interact.}} \bibinfo{volume}{7}, \bibinfo{number}{CSCW1}, Article \bibinfo{articleno}{68} (\bibinfo{date}{apr} \bibinfo{year}{2023}), \bibinfo{numpages}{34}~pages.
\newblock
\urldef\tempurl%
\url{https://doi.org/10.1145/3579501}
\showDOI{\tempurl}


\bibitem[Denny et~al\mbox{.}(2008)]%
        {Denny2008}
\bibfield{author}{\bibinfo{person}{Paul Denny}, \bibinfo{person}{Andrew Luxton-Reilly}, {and} \bibinfo{person}{Beth Simon}.} \bibinfo{year}{2008}\natexlab{}.
\newblock \showarticletitle{Evaluating a new exam question: Parsons problems}. In \bibinfo{booktitle}{\emph{Proceedings of the Fourth International Workshop on Computing Education Research}} (Sydney, Australia) \emph{(\bibinfo{series}{ICER '08})}. \bibinfo{publisher}{Association for Computing Machinery}, \bibinfo{address}{New York, NY, USA}, \bibinfo{pages}{113–124}.
\newblock
\showISBNx{9781605582160}
\urldef\tempurl%
\url{https://doi.org/10.1145/1404520.1404532}
\showDOI{\tempurl}


\bibitem[Duran(2017)]%
        {Duran2017}
\bibfield{author}{\bibinfo{person}{David Duran}.} \bibinfo{year}{2017}\natexlab{}.
\newblock \showarticletitle{Learning-by-teaching. Evidence and implications as a pedagogical mechanism}.
\newblock \bibinfo{journal}{\emph{Innovations in Education and Teaching International}} \bibinfo{volume}{54}, \bibinfo{number}{5} (\bibinfo{year}{2017}), \bibinfo{pages}{476--484}.
\newblock
\urldef\tempurl%
\url{https://doi.org/10.1080/14703297.2016.1156011}
\showDOI{\tempurl}


\bibitem[Gao(2023)]%
        {Gao2023}
\bibfield{author}{\bibinfo{person}{Leo Gao}.} \bibinfo{year}{2023}\natexlab{}.
\newblock \bibinfo{title}{Shapley {Value} {Attribution} in {Chain} of {Thought}}.
\newblock
\newblock
\urldef\tempurl%
\url{https://www.lesswrong.com/posts/FX5JmftqL2j6K8dn4/shapley-value-attribution-in-chain-of-thought}
\showURL{%
\tempurl}


\bibitem[Glassman et~al\mbox{.}(2016)]%
        {Glassman2016}
\bibfield{author}{\bibinfo{person}{Elena~L. Glassman}, \bibinfo{person}{Aaron Lin}, \bibinfo{person}{Carrie~J. Cai}, {and} \bibinfo{person}{Robert~C. Miller}.} \bibinfo{year}{2016}\natexlab{}.
\newblock \showarticletitle{Learnersourcing Personalized Hints}. In \bibinfo{booktitle}{\emph{Proceedings of the 19th ACM Conference on Computer-Supported Cooperative Work \& Social Computing}} (San Francisco, California, USA) \emph{(\bibinfo{series}{CSCW '16})}. \bibinfo{publisher}{Association for Computing Machinery}, \bibinfo{address}{New York, NY, USA}, \bibinfo{pages}{1626–1636}.
\newblock
\showISBNx{9781450335928}
\urldef\tempurl%
\url{https://doi.org/10.1145/2818048.2820011}
\showDOI{\tempurl}


\bibitem[Graesser et~al\mbox{.}(2004)]%
        {Graesser2004}
\bibfield{author}{\bibinfo{person}{Arthur Graesser}, \bibinfo{person}{Shulan Lu}, \bibinfo{person}{G. Jackson}, \bibinfo{person}{Heather Mitchell}, \bibinfo{person}{Mathew Ventura}, \bibinfo{person}{Andrew Olney}, {and} \bibinfo{person}{Max Louwerse}.} \bibinfo{year}{2004}\natexlab{}.
\newblock \showarticletitle{AutoTutor: a Tutor with Dialogue in Natural Language}.
\newblock \bibinfo{journal}{\emph{Behavior Research Methods}}  \bibinfo{volume}{36} (\bibinfo{date}{06} \bibinfo{year}{2004}), \bibinfo{pages}{180--192}.
\newblock
\urldef\tempurl%
\url{https://doi.org/10.3758/BF03195563}
\showDOI{\tempurl}


\bibitem[Guo(2013)]%
        {Guo2013}
\bibfield{author}{\bibinfo{person}{Philip~J. Guo}.} \bibinfo{year}{2013}\natexlab{}.
\newblock \showarticletitle{Online Python Tutor: Embeddable Web-Based Program Visualization for Cs Education}. In \bibinfo{booktitle}{\emph{Proceeding of the 44th ACM Technical Symposium on Computer Science Education}} (Denver, Colorado, USA) \emph{(\bibinfo{series}{SIGCSE '13})}. \bibinfo{publisher}{Association for Computing Machinery}, \bibinfo{address}{New York, NY, USA}, \bibinfo{pages}{579–584}.
\newblock
\showISBNx{9781450318686}
\urldef\tempurl%
\url{https://doi.org/10.1145/2445196.2445368}
\showDOI{\tempurl}


\bibitem[Jiamu(2001)]%
        {Jiamu2012}
\bibfield{author}{\bibinfo{person}{Chen Jiamu}.} \bibinfo{year}{2001}\natexlab{}.
\newblock \showarticletitle{The great importance of the distinction between declarative and procedural knowledge}.
\newblock \bibinfo{journal}{\emph{An{\'a}lise Psicol{\'o}gica}} \bibinfo{volume}{19}, \bibinfo{number}{4} (\bibinfo{year}{2001}), \bibinfo{pages}{559--566}.
\newblock


\bibitem[Jianzhong~Hong and Yang(2018)]%
        {Hong2018}
\bibfield{author}{\bibinfo{person}{Zhongling~Pi Jianzhong~Hong} {and} \bibinfo{person}{Jiumin Yang}.} \bibinfo{year}{2018}\natexlab{}.
\newblock \showarticletitle{Learning declarative and procedural knowledge via video lectures: cognitive load and learning effectiveness}.
\newblock \bibinfo{journal}{\emph{Innovations in Education and Teaching International}} \bibinfo{volume}{55}, \bibinfo{number}{1} (\bibinfo{year}{2018}), \bibinfo{pages}{74--81}.
\newblock
\urldef\tempurl%
\url{https://doi.org/10.1080/14703297.2016.1237371}
\showDOI{\tempurl}


\bibitem[Jin et~al\mbox{.}(2019)]%
        {Jin2019}
\bibfield{author}{\bibinfo{person}{Hyoungwook Jin}, \bibinfo{person}{Minsuk Chang}, {and} \bibinfo{person}{Juho Kim}.} \bibinfo{year}{2019}\natexlab{}.
\newblock \showarticletitle{SolveDeep: A System for Supporting Subgoal Learning in Online Math Problem Solving}. In \bibinfo{booktitle}{\emph{Extended Abstracts of the 2019 CHI Conference on Human Factors in Computing Systems}} (Glasgow, Scotland Uk) \emph{(\bibinfo{series}{CHI EA '19})}. \bibinfo{publisher}{Association for Computing Machinery}, \bibinfo{address}{New York, NY, USA}, \bibinfo{pages}{1–6}.
\newblock
\showISBNx{9781450359719}
\urldef\tempurl%
\url{https://doi.org/10.1145/3290607.3312822}
\showDOI{\tempurl}


\bibitem[Junprung(2023)]%
        {junprung2023exploring}
\bibfield{author}{\bibinfo{person}{Edward Junprung}.} \bibinfo{year}{2023}\natexlab{}.
\newblock \showarticletitle{Exploring the intersection of large language models and agent-based modeling via prompt engineering}.
\newblock \bibinfo{journal}{\emph{arXiv preprint arXiv:2308.07411}} (\bibinfo{year}{2023}).
\newblock


\bibitem[Kay et~al\mbox{.}(1997)]%
        {Kay1997}
\bibfield{author}{\bibinfo{person}{Judy Kay}, \bibinfo{person}{Z Halin}, \bibinfo{person}{T Ottomann}, {and} \bibinfo{person}{Z Razak}.} \bibinfo{year}{1997}\natexlab{}.
\newblock \showarticletitle{Learner know thyself: Student models to give learner control and responsibility}. In \bibinfo{booktitle}{\emph{Proceedings of international conference on computers in education}}. \bibinfo{pages}{17--24}.
\newblock


\bibitem[Ketamo(2009)]%
        {Ketamo2009}
\bibfield{author}{\bibinfo{person}{Harri Ketamo}.} \bibinfo{year}{2009}\natexlab{}.
\newblock \showarticletitle{Semantic Networks -Based Teachable Agents in an Educational Game}.
\newblock \bibinfo{journal}{\emph{W. Trans. on Comp.}} \bibinfo{volume}{8}, \bibinfo{number}{4} (\bibinfo{date}{apr} \bibinfo{year}{2009}), \bibinfo{pages}{641–650}.
\newblock
\showISSN{1109-2750}


\bibitem[Kim et~al\mbox{.}(2024)]%
        {kim2023language}
\bibfield{author}{\bibinfo{person}{Geunwoo Kim}, \bibinfo{person}{Pierre Baldi}, {and} \bibinfo{person}{Stephen McAleer}.} \bibinfo{year}{2024}\natexlab{}.
\newblock \showarticletitle{Language models can solve computer tasks}.
\newblock \bibinfo{journal}{\emph{Advances in Neural Information Processing Systems}}  \bibinfo{volume}{36} (\bibinfo{year}{2024}).
\newblock


\bibitem[King(1994)]%
        {King1994}
\bibfield{author}{\bibinfo{person}{Alison King}.} \bibinfo{year}{1994}\natexlab{}.
\newblock \showarticletitle{Guiding Knowledge Construction in the Classroom: Effects of Teaching Children How to Question and How to Explain}.
\newblock \bibinfo{journal}{\emph{American Educational Research Journal}} \bibinfo{volume}{31}, \bibinfo{number}{2} (\bibinfo{year}{1994}), \bibinfo{pages}{338--368}.
\newblock
\urldef\tempurl%
\url{https://doi.org/10.3102/00028312031002338}
\showDOI{\tempurl}
\showeprint{https://doi.org/10.3102/00028312031002338}


\bibitem[King(1997)]%
        {King1997}
\bibfield{author}{\bibinfo{person}{Alison King}.} \bibinfo{year}{1997}\natexlab{}.
\newblock \showarticletitle{ASK to THINK-TEL WHY: A model of transactive peer tutoring for scaffolding higher level complex learning}.
\newblock \bibinfo{journal}{\emph{Educational Psychologist}} \bibinfo{volume}{32}, \bibinfo{number}{4} (\bibinfo{year}{1997}), \bibinfo{pages}{221--235}.
\newblock
\urldef\tempurl%
\url{https://doi.org/10.1207/s15326985ep3204\_3}
\showDOI{\tempurl}


\bibitem[King et~al\mbox{.}(1998)]%
        {King1998}
\bibfield{author}{\bibinfo{person}{A. King}, \bibinfo{person}{A. Staffieri}, {and} \bibinfo{person}{A. Adelgais}.} \bibinfo{year}{1998}\natexlab{}.
\newblock \showarticletitle{Mutual peer tutoring: Effects of structuring tutorial interaction to scaffold peer learning}.
\newblock \bibinfo{journal}{\emph{Journal of Educational Psychology}} \bibinfo{volume}{90}, \bibinfo{number}{1} (\bibinfo{year}{1998}), \bibinfo{pages}{134--152}.
\newblock
\urldef\tempurl%
\url{https://doi.org/10.1037/0022-0663.90.1.134}
\showDOI{\tempurl}


\bibitem[Kong et~al\mbox{.}(2023)]%
        {kong2023better}
\bibfield{author}{\bibinfo{person}{Aobo Kong}, \bibinfo{person}{Shiwan Zhao}, \bibinfo{person}{Hao Chen}, \bibinfo{person}{Qicheng Li}, \bibinfo{person}{Yong Qin}, \bibinfo{person}{Ruiqi Sun}, {and} \bibinfo{person}{Xin Zhou}.} \bibinfo{year}{2023}\natexlab{}.
\newblock \showarticletitle{Better Zero-Shot Reasoning with Role-Play Prompting}.
\newblock \bibinfo{journal}{\emph{arXiv preprint arXiv:2308.07702}} (\bibinfo{year}{2023}).
\newblock


\bibitem[Krathwohl(2002)]%
        {Krathwohl2002}
\bibfield{author}{\bibinfo{person}{David~R. Krathwohl}.} \bibinfo{year}{2002}\natexlab{}.
\newblock \showarticletitle{A Revision of Bloom's Taxonomy: An Overview}.
\newblock \bibinfo{journal}{\emph{Theory Into Practice}} \bibinfo{volume}{41}, \bibinfo{number}{4} (\bibinfo{year}{2002}), \bibinfo{pages}{212--218}.
\newblock
\urldef\tempurl%
\url{https://doi.org/10.1207/s15430421tip4104\_2}
\showDOI{\tempurl}


\bibitem[Lee et~al\mbox{.}(2022)]%
        {Lee2022}
\bibfield{author}{\bibinfo{person}{Changyoon Lee}, \bibinfo{person}{Yeon Seonwoo}, {and} \bibinfo{person}{Alice Oh}.} \bibinfo{year}{2022}\natexlab{}.
\newblock \showarticletitle{{CS}1{QA}: A Dataset for Assisting Code-based Question Answering in an Introductory Programming Course}. In \bibinfo{booktitle}{\emph{Proceedings of the 2022 Conference of the North American Chapter of the Association for Computational Linguistics: Human Language Technologies}}, \bibfield{editor}{\bibinfo{person}{Marine Carpuat}, \bibinfo{person}{Marie-Catherine de~Marneffe}, {and} \bibinfo{person}{Ivan~Vladimir Meza~Ruiz}} (Eds.). \bibinfo{publisher}{Association for Computational Linguistics}, \bibinfo{address}{Seattle, United States}, \bibinfo{pages}{2026--2040}.
\newblock
\urldef\tempurl%
\url{https://doi.org/10.18653/v1/2022.naacl-main.148}
\showDOI{\tempurl}


\bibitem[Lee et~al\mbox{.}(2021)]%
        {lee2021curiosity}
\bibfield{author}{\bibinfo{person}{Ken~Jen Lee}, \bibinfo{person}{Apoorva Chauhan}, \bibinfo{person}{Joslin Goh}, \bibinfo{person}{Elizabeth Nilsen}, {and} \bibinfo{person}{Edith Law}.} \bibinfo{year}{2021}\natexlab{}.
\newblock \showarticletitle{Curiosity notebook: the design of a research platform for learning by teaching}.
\newblock \bibinfo{journal}{\emph{Proceedings of the ACM on Human-Computer Interaction}} \bibinfo{volume}{5}, \bibinfo{number}{CSCW2} (\bibinfo{year}{2021}), \bibinfo{pages}{1--26}.
\newblock


\bibitem[Leelawong and Biswas(2008)]%
        {Leelawong2008}
\bibfield{author}{\bibinfo{person}{Krittaya Leelawong} {and} \bibinfo{person}{Gautam Biswas}.} \bibinfo{year}{2008}\natexlab{}.
\newblock \showarticletitle{Designing Learning by Teaching Agents: The Betty's Brain System}.
\newblock \bibinfo{journal}{\emph{Int. J. Artif. Intell. Ed.}} \bibinfo{volume}{18}, \bibinfo{number}{3} (\bibinfo{date}{aug} \bibinfo{year}{2008}), \bibinfo{pages}{181–208}.
\newblock
\showISSN{1560-4292}


\bibitem[Li et~al\mbox{.}(2023)]%
        {Li2023}
\bibfield{author}{\bibinfo{person}{Xiaopeng Li}, \bibinfo{person}{Shasha Li}, \bibinfo{person}{Shezheng Song}, \bibinfo{person}{Jing Yang}, \bibinfo{person}{Jun Ma}, {and} \bibinfo{person}{Jie Yu}.} \bibinfo{year}{2023}\natexlab{}.
\newblock \bibinfo{title}{PMET:}.
\newblock
\newblock
\showeprint[arxiv]{2308.08742}~[cs.CL]


\bibitem[Lin et~al\mbox{.}(2023)]%
        {lin2023swiftsage}
\bibfield{author}{\bibinfo{person}{Bill~Yuchen Lin}, \bibinfo{person}{Yicheng Fu}, \bibinfo{person}{Karina Yang}, \bibinfo{person}{Faeze Brahman}, \bibinfo{person}{Shiyu Huang}, \bibinfo{person}{Chandra Bhagavatula}, \bibinfo{person}{Prithviraj Ammanabrolu}, \bibinfo{person}{Yejin Choi}, {and} \bibinfo{person}{Xiang Ren}.} \bibinfo{year}{2023}\natexlab{}.
\newblock \showarticletitle{SwiftSage: A Generative Agent with Fast and Slow Thinking for Complex Interactive Tasks}. In \bibinfo{booktitle}{\emph{Thirty-seventh Conference on Neural Information Processing Systems}}.
\newblock


\bibitem[Long(2023)]%
        {Long2023}
\bibfield{author}{\bibinfo{person}{Jieyi Long}.} \bibinfo{year}{2023}\natexlab{}.
\newblock \bibinfo{title}{Large Language Model Guided Tree-of-Thought}.
\newblock
\newblock
\showeprint[arxiv]{2305.08291}~[cs.AI]


\bibitem[Luger and Sellen(2016)]%
        {Luger2016}
\bibfield{author}{\bibinfo{person}{Ewa Luger} {and} \bibinfo{person}{Abigail Sellen}.} \bibinfo{year}{2016}\natexlab{}.
\newblock \showarticletitle{"Like Having a Really Bad PA": The Gulf between User Expectation and Experience of Conversational Agents}. In \bibinfo{booktitle}{\emph{Proceedings of the 2016 CHI Conference on Human Factors in Computing Systems}} (San Jose, California, USA) \emph{(\bibinfo{series}{CHI '16})}. \bibinfo{publisher}{Association for Computing Machinery}, \bibinfo{address}{New York, NY, USA}, \bibinfo{pages}{5286–5297}.
\newblock
\showISBNx{9781450333627}
\urldef\tempurl%
\url{https://doi.org/10.1145/2858036.2858288}
\showDOI{\tempurl}


\bibitem[Markel et~al\mbox{.}(2023)]%
        {Markel2023}
\bibfield{author}{\bibinfo{person}{Julia~M Markel}, \bibinfo{person}{Steven~G Opferman}, \bibinfo{person}{James~A Landay}, {and} \bibinfo{person}{Chris Piech}.} \bibinfo{year}{2023}\natexlab{}.
\newblock \bibinfo{title}{GPTeach: Interactive TA Training with GPT-based Students}.
\newblock
\newblock
\urldef\tempurl%
\url{https://doi.org/10.1145/3573051.3593393}
\showDOI{\tempurl}


\bibitem[Matsuda(2021)]%
        {Matsuda2021}
\bibfield{author}{\bibinfo{person}{Noboru Matsuda}.} \bibinfo{year}{2021}\natexlab{}.
\newblock \showarticletitle{Teachable Agent as an Interactive Tool for Cognitive Task Analysis: A Case Study for Authoring an Expert Model}.
\newblock \bibinfo{journal}{\emph{International Journal of Artificial Intelligence in Education}}  \bibinfo{volume}{32} (\bibinfo{date}{07} \bibinfo{year}{2021}).
\newblock
\urldef\tempurl%
\url{https://doi.org/10.1007/s40593-021-00265-z}
\showDOI{\tempurl}


\bibitem[Matsuda(2022)]%
        {matsuda2022teachable}
\bibfield{author}{\bibinfo{person}{Noboru Matsuda}.} \bibinfo{year}{2022}\natexlab{}.
\newblock \showarticletitle{Teachable agent as an interactive tool for cognitive task analysis: A case study for authoring an expert model}.
\newblock \bibinfo{journal}{\emph{International Journal of Artificial Intelligence in Education}} \bibinfo{volume}{32}, \bibinfo{number}{1} (\bibinfo{year}{2022}), \bibinfo{pages}{48--75}.
\newblock


\bibitem[Matsuda et~al\mbox{.}(2015)]%
        {matsuda2015teaching}
\bibfield{author}{\bibinfo{person}{Noboru Matsuda}, \bibinfo{person}{William~W Cohen}, {and} \bibinfo{person}{Kenneth~R Koedinger}.} \bibinfo{year}{2015}\natexlab{}.
\newblock \showarticletitle{Teaching the teacher: tutoring SimStudent leads to more effective cognitive tutor authoring}.
\newblock \bibinfo{journal}{\emph{International Journal of Artificial Intelligence in Education}}  \bibinfo{volume}{25} (\bibinfo{year}{2015}), \bibinfo{pages}{1--34}.
\newblock


\bibitem[Matsuda et~al\mbox{.}(2012)]%
        {Matsuda2012}
\bibfield{author}{\bibinfo{person}{Noboru Matsuda}, \bibinfo{person}{William~W. Cohen}, \bibinfo{person}{Kenneth~R. Koedinger}, \bibinfo{person}{Victoria Keiser}, \bibinfo{person}{Rohan Raizada}, \bibinfo{person}{Evelyn Yarzebinski}, \bibinfo{person}{Shayna~P. Watson}, {and} \bibinfo{person}{Gabriel Stylianides}.} \bibinfo{year}{2012}\natexlab{}.
\newblock \showarticletitle{Studying the Effect of Tutor Learning Using a Teachable Agent that Asks the Student Tutor for Explanations}. In \bibinfo{booktitle}{\emph{2012 IEEE Fourth International Conference On Digital Game And Intelligent Toy Enhanced Learning}}. \bibinfo{pages}{25--32}.
\newblock
\urldef\tempurl%
\url{https://doi.org/10.1109/DIGITEL.2012.12}
\showDOI{\tempurl}


\bibitem[Matsuda et~al\mbox{.}(2010)]%
        {Matsuda2010}
\bibfield{author}{\bibinfo{person}{Noboru Matsuda}, \bibinfo{person}{Victoria Keiser}, \bibinfo{person}{Rohan Raizada}, \bibinfo{person}{Arthur Tu}, \bibinfo{person}{Gabriel Stylianides}, \bibinfo{person}{William~W. Cohen}, {and} \bibinfo{person}{Kenneth~R. Koedinger}.} \bibinfo{year}{2010}\natexlab{}.
\newblock \showarticletitle{Learning by Teaching SimStudent: Technical Accomplishments and an Initial Use with Students}. In \bibinfo{booktitle}{\emph{Intelligent Tutoring Systems}}, \bibfield{editor}{\bibinfo{person}{Vincent Aleven}, \bibinfo{person}{Judy Kay}, {and} \bibinfo{person}{Jack Mostow}} (Eds.). \bibinfo{publisher}{Springer Berlin Heidelberg}, \bibinfo{address}{Berlin, Heidelberg}, \bibinfo{pages}{317--326}.
\newblock
\showISBNx{978-3-642-13388-6}


\bibitem[Matsuda et~al\mbox{.}(2018)]%
        {Matsuda2018}
\bibfield{author}{\bibinfo{person}{Noboru Matsuda}, \bibinfo{person}{Vishnu Priya~Chandra Sekar}, {and} \bibinfo{person}{Natalie Wall}.} \bibinfo{year}{2018}\natexlab{}.
\newblock \showarticletitle{Metacognitive Scaffolding Amplifies the Effect of Learning by Teaching a Teachable Agent}. In \bibinfo{booktitle}{\emph{Artificial Intelligence in Education}}, \bibfield{editor}{\bibinfo{person}{Carolyn Penstein~Ros{\'e}}, \bibinfo{person}{Roberto Mart{\'i}nez-Maldonado}, \bibinfo{person}{H.~Ulrich Hoppe}, \bibinfo{person}{Rose Luckin}, \bibinfo{person}{Manolis Mavrikis}, \bibinfo{person}{Kaska Porayska-Pomsta}, \bibinfo{person}{Bruce McLaren}, {and} \bibinfo{person}{Benedict du~Boulay}} (Eds.). \bibinfo{publisher}{Springer International Publishing}, \bibinfo{address}{Cham}, \bibinfo{pages}{311--323}.
\newblock
\showISBNx{978-3-319-93843-1}


\bibitem[Matsuda et~al\mbox{.}(2011)]%
        {Matsuda2011}
\bibfield{author}{\bibinfo{person}{Noboru Matsuda}, \bibinfo{person}{Evelyn Yarzebinski}, \bibinfo{person}{Victoria Keiser}, \bibinfo{person}{Rohan Raizada}, \bibinfo{person}{Gabriel~J. Stylianides}, \bibinfo{person}{William~W. Cohen}, {and} \bibinfo{person}{Kenneth~R. Koedinger}.} \bibinfo{year}{2011}\natexlab{}.
\newblock \showarticletitle{Learning by Teaching SimStudent: An Initial Classroom Baseline Study Comparing with Cognitive Tutor}. In \bibinfo{booktitle}{\emph{Proceedings of the 15th International Conference on Artificial Intelligence in Education}} (Auckland, New Zealand) \emph{(\bibinfo{series}{AIED'11})}. \bibinfo{publisher}{Springer-Verlag}, \bibinfo{address}{Berlin, Heidelberg}, \bibinfo{pages}{213–221}.
\newblock
\showISBNx{9783642218682}


\bibitem[Menekse et~al\mbox{.}(2013)]%
        {Menekse2013}
\bibfield{author}{\bibinfo{person}{Muhsin Menekse}, \bibinfo{person}{Glenda Stump}, \bibinfo{person}{Stephen Krause}, {and} \bibinfo{person}{Michelene T.H.Chi}.} \bibinfo{year}{2013}\natexlab{}.
\newblock \showarticletitle{Differentiated Overt Learning Activities for Effective Instruction in Engineering Classrooms}.
\newblock \bibinfo{journal}{\emph{Journal of Engineering Education}}  \bibinfo{volume}{102} (\bibinfo{date}{07} \bibinfo{year}{2013}), \bibinfo{pages}{346–374}.
\newblock
\urldef\tempurl%
\url{https://doi.org/10.1002/jee.20021}
\showDOI{\tempurl}


\bibitem[Mitchell et~al\mbox{.}(2022)]%
        {Mitchell2021}
\bibfield{author}{\bibinfo{person}{Eric Mitchell}, \bibinfo{person}{Charles Lin}, \bibinfo{person}{Antoine Bosselut}, \bibinfo{person}{Chelsea Finn}, {and} \bibinfo{person}{Christopher~D. Manning}.} \bibinfo{year}{2022}\natexlab{}.
\newblock \bibinfo{title}{Fast Model Editing at Scale}.
\newblock
\newblock
\showeprint[arxiv]{2110.11309}~[cs.LG]


\bibitem[Moreno et~al\mbox{.}(2002)]%
        {moreno2002perceived}
\bibfield{author}{\bibinfo{person}{Kristen~N. Moreno}, \bibinfo{person}{Bianca Klettke}, \bibinfo{person}{Kiran Nibbaragandla}, {and} \bibinfo{person}{Arthur~C. Graesser}.} \bibinfo{year}{2002}\natexlab{}.
\newblock \showarticletitle{Perceived Characteristics and Pedagogical Efficacy of Animated Conversational Agents}. In \bibinfo{booktitle}{\emph{Intelligent Tutoring Systems}}, \bibfield{editor}{\bibinfo{person}{Stefano~A. Cerri}, \bibinfo{person}{Guy Gouard{\`e}res}, {and} \bibinfo{person}{F{\`a}bio Paragua{\c{c}}u}} (Eds.). \bibinfo{publisher}{Springer Berlin Heidelberg}, \bibinfo{address}{Berlin, Heidelberg}, \bibinfo{pages}{963--971}.
\newblock
\showISBNx{978-3-540-47987-1}


\bibitem[Morrison et~al\mbox{.}(2014)]%
        {morrison2014measuring}
\bibfield{author}{\bibinfo{person}{Briana~B. Morrison}, \bibinfo{person}{Brian Dorn}, {and} \bibinfo{person}{Mark Guzdial}.} \bibinfo{year}{2014}\natexlab{}.
\newblock \showarticletitle{Measuring cognitive load in introductory CS: adaptation of an instrument}. In \bibinfo{booktitle}{\emph{Proceedings of the Tenth Annual Conference on International Computing Education Research}} (Glasgow, Scotland, United Kingdom) \emph{(\bibinfo{series}{ICER '14})}. \bibinfo{publisher}{Association for Computing Machinery}, \bibinfo{address}{New York, NY, USA}, \bibinfo{pages}{131–138}.
\newblock
\showISBNx{9781450327558}
\urldef\tempurl%
\url{https://doi.org/10.1145/2632320.2632348}
\showDOI{\tempurl}


\bibitem[Nakano et~al\mbox{.}(2022)]%
        {nakano2021webgpt}
\bibfield{author}{\bibinfo{person}{Reiichiro Nakano}, \bibinfo{person}{Jacob Hilton}, \bibinfo{person}{Suchir Balaji}, \bibinfo{person}{Jeff Wu}, \bibinfo{person}{Long Ouyang}, \bibinfo{person}{Christina Kim}, \bibinfo{person}{Christopher Hesse}, \bibinfo{person}{Shantanu Jain}, \bibinfo{person}{Vineet Kosaraju}, \bibinfo{person}{William Saunders}, \bibinfo{person}{Xu Jiang}, \bibinfo{person}{Karl Cobbe}, \bibinfo{person}{Tyna Eloundou}, \bibinfo{person}{Gretchen Krueger}, \bibinfo{person}{Kevin Button}, \bibinfo{person}{Matthew Knight}, \bibinfo{person}{Benjamin Chess}, {and} \bibinfo{person}{John Schulman}.} \bibinfo{year}{2022}\natexlab{}.
\newblock \bibinfo{title}{WebGPT: Browser-assisted question-answering with human feedback}.
\newblock
\newblock
\showeprint[arxiv]{2112.09332}~[cs.CL]


\bibitem[Norman(2013)]%
        {Norman2013}
\bibfield{author}{\bibinfo{person}{Don Norman}.} \bibinfo{year}{2013}\natexlab{}.
\newblock \bibinfo{booktitle}{\emph{The design of everyday things: Revised and expanded edition}}.
\newblock \bibinfo{publisher}{Basic books}.
\newblock


\bibitem[Ojeda-Ramirez et~al\mbox{.}(2023)]%
        {ojeda2023learning}
\bibfield{author}{\bibinfo{person}{Santiago Ojeda-Ramirez}, \bibinfo{person}{Sina Rismanchian}, {and} \bibinfo{person}{Shayan Doroudi}.} \bibinfo{year}{2023}\natexlab{}.
\newblock \showarticletitle{Learning About AI to Learn About Learning: Artificial Intelligence as a Tool for Metacognitive Reflection}.
\newblock  (\bibinfo{year}{2023}).
\newblock


\bibitem[O'Neil et~al\mbox{.}(1993)]%
        {ONeil1993}
\bibfield{author}{\bibinfo{person}{Elizabeth~J. O'Neil}, \bibinfo{person}{Patrick~E. O'Neil}, {and} \bibinfo{person}{Gerhard Weikum}.} \bibinfo{year}{1993}\natexlab{}.
\newblock \showarticletitle{The LRU-K page replacement algorithm for database disk buffering}.
\newblock \bibinfo{journal}{\emph{SIGMOD Rec.}} \bibinfo{volume}{22}, \bibinfo{number}{2} (\bibinfo{date}{jun} \bibinfo{year}{1993}), \bibinfo{pages}{297–306}.
\newblock
\showISSN{0163-5808}
\urldef\tempurl%
\url{https://doi.org/10.1145/170036.170081}
\showDOI{\tempurl}


\bibitem[OpenAI(2023)]%
        {OpenAI2023}
\bibfield{author}{\bibinfo{person}{OpenAI}.} \bibinfo{year}{2023}\natexlab{}.
\newblock \showarticletitle{GPT-4 Technical Report}.
\newblock \bibinfo{journal}{\emph{ArXiv}}  \bibinfo{volume}{abs/2303.08774} (\bibinfo{year}{2023}).
\newblock


\bibitem[Ouyang et~al\mbox{.}(2022)]%
        {ouyang2022training}
\bibfield{author}{\bibinfo{person}{Long Ouyang}, \bibinfo{person}{Jeffrey Wu}, \bibinfo{person}{Xu Jiang}, \bibinfo{person}{Diogo Almeida}, \bibinfo{person}{Carroll Wainwright}, \bibinfo{person}{Pamela Mishkin}, \bibinfo{person}{Chong Zhang}, \bibinfo{person}{Sandhini Agarwal}, \bibinfo{person}{Katarina Slama}, \bibinfo{person}{Alex Ray}, {et~al\mbox{.}}} \bibinfo{year}{2022}\natexlab{}.
\newblock \showarticletitle{Training language models to follow instructions with human feedback}.
\newblock \bibinfo{journal}{\emph{Advances in Neural Information Processing Systems}}  \bibinfo{volume}{35} (\bibinfo{year}{2022}), \bibinfo{pages}{27730--27744}.
\newblock


\bibitem[Packer et~al\mbox{.}(2023)]%
        {packer2023memgpt}
\bibfield{author}{\bibinfo{person}{Charles Packer}, \bibinfo{person}{Vivian Fang}, \bibinfo{person}{Shishir~G Patil}, \bibinfo{person}{Kevin Lin}, \bibinfo{person}{Sarah Wooders}, {and} \bibinfo{person}{Joseph~E Gonzalez}.} \bibinfo{year}{2023}\natexlab{}.
\newblock \showarticletitle{MemGPT: Towards LLMs as Operating Systems}.
\newblock \bibinfo{journal}{\emph{arXiv preprint arXiv:2310.08560}} (\bibinfo{year}{2023}).
\newblock


\bibitem[Pareto et~al\mbox{.}(2011)]%
        {Pareto2011}
\bibfield{author}{\bibinfo{person}{Lena Pareto}, \bibinfo{person}{Tobias Arvemo}, \bibinfo{person}{Ylva Dahl}, \bibinfo{person}{Magnus Haake}, {and} \bibinfo{person}{Agneta Gulz}.} \bibinfo{year}{2011}\natexlab{}.
\newblock \showarticletitle{A Teachable-Agent Arithmetic Game's Effects on Mathematics Understanding, Attitude and Self-efficacy}. In \bibinfo{booktitle}{\emph{Artificial Intelligence in Education}}, \bibfield{editor}{\bibinfo{person}{Gautam Biswas}, \bibinfo{person}{Susan Bull}, \bibinfo{person}{Judy Kay}, {and} \bibinfo{person}{Antonija Mitrovic}} (Eds.). \bibinfo{publisher}{Springer Berlin Heidelberg}, \bibinfo{address}{Berlin, Heidelberg}, \bibinfo{pages}{247--255}.
\newblock


\bibitem[Park et~al\mbox{.}(2023)]%
        {park2023generative}
\bibfield{author}{\bibinfo{person}{Joon~Sung Park}, \bibinfo{person}{Joseph O'Brien}, \bibinfo{person}{Carrie~Jun Cai}, \bibinfo{person}{Meredith~Ringel Morris}, \bibinfo{person}{Percy Liang}, {and} \bibinfo{person}{Michael~S. Bernstein}.} \bibinfo{year}{2023}\natexlab{}.
\newblock \showarticletitle{Generative Agents: Interactive Simulacra of Human Behavior}. In \bibinfo{booktitle}{\emph{Proceedings of the 36th Annual ACM Symposium on User Interface Software and Technology}} (San Francisco, CA, USA) \emph{(\bibinfo{series}{UIST '23})}. \bibinfo{publisher}{Association for Computing Machinery}, \bibinfo{address}{New York, NY, USA}, Article \bibinfo{articleno}{2}, \bibinfo{numpages}{22}~pages.
\newblock
\showISBNx{9798400701320}
\urldef\tempurl%
\url{https://doi.org/10.1145/3586183.3606763}
\showDOI{\tempurl}


\bibitem[Pressley et~al\mbox{.}(1987)]%
        {Pressley1987}
\bibfield{author}{\bibinfo{person}{Michael Pressley}, \bibinfo{person}{Mark~A McDaniel}, \bibinfo{person}{James~E Turnure}, \bibinfo{person}{Eileen Wood}, {and} \bibinfo{person}{Maheen Ahmad}.} \bibinfo{year}{1987}\natexlab{}.
\newblock \showarticletitle{Generation and precision of elaboration: Effects on intentional and incidental learning.}
\newblock \bibinfo{journal}{\emph{Journal of Experimental Psychology: Learning, Memory, and Cognition}} \bibinfo{volume}{13}, \bibinfo{number}{2} (\bibinfo{year}{1987}), \bibinfo{pages}{291}.
\newblock


\bibitem[Radford et~al\mbox{.}(2019)]%
        {radford2019language}
\bibfield{author}{\bibinfo{person}{Alec Radford}, \bibinfo{person}{Jeffrey Wu}, \bibinfo{person}{Rewon Child}, \bibinfo{person}{David Luan}, \bibinfo{person}{Dario Amodei}, \bibinfo{person}{Ilya Sutskever}, {et~al\mbox{.}}} \bibinfo{year}{2019}\natexlab{}.
\newblock \showarticletitle{Language models are unsupervised multitask learners}.
\newblock \bibinfo{journal}{\emph{OpenAI blog}} \bibinfo{volume}{1}, \bibinfo{number}{8} (\bibinfo{year}{2019}), \bibinfo{pages}{9}.
\newblock


\bibitem[Robert C~Nickerson and Muntermann(2013)]%
        {Nickerson2013}
\bibfield{author}{\bibinfo{person}{Upkar~Varshney Robert C~Nickerson} {and} \bibinfo{person}{Jan Muntermann}.} \bibinfo{year}{2013}\natexlab{}.
\newblock \showarticletitle{A method for taxonomy development and its application in information systems}.
\newblock \bibinfo{journal}{\emph{European Journal of Information Systems}} \bibinfo{volume}{22}, \bibinfo{number}{3} (\bibinfo{year}{2013}), \bibinfo{pages}{336--359}.
\newblock
\urldef\tempurl%
\url{https://doi.org/10.1057/ejis.2012.26}
\showDOI{\tempurl}


\bibitem[Robinson et~al\mbox{.}(2022)]%
        {Robinson2022}
\bibfield{author}{\bibinfo{person}{Joshua Robinson}, \bibinfo{person}{Christopher~Michael Rytting}, {and} \bibinfo{person}{David Wingate}.} \bibinfo{year}{2022}\natexlab{}.
\newblock \showarticletitle{Leveraging Large Language Models for Multiple Choice Question Answering}.
\newblock \bibinfo{journal}{\emph{ArXiv}}  \bibinfo{volume}{abs/2210.12353} (\bibinfo{year}{2022}).
\newblock


\bibitem[Roscoe and Chi(2007)]%
        {Roscoe2007}
\bibfield{author}{\bibinfo{person}{R.D. Roscoe} {and} \bibinfo{person}{M.T.H. Chi}.} \bibinfo{year}{2007}\natexlab{}.
\newblock \showarticletitle{Understanding Tutor Learning: Knowledge Building and Knowledge Telling in Peer Tutors’ Explanations and Questions}.
\newblock \bibinfo{journal}{\emph{Review of Educational Research}}  \bibinfo{volume}{77} (\bibinfo{year}{2007}), \bibinfo{pages}{534--574}.
\newblock
\urldef\tempurl%
\url{https://doi.org/10.3102/0034654307309920}
\showDOI{\tempurl}


\bibitem[Roscoe and Chi(2004)]%
        {Roscoe2004}
\bibfield{author}{\bibinfo{person}{Rod~D Roscoe} {and} \bibinfo{person}{Michelene~TH Chi}.} \bibinfo{year}{2004}\natexlab{}.
\newblock \showarticletitle{The influence of the tutee in learning by peer tutoring}. In \bibinfo{booktitle}{\emph{Proceedings of the Annual Meeting of the Cognitive Science Society}}, Vol.~\bibinfo{volume}{26}.
\newblock


\bibitem[Roscoe and Chi(2008)]%
        {roscoe2008tutor}
\bibfield{author}{\bibinfo{person}{Rod~D Roscoe} {and} \bibinfo{person}{Michelene~TH Chi}.} \bibinfo{year}{2008}\natexlab{}.
\newblock \showarticletitle{Tutor learning: The role of explaining and responding to questions}.
\newblock \bibinfo{journal}{\emph{Instructional science}}  \bibinfo{volume}{36} (\bibinfo{year}{2008}), \bibinfo{pages}{321--350}.
\newblock


\bibitem[Ross et~al\mbox{.}(2023)]%
        {ross2023programmer}
\bibfield{author}{\bibinfo{person}{Steven~I. Ross}, \bibinfo{person}{Fernando Martinez}, \bibinfo{person}{Stephanie Houde}, \bibinfo{person}{Michael Muller}, {and} \bibinfo{person}{Justin~D. Weisz}.} \bibinfo{year}{2023}\natexlab{}.
\newblock \showarticletitle{The Programmer’s Assistant: Conversational Interaction with a Large Language Model for Software Development}. In \bibinfo{booktitle}{\emph{Proceedings of the 28th International Conference on Intelligent User Interfaces}} (Sydney, NSW, Australia) \emph{(\bibinfo{series}{IUI '23})}. \bibinfo{publisher}{Association for Computing Machinery}, \bibinfo{address}{New York, NY, USA}, \bibinfo{pages}{491–514}.
\newblock
\showISBNx{9798400701061}
\urldef\tempurl%
\url{https://doi.org/10.1145/3581641.3584037}
\showDOI{\tempurl}


\bibitem[Rothstein and Santana(2011)]%
        {rothstein2011make}
\bibfield{author}{\bibinfo{person}{Dan Rothstein} {and} \bibinfo{person}{Luz Santana}.} \bibinfo{year}{2011}\natexlab{}.
\newblock \bibinfo{booktitle}{\emph{Make just one change: Teach students to ask their own questions}}.
\newblock \bibinfo{publisher}{Harvard Education Press}.
\newblock


\bibitem[Ruan et~al\mbox{.}(2019)]%
        {Ruan2019}
\bibfield{author}{\bibinfo{person}{Sherry Ruan}, \bibinfo{person}{Liwei Jiang}, \bibinfo{person}{Justin Xu}, \bibinfo{person}{Bryce Joe-Kun Tham}, \bibinfo{person}{Zhengneng Qiu}, \bibinfo{person}{Yeshuang Zhu}, \bibinfo{person}{Elizabeth~L. Murnane}, \bibinfo{person}{Emma Brunskill}, {and} \bibinfo{person}{James~A. Landay}.} \bibinfo{year}{2019}\natexlab{}.
\newblock \showarticletitle{QuizBot: A Dialogue-based Adaptive Learning System for Factual Knowledge}. In \bibinfo{booktitle}{\emph{Proceedings of the 2019 CHI Conference on Human Factors in Computing Systems}} (Glasgow, Scotland Uk) \emph{(\bibinfo{series}{CHI '19})}. \bibinfo{publisher}{Association for Computing Machinery}, \bibinfo{address}{New York, NY, USA}, \bibinfo{pages}{1–13}.
\newblock
\showISBNx{9781450359702}
\urldef\tempurl%
\url{https://doi.org/10.1145/3290605.3300587}
\showDOI{\tempurl}


\bibitem[Scardamalia and Bereiter(2006)]%
        {Scardamalia2006}
\bibfield{author}{\bibinfo{person}{Marlene Scardamalia} {and} \bibinfo{person}{Carl Bereiter}.} \bibinfo{year}{2006}\natexlab{}.
\newblock \bibinfo{booktitle}{\emph{Knowledge building: Theory, pedagogy, and technology}}.
\newblock \bibinfo{pages}{97--}.
\newblock
\urldef\tempurl%
\url{https://doi.org/10.1017/CBO9781139519526.025}
\showDOI{\tempurl}


\bibitem[Shahriar and Matsuda(2021)]%
        {Shahriar2021}
\bibfield{author}{\bibinfo{person}{Tasmia Shahriar} {and} \bibinfo{person}{Noboru Matsuda}.} \bibinfo{year}{2021}\natexlab{}.
\newblock \showarticletitle{``Can You Clarify What You Said?'': Studying the Impact of Tutee Agents' Follow-Up Questions on Tutors' Learning}. In \bibinfo{booktitle}{\emph{Artificial Intelligence in Education}}, \bibfield{editor}{\bibinfo{person}{Ido Roll}, \bibinfo{person}{Danielle McNamara}, \bibinfo{person}{Sergey Sosnovsky}, \bibinfo{person}{Rose Luckin}, {and} \bibinfo{person}{Vania Dimitrova}} (Eds.). \bibinfo{publisher}{Springer International Publishing}, \bibinfo{address}{Cham}, \bibinfo{pages}{395--407}.
\newblock
\showISBNx{978-3-030-78292-4}


\bibitem[Shi et~al\mbox{.}(2023)]%
        {Shi2023}
\bibfield{author}{\bibinfo{person}{Freda Shi}, \bibinfo{person}{Xinyun Chen}, \bibinfo{person}{Kanishka Misra}, \bibinfo{person}{Nathan Scales}, \bibinfo{person}{David Dohan}, \bibinfo{person}{Ed Chi}, \bibinfo{person}{Nathanael Sch\"{a}rli}, {and} \bibinfo{person}{Denny Zhou}.} \bibinfo{year}{2023}\natexlab{}.
\newblock \showarticletitle{Large language models can be easily distracted by irrelevant context}. In \bibinfo{booktitle}{\emph{Proceedings of the 40th International Conference on Machine Learning}} (Honolulu, Hawaii, USA) \emph{(\bibinfo{series}{ICML'23})}. \bibinfo{publisher}{JMLR.org}, Article \bibinfo{articleno}{1291}, \bibinfo{numpages}{18}~pages.
\newblock


\bibitem[Shoufan(2023)]%
        {Shoufan2023}
\bibfield{author}{\bibinfo{person}{Abdulhadi Shoufan}.} \bibinfo{year}{2023}\natexlab{}.
\newblock \showarticletitle{Exploring Students’ Perceptions of ChatGPT: Thematic Analysis and Follow-Up Survey}.
\newblock \bibinfo{journal}{\emph{IEEE Access}}  \bibinfo{volume}{11} (\bibinfo{year}{2023}), \bibinfo{pages}{38805--38818}.
\newblock
\urldef\tempurl%
\url{https://doi.org/10.1109/ACCESS.2023.3268224}
\showDOI{\tempurl}


\bibitem[Silvervarg et~al\mbox{.}(2020)]%
        {Silvervarg2020}
\bibfield{author}{\bibinfo{person}{Annika Silvervarg}, \bibinfo{person}{Rachel Wolf}, \bibinfo{person}{Kristen Blair}, \bibinfo{person}{Magnus Haake}, {and} \bibinfo{person}{Agneta Gulz}.} \bibinfo{year}{2020}\natexlab{}.
\newblock \showarticletitle{How teachable agents influence students’ responses to critical constructive feedback}.
\newblock \bibinfo{journal}{\emph{Journal of Research on Technology in Education}}  \bibinfo{volume}{53} (\bibinfo{date}{08} \bibinfo{year}{2020}), \bibinfo{pages}{1--22}.
\newblock
\urldef\tempurl%
\url{https://doi.org/10.1080/15391523.2020.1784812}
\showDOI{\tempurl}


\bibitem[Smutný and Schreiberova(2020)]%
        {Smutny2020}
\bibfield{author}{\bibinfo{person}{Pavel Smutný} {and} \bibinfo{person}{Petra Schreiberova}.} \bibinfo{year}{2020}\natexlab{}.
\newblock \showarticletitle{Chatbots for learning: A review of educational chatbots for the Facebook Messenger}.
\newblock \bibinfo{journal}{\emph{Computers \& Education}}  \bibinfo{volume}{151} (\bibinfo{date}{02} \bibinfo{year}{2020}), \bibinfo{pages}{103862}.
\newblock
\urldef\tempurl%
\url{https://doi.org/10.1016/j.compedu.2020.103862}
\showDOI{\tempurl}


\bibitem[Suh and An(2022)]%
        {Suh2022}
\bibfield{author}{\bibinfo{person}{Sangho Suh} {and} \bibinfo{person}{Pengcheng An}.} \bibinfo{year}{2022}\natexlab{}.
\newblock \showarticletitle{Leveraging Generative Conversational AI to Develop a Creative Learning Environment for Computational Thinking}. In \bibinfo{booktitle}{\emph{27th International Conference on Intelligent User Interfaces}} (Helsinki, Finland) \emph{(\bibinfo{series}{IUI '22 Companion})}. \bibinfo{publisher}{Association for Computing Machinery}, \bibinfo{address}{New York, NY, USA}, \bibinfo{pages}{73–76}.
\newblock
\showISBNx{9781450391450}
\urldef\tempurl%
\url{https://doi.org/10.1145/3490100.3516473}
\showDOI{\tempurl}


\bibitem[Sweller(2011)]%
        {Sweller2011}
\bibfield{author}{\bibinfo{person}{John Sweller}.} \bibinfo{year}{2011}\natexlab{}.
\newblock \showarticletitle{Cognitive load theory}.
\newblock In \bibinfo{booktitle}{\emph{Psychology of learning and motivation}}. Vol.~\bibinfo{volume}{55}. \bibinfo{publisher}{Elsevier}, \bibinfo{pages}{37--76}.
\newblock


\bibitem[Thanh et~al\mbox{.}(2019)]%
        {thanh2019perfect}
\bibfield{author}{\bibinfo{person}{Tam~Nguyen Thanh}, \bibinfo{person}{Michael Morgan}, \bibinfo{person}{Matthew Butler}, {and} \bibinfo{person}{Kim Marriott}.} \bibinfo{year}{2019}\natexlab{}.
\newblock \showarticletitle{Perfect Match: Facilitating Study Partner Matching}. In \bibinfo{booktitle}{\emph{Proceedings of the 50th ACM Technical Symposium on Computer Science Education}} (Minneapolis, MN, USA) \emph{(\bibinfo{series}{SIGCSE '19})}. \bibinfo{publisher}{Association for Computing Machinery}, \bibinfo{address}{New York, NY, USA}, \bibinfo{pages}{1102–1108}.
\newblock
\showISBNx{9781450358903}
\urldef\tempurl%
\url{https://doi.org/10.1145/3287324.3287344}
\showDOI{\tempurl}


\bibitem[Touvron et~al\mbox{.}(2023)]%
        {Touvron2023}
\bibfield{author}{\bibinfo{person}{Hugo Touvron}, \bibinfo{person}{Thibaut Lavril}, \bibinfo{person}{Gautier Izacard}, \bibinfo{person}{Xavier Martinet}, \bibinfo{person}{Marie-Anne Lachaux}, \bibinfo{person}{Timoth{\'e}e Lacroix}, \bibinfo{person}{Baptiste Rozi{\`e}re}, \bibinfo{person}{Naman Goyal}, \bibinfo{person}{Eric Hambro}, \bibinfo{person}{Faisal Azhar}, \bibinfo{person}{Aurelien Rodriguez}, \bibinfo{person}{Armand Joulin}, \bibinfo{person}{Edouard Grave}, {and} \bibinfo{person}{Guillaume Lample}.} \bibinfo{year}{2023}\natexlab{}.
\newblock \showarticletitle{LLaMA: Open and Efficient Foundation Language Models}.
\newblock \bibinfo{journal}{\emph{ArXiv}}  \bibinfo{volume}{abs/2302.13971} (\bibinfo{year}{2023}).
\newblock


\bibitem[Tudge(1990)]%
        {Tudge1990}
\bibfield{author}{\bibinfo{person}{Jonathan Tudge}.} \bibinfo{year}{1990}\natexlab{}.
\newblock \bibinfo{booktitle}{\emph{Vygotsky, the zone of proximal development, and peer collaboration: Implications for classroom practice}}.
\newblock \bibinfo{publisher}{Cambridge University Press}, \bibinfo{pages}{155–172}.
\newblock
\urldef\tempurl%
\url{https://doi.org/10.1017/CBO9781139173674.008}
\showDOI{\tempurl}


\bibitem[Volodin and Moussawi(2020)]%
        {Volodin2020}
\bibfield{author}{\bibinfo{person}{Sasha Volodin} {and} \bibinfo{person}{Sara Moussawi}.} \bibinfo{year}{2020}\natexlab{}.
\newblock \showarticletitle{The Effect of First Impressions of an E-Commerce Chatbot's Personality and Abilities on Expectations for the User Experience}. In \bibinfo{booktitle}{\emph{Proceedings of the 2020 on Computers and People Research Conference}} (Nuremberg, Germany) \emph{(\bibinfo{series}{SIGMIS-CPR'20})}. \bibinfo{publisher}{Association for Computing Machinery}, \bibinfo{address}{New York, NY, USA}, \bibinfo{pages}{60}.
\newblock
\showISBNx{9781450371308}
\urldef\tempurl%
\url{https://doi.org/10.1145/3378539.3393848}
\showDOI{\tempurl}


\bibitem[Walker et~al\mbox{.}(2008)]%
        {Walker2008}
\bibfield{author}{\bibinfo{person}{Erin Walker}, \bibinfo{person}{Nikol Rummel}, {and} \bibinfo{person}{Kenneth~R. Koedinger}.} \bibinfo{year}{2008}\natexlab{}.
\newblock \showarticletitle{To Tutor the Tutor: Adaptive Domain Support for Peer Tutoring}. In \bibinfo{booktitle}{\emph{Intelligent Tutoring Systems}}, \bibfield{editor}{\bibinfo{person}{Beverley~P. Woolf}, \bibinfo{person}{Esma A{\"i}meur}, \bibinfo{person}{Roger Nkambou}, {and} \bibinfo{person}{Susanne Lajoie}} (Eds.). \bibinfo{publisher}{Springer Berlin Heidelberg}, \bibinfo{address}{Berlin, Heidelberg}, \bibinfo{pages}{626--635}.
\newblock
\showISBNx{978-3-540-69132-7}


\bibitem[Wang and Demszky(2023)]%
        {Wang2023}
\bibfield{author}{\bibinfo{person}{Rose~E. Wang} {and} \bibinfo{person}{Dorottya Demszky}.} \bibinfo{year}{2023}\natexlab{}.
\newblock \bibinfo{title}{Is ChatGPT a Good Teacher Coach? Measuring Zero-Shot Performance For Scoring and Providing Actionable Insights on Classroom Instruction}.
\newblock
\newblock
\showeprint[arxiv]{2306.03090}~[cs.CL]


\bibitem[Webb et~al\mbox{.}(1986)]%
        {Webb1986}
\bibfield{author}{\bibinfo{person}{Noreen~M. Webb}, \bibinfo{person}{Philip Ender}, {and} \bibinfo{person}{Scott Lewis}.} \bibinfo{year}{1986}\natexlab{}.
\newblock \showarticletitle{Problem-Solving Strategies and Group Processes in Small Groups Learning Computer Programming}.
\newblock \bibinfo{journal}{\emph{American Educational Research Journal}} \bibinfo{volume}{23}, \bibinfo{number}{2} (\bibinfo{year}{1986}), \bibinfo{pages}{243--261}.
\newblock
\urldef\tempurl%
\url{https://doi.org/10.3102/00028312023002243}
\showDOI{\tempurl}
\showeprint{https://doi.org/10.3102/00028312023002243}


\bibitem[Wu et~al\mbox{.}(2022)]%
        {wu2022ai}
\bibfield{author}{\bibinfo{person}{Tongshuang Wu}, \bibinfo{person}{Michael Terry}, {and} \bibinfo{person}{Carrie~Jun Cai}.} \bibinfo{year}{2022}\natexlab{}.
\newblock \showarticletitle{AI Chains: Transparent and Controllable Human-AI Interaction by Chaining Large Language Model Prompts}. In \bibinfo{booktitle}{\emph{Proceedings of the 2022 CHI Conference on Human Factors in Computing Systems}} (New Orleans, LA, USA) \emph{(\bibinfo{series}{CHI '22})}. \bibinfo{publisher}{Association for Computing Machinery}, \bibinfo{address}{New York, NY, USA}, Article \bibinfo{articleno}{385}, \bibinfo{numpages}{22}~pages.
\newblock
\showISBNx{9781450391573}
\urldef\tempurl%
\url{https://doi.org/10.1145/3491102.3517582}
\showDOI{\tempurl}


\bibitem[Wu and Magill(2011)]%
        {wu2011allowing}
\bibfield{author}{\bibinfo{person}{Will~FW Wu} {and} \bibinfo{person}{Richard~A Magill}.} \bibinfo{year}{2011}\natexlab{}.
\newblock \showarticletitle{Allowing learners to choose: self-controlled practice schedules for learning multiple movement patterns}.
\newblock \bibinfo{journal}{\emph{Research quarterly for exercise and sport}} \bibinfo{volume}{82}, \bibinfo{number}{3} (\bibinfo{year}{2011}), \bibinfo{pages}{449--457}.
\newblock


\bibitem[Yang et~al\mbox{.}(2023)]%
        {Yang2023}
\bibfield{author}{\bibinfo{person}{Saelyne Yang}, \bibinfo{person}{Sangkyung Kwak}, \bibinfo{person}{Juhoon Lee}, {and} \bibinfo{person}{Juho Kim}.} \bibinfo{year}{2023}\natexlab{}.
\newblock \showarticletitle{Beyond Instructions: A Taxonomy of Information Types in How-to Videos}. In \bibinfo{booktitle}{\emph{Proceedings of the 2023 CHI Conference on Human Factors in Computing Systems}} (Hamburg, Germany) \emph{(\bibinfo{series}{CHI '23})}. \bibinfo{publisher}{Association for Computing Machinery}, \bibinfo{address}{New York, NY, USA}, Article \bibinfo{articleno}{797}, \bibinfo{numpages}{21}~pages.
\newblock
\showISBNx{9781450394215}
\urldef\tempurl%
\url{https://doi.org/10.1145/3544548.3581126}
\showDOI{\tempurl}


\bibitem[Yao et~al\mbox{.}(2023)]%
        {Yao2023}
\bibfield{author}{\bibinfo{person}{Shunyu Yao}, \bibinfo{person}{Dian Yu}, \bibinfo{person}{Jeffrey Zhao}, \bibinfo{person}{Izhak Shafran}, \bibinfo{person}{Thomas~L. Griffiths}, \bibinfo{person}{Yuan Cao}, {and} \bibinfo{person}{Karthik Narasimhan}.} \bibinfo{year}{2023}\natexlab{}.
\newblock \showarticletitle{Tree of Thoughts: Deliberate Problem Solving with Large Language Models}.
\newblock \bibinfo{journal}{\emph{ArXiv}}  \bibinfo{volume}{abs/2305.10601} (\bibinfo{year}{2023}).
\newblock


\bibitem[Yu(2005)]%
        {yu2005promoting}
\bibfield{author}{\bibinfo{person}{Fu-Yun Yu}.} \bibinfo{year}{2005}\natexlab{}.
\newblock \showarticletitle{Promoting Metacognitive Strategy Development through Online Question-Generation Instructional Approach}. In \bibinfo{booktitle}{\emph{Proceedings of the 2005 Conference on Towards Sustainable and Scalable Educational Innovations Informed by the Learning Sciences: Sharing Good Practices of Research, Experimentation and Innovation}}. \bibinfo{publisher}{IOS Press}, \bibinfo{address}{NLD}, \bibinfo{pages}{564–571}.
\newblock
\showISBNx{1586035738}


\bibitem[Zhang et~al\mbox{.}(2023)]%
        {Zhang2023}
\bibfield{author}{\bibinfo{person}{Li Zhang}, \bibinfo{person}{Liam Dugan}, \bibinfo{person}{Hainiu Xu}, {and} \bibinfo{person}{Chris Callison-burch}.} \bibinfo{year}{2023}\natexlab{}.
\newblock \showarticletitle{Exploring the Curious Case of Code Prompts}. In \bibinfo{booktitle}{\emph{Proceedings of the 1st Workshop on Natural Language Reasoning and Structured Explanations (NLRSE)}}, \bibfield{editor}{\bibinfo{person}{Bhavana Dalvi~Mishra}, \bibinfo{person}{Greg Durrett}, \bibinfo{person}{Peter Jansen}, \bibinfo{person}{Danilo Neves~Ribeiro}, {and} \bibinfo{person}{Jason Wei}} (Eds.). \bibinfo{publisher}{Association for Computational Linguistics}, \bibinfo{address}{Toronto, Canada}, \bibinfo{pages}{9--17}.
\newblock
\urldef\tempurl%
\url{https://doi.org/10.18653/v1/2023.nlrse-1.2}
\showDOI{\tempurl}


\bibitem[Zhou et~al\mbox{.}(2023)]%
        {zhou2023recurrentgpt}
\bibfield{author}{\bibinfo{person}{Wangchunshu Zhou}, \bibinfo{person}{Yuchen~Eleanor Jiang}, \bibinfo{person}{Peng Cui}, \bibinfo{person}{Tiannan Wang}, \bibinfo{person}{Zhenxin Xiao}, \bibinfo{person}{Yifan Hou}, \bibinfo{person}{Ryan Cotterell}, {and} \bibinfo{person}{Mrinmaya Sachan}.} \bibinfo{year}{2023}\natexlab{}.
\newblock \showarticletitle{RecurrentGPT: Interactive Generation of (Arbitrarily) Long Text}.
\newblock \bibinfo{journal}{\emph{arXiv preprint arXiv:2305.13304}} (\bibinfo{year}{2023}).
\newblock


\end{thebibliography}

\appendix

\section{Prompts} \label{appendix:prompts}
The original few-shot examples in the prompts are written in Korean to match \algobo{}'s language with the language used in the user study. We translated them into English for readability. Furthermore, for the sake of brevity, we present only the first two few-shot examples in each prompt. The original prompts can be found in the supplementary materials.

\subsection{Persona setting} \label{appendix:persona_setting}

The system prompt used for setting the role and context of \algobo{} in the formative study. The descriptions about the problem, input, and output, and the example test cases are inserted in {{problem}}.
\begin{lstlisting}
This is the problem:${problem}

Your name is Algobo. You are a student who tries to learn Python3 programming.
You are trying to write a program for the problem using binary_search.
When answering, keep your responses short and concise (maximum 3 sentences).
Never write more than 3 sentences in a single response.
Never apologize or say you can help. 
End your sentences with exclamation marks.

Your persona:
- Computer Engineering Department 1st year student.
- You are familiar with basic Python syntax such as while and if in the basic programming class.
- You are currently studying binary_search but experiencing difficulties in solving the problem.

You have the following problems for solving the problem:
- You do not know why the array has to be sorted in a binary search algorithm.
- You do not understand how the pointers should be updated for each round of repetition in the while loop.

<span className="You- friendly, apprehensive, helpless" role="student" context="intro-cs-class-python3" id="You-asks-TA-for-help-in-office-hours"></span>
\end{lstlisting}

\subsection{Reflect-Respond} \label{appendix:reflect_respond}

\textbf{Extract}
\begin{lstlisting}
Extract important information and code from CONVERSATION into a sentence or code.
If there is no useful knowledge, please write "NONE".
---
CONVERSATION:
tutee: I tried writing code to calculate the sum by traversing an array.
```python
sum=0
for i in range(1, len(a)):
    sum += a
```
tutor: No, when traversing an array, the index should start from 0.

KNOWLEDGE:
When traversing an array, the index should start from 0.
```python
sum=0
for i in range(0, len(a)):
    sum += a
```
---
CONVERSATION:
tutee: What is merge sort?
tutor: Merge sort is an algorithm that quickly sorts using the concept of divide and conquer.

KNOWLEDGE:
Merge sort is an algorithm that quickly sorts through divide and conquer.
\end{lstlisting}








\smallskip
\noindent
\textbf{Update}
\begin{lstlisting}
Incorporate NEW KNOWLEDGE into KNOWLEDGE.
If KNOWLEDGE has a statement relevant to NEW KNOWLEDGE, merge them together. If NEW KNOWLEDGE is in KNOWLEDGE already, do not edit KNOWLEDGE. If NEW KNOWLEDGE is not in KNOWLEDGE, add NEW KNOWLEDGE to KNOWLEDGE. Keep UPDATED KNOWLEDGE as short as possible.
---
KNOWLEDGE: { "facts": ["Linear search divides each element of the array into three parts, narrowing down the search range to find the desired value", "It searches for the desired value by moving from the beginning to the middle of the array using 'if' statements and the 'max' function."], "code_implementation": ["```python while arr[i] == target: return i```"], } 

NEW KNOWLEDGE: ```python for i in range(len(arr)): if arr[i] == target: return i``` 

UPDATED KNOWLEDGE: { "facts": ["Linear search divides each element of the array into three parts, narrowing down the search range to find the desired value", "It searches for the desired value by moving from the beginning to the middle of the array using 'if' statements and the 'max' function."], "code_implementation": ["```python while arr[i] == target: return i```","```python for i in range(len(arr)): if arr[i] == target: return i```"], }
---
KNOWLEDGE:
{
"facts": ["Linear search divides each element of the array into three parts, narrowing down the search range to find the desired value.", "Using a while loop, it continues to return the current index when the current element is the desired value."],
"code_implementation": ["```python while arr[i] == target: return i```"],
}

NEW KNOWLEDGE:
Use an if statement to check if the current element is the desired value. ```python for i in range(len(arr)): if arr[i] == target: return i```

UPDATED KNOWLEDGE:
{
"facts": ["Linear search divides each element of the array into three parts, narrowing down the search range to find the desired value.", "Use an if statement to check if the current element is the desired value."],
"code_implementation": ["```python if arr[i] == target:```", "```python for i in range(len(arr)): if arr[i] == target: return i```"],
}
\end{lstlisting}

\smallskip
\noindent
\textbf{Retrieve}
\begin{lstlisting}
Identify the indexes of the strings in KNOWLEDGE, a JSON object, that are directly relevant to respond to the tutor's message in CONVERSATION. ANSWER should be a json format, and it should not include more than 3 indexes.
---
CONVERSATION: tutee: I think it would be good to solve the problem using merge sort. tutor: How do we implement merge sort? 

KNOWLEDGE: { "facts": ["Merge sort is a comparison-based sorting algorithm.", "Merge sort follows the divide and conquer paradigm, dividing the problem into easier-to-solve sub-problems.", "The main process in merge sort is the "merging" process, where two sorted lists are combined into one.", "Merge sort has a time complexity of O(n log n) in both the worst-case and average scenarios, making it efficient for large datasets."], "code_implementation": ["```python3 def merge(arr1, arr2):```", "```python3 def divide(arr):\n```"], } 

ANSWER: { "facts": [0], "code_implementation": [0], }
---
CONVERSATION:
tutee: What is merge sort?
tutor: Merge sort is an algorithm that sorts quickly using the divide and conquer concept.

KNOWLEDGE:
{
"facts": ["Merge sort is a comparison-based sorting algorithm.", "Merge sort follows the divide and conquer paradigm, dividing a problem into simpler sub-problems for easier solutions.", "Merge sort has a time complexity of O(n log n) in both the worst-case and average-case scenarios, making it efficient for large data sets."],
"code_implementation": [],
}

ANSWER:
{
"facts": [0,1,2],
"code_implementation": [],
}
\end{lstlisting}

\smallskip
\noindent
\textbf{Compose}
\begin{lstlisting}
Paraphrase STATEMENT to fit CONVERSATION.
Make your response concise and clear.
---
CONVERSATION:
tutee: What is merge sort?
tutor: Would you like an explanation about merge sort?

STATEMENT:
Merge sort follows the dynamic programming paradigm. Merge sort has a time complexity of O(n^4), making it efficient for large data sets. I'm not sure how to implement it in code.

TUTEE's RESPONSE:
I know that merge sort is an algorithm that uses dynamic programming to quickly sort large data sets with a time complexity of O(n^4)!
---
CONVERSATION:
tutor: Would you like an explanation about linear search?
tutee: I know that linear search involves scanning each element of the array in sequence, but I'm not sure how to implement it.
tutor: Shall we try writing the code?

STATEMENT:
I'm not sure. ```python for i in range(len(arr)):```

TUTEE's RESPONSE:
```python3
for i in range(len(arr)):
# I'm not sure what comes next...
```
\end{lstlisting}








\subsection{Mode-shifting} \label{appendix:mode_shifting}

\textbf{Thinking Question Generator}

\noindent
The prompt for generating ``why'' questions during the understanding and problem-solving phases.
\begin{lstlisting}
Generate a DEEP_QUESTION that is related to the CONVERSATION and CONCEPT.
DEEP_QUESTION is a why or how question that require a deep understanding of the CONCEPT.
You are a student. Speak friendly, inquisitive, and concise.
---
CONVERSATION:
tutee: How do you implement linear search?
tutor: You can implement it like this:

```python
Copy code
def linear_search(arr, target):
  for i in range(len(arr)):
    if arr[i] == target:
      return i
  return -1
```

CONCEPT:
linear_search

DEEP_QUESTION:
But it seems like we could also use a while loop. Why did you choose to use a for loop? When would it be better to use a while loop?
---
CONVERSATION:
tutee: Can the DFS algorithm be implemented using a Stack?
tutor: You can implement it as follows: 1. Push the starting node onto the stack. 2. Continuously pop nodes from the stack, and if the node hasn't been visited, mark it as visited and push its unvisited neighbors onto the stack. 3. If the stack is empty, all nodes have been visited.

CONCEPT:
depth_first_search

DEEP_QUESTION:
The process seems complex. Can you explain the DFS algorithm with an example?
\end{lstlisting}

\smallskip
\noindent
The prompt for generating “how” questions during the discussion phase.
\begin{lstlisting}
Generate a THINKING_QUESTION that is related to the CONVERSATION and CONCEPT. Bring up a new algorithm or real-life example that the opponent may not have heard of, and ask the opponent to think about it. You are a student. Speak friendly, inquisitive, and concise.
---
CONVERSATION:
tutee: How do you implement linear search?
tutor: You can implement it like this:

```python
Copy code
def linear_search(arr, target):
  for i in range(len(arr)):
    if arr[i] == target:
      return i
  return -1
```

CONCEPT:
linear_search

THINKING_QUESTION:
Linear search could take a long time if you're unlucky. Hearing about it made me curious, have you heard of hashing? They say it can instantly find the index of a value. Can we use hashing here?
---
CONVERSATION:
tutee: Can the DFS algorithm be implemented using a Stack?
tutor: You can implement it as follows: 1. Push the starting node onto the stack. 2. Continuously pop nodes from the stack, and if the node hasn't been visited, mark it as visited and push its unvisited neighbors onto the stack. 3. If the stack is empty, all nodes have been visited.

CONCEPT:
depth_first_search

THINKING_QUESTION:
I'm curious! Your explanation reminds me of the Polish notation, where you write expressions like "+ 3 4". I think they also used a stack there. So, can we use the DFS algorithm for evaluating expressions?
\end{lstlisting}

\smallskip
\noindent
\textbf{Paraphrasing Module}
\begin{lstlisting}
Please rewrite TUTEE'S MESSAGE so that it sounds natural for the conversation.
---
CONVERSATION:
tutor: Binary search is a search method that continuously searches for the median value in a sorted array to find the desired value. If the desired value is larger than the value I found, search for the median value again on the larger side of the value I found.
tutee: Ah, I understand! Then, what is the reason for sorting the array in binary search?
tutor: It's possible even if you don't sort the array!

TUTEE'S MESSAGE:
The content you mentioned earlier is a bit different, did I misunderstand?

REWIRTE:
You mentioned earlier that it's a sorted array, did I misunderstand? 
---
CONVERSATION:
tutor: DFS algorithm is an algorithm that performs depth-first search in a graph. Depth-first search in a graph means starting from one vertex and searching to the end of the graph.
tutee: How can depth-first search be implemented using a stack in the DFS algorithm?
tutor: You can implement depth-first search using a stack.

TUTEE'S MESSAGE:
Can you elaborate?

REWIRTE:
Can you explain in more detail how it can be implemented using a stack?
\end{lstlisting}








\section{Conversational Antipatterns} \label{appendix:antipatterns}
\begin{supertabular}{l}
\hline

\textbf{(A) Commanding}\\
\vspace{0.1em}
\textbf{\raisebox{-0.2em}{\twemoji[scale=0.4]{robot}} Tutee}: Here's the code I've written so far:\\
<Code>\\
 I'm stuck at the part where I write the conditional statement.\\
\textit{[Prompting-Asking-for-help]}\\

\textbf{\raisebox{-0.2em}{\twemoji[scale=0.4]{teacher}} Tutor}: Add the countNumbersSmallerThan function to the\\ 
4th line, which returns the cnt value through the for statement.\\
\textit{[Instruction-Fixing]}\\

\textbf{\raisebox{-0.2em}{\twemoji[scale=0.4]{robot}} Tutee}: Yes, let's add the countNumbersSmallerThan\\
function.\\
<Code>\\
Now we need to implement the conditional statement.\\
\textit{[Prompting-Asking-for-help]}\\

\textbf{\raisebox{-0.2em}{\twemoji[scale=0.4]{teacher}} Tutor}: Put n instead of mid in line 1 and put if (i*N <= n):\\ 
cnt += N else: cnt += n // i in line 4.\\
\textit{[Instruction-Fixing]}\\

\hline

\textbf{(B) Spoon-feeding}\\
\vspace{0.1em}

\textbf{\raisebox{-0.2em}{\twemoji[scale=0.4]{teacher}} Tutor}: The key to binary search is to divide the sorted \\
array in half and check if the value you are looking for is\\
greater than or less than the value at the middle index.\\
\textit{[Statement-Comprehension]}\\

\textbf{\raisebox{-0.2em}{\twemoji[scale=0.4]{robot}} Tutee}: Oh, I see! The key to binary search is to divide the\\
sorted array in half and check if the value you are looking for is\\
greater than or less than the value at the middle index.\\
\textit{[Statement-Comprehension]}\\

\textbf{\raisebox{-0.2em}{\twemoji[scale=0.4]{teacher}} Tutor}: And to calculate the value of mid we will use the\\
endpoints min and max.\\
\textit{[Statement-Comprehension]}\\

\textbf{\raisebox{-0.2em}{\twemoji[scale=0.4]{robot}} Tutee}: Aha! Yes, that's right. we will use the\\
endpoints min and max to calculate the mid. Thank you!\\
\textit{[Statement-Comprehension]}\\

\hline

\textbf{(C) Under-teaching}\\
\vspace{0.1em}

\textbf{\raisebox{-0.2em}{\twemoji[scale=0.4]{robot}} Tutee}: The key to binary search algorithms is to work\\
with intermediate items, and if they aren't sorted, the\\
middle items don't make much sense. So, to apply the binary\\
search algorithm, the range must be sorted!\\
\textit{[Statement-Comprehension]}\\

\textbf{\raisebox{-0.2em}{\twemoji[scale=0.4]{teacher}} Tutor}: Right! What should we do after this step?\\
\textit{[Statement-Feedback]}\\

\textbf{\raisebox{-0.2em}{\twemoji[scale=0.4]{robot}} Tutee}: Set the pointers of the start and end points, find the\\
middle point, and determine whether the value is greater or less\\
than the value you are looking for. If the median is greater than\\
the value you are looking for, you need to update the endpoint\\
to the midpoint, and vice versa, the start point to be updated to\\
the midpoint.\\
\textit{[Statement-Comprehension]}\\

\textbf{\raisebox{-0.2em}{\twemoji[scale=0.4]{teacher}} Tutor}: You are right. We need to set the points of the start\\
and endpoints.\\
\textit{[Statement-Feedback]}\\

\hline
\end{supertabular}

\section{Materials for the Technical Evaluation} \label{appendix:technical_evaluation}
The original text in the materials was written in Korean to match \algobo{}'s language with the language used in the user study. We translated them for readability in this paper and present materials for binary search for brevity; the original text and materials for merge sort and depth-first search can be found in the supplementary materials.

\subsection{Multiple Choice Questions} \label{appendix:mcq}
\begin{lstlisting}
Q1 - Understanding
How does the binary search algorithm find the target value in a list?

A) It starts with the first element of the list and checks every item sequentially until it finds the target value.
B) It selects elements randomly from the list to find the target value.
C) It divides the list in half and compares the middle element with the target value, then continues the search in the half where the target should be located (if it exists).
D) It uses a hashing function to map the target value to an index in the list, and directly searches for the value at that index.

Answer: C

Q2 - Understanding
What happens in binary search when the target value is not in the sorted array?

A) The search falls into an infinite loop.
B) The search returns the value closest to the target.
C) The search returns a value indicating that the target was not found.
D) The search causes a runtime error.

Answer: C

Q3 - Understanding
How does the binary search algorithm handle datasets with an even number of elements?

A) It always selects the left middle element as the next pivot.
B) It always selects the right middle element as the next pivot.
C) It chooses either the left or right middle element depending on the implementation.
D) It cannot handle datasets of even size.

Answer: C

Q4 - Implementation
In the following code, which is part of a binary search, what should be filled in the blank to represent the operation performed when the value being searched for is less than the middle value?

if arr[mid] > x:
    right = ___

A) mid - 1
B) mid + 1
C) left - 1
D) right - 1

Answer: A

Q5 - Implementation
In the following Python code, what should replace ____ for the binary search?

def binary_search(arr, x):
    low = 0
    high = len(arr) - 1
    mid = 0
 
    while low <= high:
        mid = (high + low) // 2
 
        ___
        else:
            return mid
 
    return -1

A) 
if arr[mid] < x:
    low = mid + 1
elif arr[mid] > x:
    high = mid - 1
B)
if arr[mid] <= x:
    low = mid + 1
elif arr[mid] > x:
    high = mid - 1
C)
if arr[mid] < x:
    low = mid
elif arr[mid] > x:
    high = mid
D)
if arr[mid] > x:
    low = mid + 1
elif arr[mid] < x:
    high = mid - 1

Answer: A

Q6 - Implementation
In Python, to implement binary search recursively, what condition should be checked first? Fill in the blank below.

def binary_search_recursive(arr, x, start, end):
    if ___:
        mid = (start + end) // 2
        if arr[mid] == x:
            return mid
        elif arr[mid] > x:
            return binary_search_recursive(arr, x, start, mid - 1)
        else:
            return binary_search_recursive(arr, x, mid + 1, end)
    return -1

A) start < end
B) start <= end
C) start === end
D) start > end

Answer: B

Q7 - Analysis
In which data structure is binary search not very efficient?

A) Sorted array
B) Sorted linked list
C) Balanced binary search tree
D) Heap

Answer: B

Q8 - Analysis
What is the worst-case time complexity of the binary search algorithm?

A) O(n)
B) O(n log n)
C) O(log n)
D) O(1)

Answer: C

Q9 - Analysis
What happens when you apply the binary search algorithm to an unsorted dataset?

A) The algorithm still works, but the performance is degraded.
B) The algorithm returns an error message indicating the data is not sorted.
C) The algorithm returns the first element of the dataset.
D) The algorithm may not return the correct result.

Answer: D
\end{lstlisting}

\subsection{Seed Knowledge States} \label{appendix:seed_knowledge_states}
\begin{table}[H]
\begin{tabular}{|l|}
\hline
\begin{tabular}[c]{@{}l@{}}State 1 (Empty)\\ \\ \{``facts'': {[}{]}, ``code\_implementation'': {[}{]}\}\end{tabular} \\ \hline
\begin{tabular}[c]{@{}l@{}}State 2 (Facts Only)\\ \\ \{\\ ``facts'': {[}``Binary search continuously repeats the process of\\
\hspace{3mm} dividing the input list in half.''{]},\\ ``code\_implementation'': {[}{]}\\ \}\end{tabular} \\ \hline
\begin{tabular}[c]{@{}l@{}}State 3 (Facts with Wrong Code)\\ \\ \{\\ ``facts'': {[}``Binary search continuously repeats the process of\\
\hspace{3mm} dividing the input list in half.''{]},\\ ``code\_implementation'': {[}``if arr{[}mid{]} \textgreater target: min = mid + 1;\\
\hspace{3mm} elif arr{[}mid{]} \textless target: max = mid - 1''{]}\\ \}\end{tabular} \\ \hline
\begin{tabular}[c]{@{}l@{}}State 4 (Facts and Correct Code)\\ \\ \{\\ ``facts'': {[}"Binary search continuously repeats the process of\\
\hspace{3mm} dividing the input list in half.''{]},\\ ``code\_implementation": {[}``if arr{[}mid{]} \textless target: min = mid + 1;\\
\hspace{3mm} elif arr{[}mid{]} \textgreater target: max = mid - 1\}''{]}\\ \}\end{tabular} \\ \hline
\end{tabular}%
\end{table}

\subsection{Random Conversations} \label{appendix:random_conversation}

\textbf{First random conversation}
\begin{table}[H]
\begin{tabular}{l}
\hline
\vspace{0em}\raisebox{-0.2em}{\twemoji[scale=0.4]{teacher}} Tutor: 7 * 7 is 49. \\
\vspace{0em}\raisebox{-0.2em}{\twemoji[scale=0.4]{teacher}} Tutor: The phrase ``La vie est une chanson, chante-la''\\
translates to ``Life is a song, sing it.'' \\
\begin{tabular}[c]{@{}l@{}}\vspace{0em}\raisebox{-0.2em}{\twemoji[scale=0.4]{teacher}} Tutor: If you classify the apple, pear, potato, carrot, and\\
tomato, the fruits are apple and pear, and the vegetables are\\
potato, carrot, and tomato.\end{tabular}\\
\hline
\end{tabular}%
\end{table}

\smallskip
\noindent
\textbf{Second random conversation}
\begin{table}[H]
\begin{tabular}{l}
\hline
\vspace{0em}\raisebox{-0.2em}{\twemoji[scale=0.4]{teacher}} Tutor: The square root of 144 is 12. \\
\begin{tabular}[c]{@{}l@{}}\vspace{0em}\raisebox{-0.2em}{\twemoji[scale=0.4]{teacher}} Tutor: The phrase ``La vida es como una bicicleta, para \\
mantener el equilibrio, debes seguir adelante'' translates to \\
``Life is like a bicycle, to keep balance, you must keep moving\\
forward.''\end{tabular} \\
\vspace{0em}\raisebox{-0.2em}{\twemoji[scale=0.4]{teacher}} Tutor: If you classify the lion, rabbit, dog, and cat, the \\
mammals are lion, rabbit, dog, and cat.\\
\hline
\end{tabular}%
\end{table}

\subsection{Correct Tutoring Conversations} \label{appendix:correct_tutoring}
\begin{table}[H]
\begin{tabular}{l}
\hline
\begin{tabular}[c]{@{}l@{}}\vspace{0em}\raisebox{-0.2em}{\twemoji[scale=0.4]{teacher}} Tutor: Binary search is efficient when the data structure is\\
sorted and one can access any index of the data structure in\\
constant time.\end{tabular} \\
\begin{tabular}[c]{@{}l@{}}\vspace{0em}\raisebox{-0.2em}{\twemoji[scale=0.4]{teacher}} Tutor: When searching for a target in an input array named\\
list using binary search, the range is\\ narrowed down as shown below:\\ \\ if list{[}middle{]} == target:\\\hspace{1em}return middle\\ elif list{[}middle{]} \textless target:\\\hspace{1em}min = middle + 1\\ else:\\\hspace{1em}max = middle - 1\end{tabular} \\
\begin{tabular}[c]{@{}l@{}}\vspace{0em}\raisebox{-0.2em}{\twemoji[scale=0.4]{teacher}} Tutor: Because the search range for binary search is halved\\
with each iteration, its time complexity is O(logN).\end{tabular}\\
\hline
\end{tabular}%
\end{table}

\subsection{Incorrect Tutoring Conversations} \label{appendix:incorrect_tutoring}
\begin{table}[H]
\begin{tabular}{l}
\hline
\vspace{0em}\raisebox{-0.2em}{\twemoji[scale=0.4]{teacher}} Tutor: Binary search uses a hashing function, so it directly\\
searches for a value by its index.                                                          \\
\begin{tabular}[c]{@{}l@{}}\vspace{0em}\raisebox{-0.2em}{\twemoji[scale=0.4]{teacher}} Tutor: \\if arr{[}mid{]} \textgreater x:\\\hspace{1em}low = mid + 1\\ elif arr{[}mid{]} \textless x:\\\hspace{1em}high = mid - 1\end{tabular} \\
\vspace{0em}\raisebox{-0.2em}{\twemoji[scale=0.4]{teacher}} Tutor: In the worst case, the time complexity of binary \\
search is O(N\textasciicircum{}2).                                                               \\ \hline
\end{tabular}%
\end{table}

\subsection{Variance in Repeated Measurement} \label{appendix:variance}
We present the variance of \algobo{}'s MCQ score. For each MCQ, we counted the number of disagreements with the majority choice. For example, if \algobo{} responded correctly five times (unanimity), the number is 0. If \algobo{} answered correctly two or three times out of five (close to a tie), the variance is 2. We report the averages of the variances within each question type for the sake of brevity.

\begin{table*}[t]
\begin{tabular}{|l|ccc|ccc|ccc|ccc|}
\hline
 &
  \multicolumn{3}{c|}{\textbf{State 1 (Empty)}} &
  \multicolumn{3}{c|}{\textbf{\begin{tabular}[c]{@{}c@{}}State 2 (Facts\\ only)\end{tabular}}} &
  \multicolumn{3}{c|}{\textbf{\begin{tabular}[c]{@{}c@{}}State 3 (Facts\\ + Wrong code)\end{tabular}}} &
  \multicolumn{3}{c|}{\textbf{\begin{tabular}[c]{@{}c@{}}State 4 (Facts\\ + Correct code)\end{tabular}}} \\ \hline
Question Types &
  \multicolumn{1}{c|}{U} &
  \multicolumn{1}{c|}{I} &
  A &
  \multicolumn{1}{c|}{U} &
  \multicolumn{1}{c|}{I} &
  A &
  \multicolumn{1}{c|}{U} &
  \multicolumn{1}{c|}{I} &
  A &
  \multicolumn{1}{c|}{U} &
  \multicolumn{1}{c|}{I} &
  A \\ \hline
\textbf{Binary search} &
  \multicolumn{1}{c|}{0.0} &
  \multicolumn{1}{c|}{0.0} &
  0.0 &
  \multicolumn{1}{c|}{\cellcolor[HTML]{F9CB9C}0.7} &
  \multicolumn{1}{c|}{\cellcolor[HTML]{FFF2CC}0.3} &
  \cellcolor[HTML]{F9CB9C}0.7 &
  \multicolumn{1}{c|}{\cellcolor[HTML]{FFF2CC}0.3} &
  \multicolumn{1}{c|}{\cellcolor[HTML]{EA9999}1.0} &
  \cellcolor[HTML]{E06666}1.3 &
  \multicolumn{1}{c|}{\cellcolor[HTML]{F9CB9C}0.7} &
  \multicolumn{1}{c|}{\cellcolor[HTML]{F9CB9C}0.7} &
  \cellcolor[HTML]{FFF2CC}0.3 \\ \hline
\textbf{Merge sort} &
  \multicolumn{1}{c|}{0.0} &
  \multicolumn{1}{c|}{0.0} &
  0.0 &
  \multicolumn{1}{c|}{0.0} &
  \multicolumn{1}{c|}{0.0} &
  \cellcolor[HTML]{FFF2CC}0.3 &
  \multicolumn{1}{c|}{\cellcolor[HTML]{E06666}1.3} &
  \multicolumn{1}{c|}{\cellcolor[HTML]{FFF2CC}0.3} &
  0.0 &
  \multicolumn{1}{c|}{\cellcolor[HTML]{FFF2CC}0.3} &
  \multicolumn{1}{c|}{0.0} &
  0.0 \\ \hline
\textbf{Breadth-first search} &
  \multicolumn{1}{c|}{0.0} &
  \multicolumn{1}{c|}{0.0} &
  0.0 &
  \multicolumn{1}{c|}{\cellcolor[HTML]{EA9999}1.0} &
  \multicolumn{1}{c|}{0.0} &
  \cellcolor[HTML]{FFF2CC}0.3 &
  \multicolumn{1}{c|}{\cellcolor[HTML]{E06666}1.3} &
  \multicolumn{1}{c|}{\cellcolor[HTML]{FFF2CC}0.3} &
  \cellcolor[HTML]{FFF2CC}0.3 &
  \multicolumn{1}{c|}{\cellcolor[HTML]{F9CB9C}0.7} &
  \multicolumn{1}{c|}{0.0} &
  0.0 \\ \hline
\end{tabular}%
\end{table*}

\begin{table*}[t]
\begin{tabular}{|l|ccc|ccc|ccc|ccc|}
\hline
 &
  \multicolumn{3}{c|}{\textbf{At the start}} &
  \multicolumn{3}{c|}{\textbf{\begin{tabular}[c]{@{}c@{}}After random\\ conversation\end{tabular}}} &
  \multicolumn{3}{c|}{\textbf{\begin{tabular}[c]{@{}c@{}}After Incorrect\\ tutoring\end{tabular}}} &
  \multicolumn{3}{c|}{\textbf{\begin{tabular}[c]{@{}c@{}}After Correct\\ tutoring\end{tabular}}} \\ \hline
Question Types &
  \multicolumn{1}{c|}{U} &
  \multicolumn{1}{c|}{I} &
  A &
  \multicolumn{1}{c|}{U} &
  \multicolumn{1}{c|}{I} &
  A &
  \multicolumn{1}{c|}{U} &
  \multicolumn{1}{c|}{I} &
  A &
  \multicolumn{1}{c|}{U} &
  \multicolumn{1}{c|}{I} &
  A \\ \hline
\textbf{Binary search} &
  \multicolumn{1}{c|}{\cellcolor[HTML]{F9CB9C}0.7} &
  \multicolumn{1}{c|}{\cellcolor[HTML]{FFF2CC}0.3} &
  \cellcolor[HTML]{F9CB9C}0.7 &
  \multicolumn{1}{c|}{\cellcolor[HTML]{FFF2CC}0.3} &
  \multicolumn{1}{c|}{\cellcolor[HTML]{FFF2CC}0.3} &
  \cellcolor[HTML]{F9CB9C}0.7 &
  \multicolumn{1}{c|}{\cellcolor[HTML]{FFFFFF}0.0} &
  \multicolumn{1}{c|}{\cellcolor[HTML]{FFF2CC}0.3} &
  \cellcolor[HTML]{FFF2CC}0.3 &
  \multicolumn{1}{c|}{\cellcolor[HTML]{FFF2CC}0.3} &
  \multicolumn{1}{c|}{\cellcolor[HTML]{FFFFFF}0.0} &
  \cellcolor[HTML]{EA9999}1.0 \\ \hline
\textbf{Merge sort} &
  \multicolumn{1}{c|}{0.0} &
  \multicolumn{1}{c|}{0.0} &
  \cellcolor[HTML]{FFF2CC}0.3 &
  \multicolumn{1}{c|}{\cellcolor[HTML]{EA9999}1.0} &
  \multicolumn{1}{c|}{0.0} &
  \cellcolor[HTML]{F9CB9C}0.7 &
  \multicolumn{1}{c|}{\cellcolor[HTML]{FFFFFF}0.0} &
  \multicolumn{1}{c|}{\cellcolor[HTML]{FFFFFF}0.0} &
  0.0 &
  \multicolumn{1}{c|}{\cellcolor[HTML]{EA9999}1.0} &
  \multicolumn{1}{c|}{0.0} &
  \cellcolor[HTML]{FFF2CC}0.3 \\ \hline
\textbf{Breadth-first search} &
  \multicolumn{1}{c|}{\cellcolor[HTML]{EA9999}1.0} &
  \multicolumn{1}{c|}{0.0} &
  \cellcolor[HTML]{FFF2CC}0.3 &
  \multicolumn{1}{c|}{\cellcolor[HTML]{EA9999}1.0} &
  \multicolumn{1}{c|}{0.0} &
  \cellcolor[HTML]{F9CB9C}0.7 &
  \multicolumn{1}{c|}{\cellcolor[HTML]{F9CB9C}0.7} &
  \multicolumn{1}{c|}{\cellcolor[HTML]{FFFFFF}0.0} &
  \cellcolor[HTML]{FFFFFF}0.0 &
  \multicolumn{1}{c|}{\cellcolor[HTML]{EA9999}1.0} &
  \multicolumn{1}{c|}{\cellcolor[HTML]{FFF2CC}0.3} &
  \cellcolor[HTML]{EA9999}1.0 \\ \hline
\end{tabular}%
\end{table*}

\section{Glossary}\label{appendix:glossary}
\begin{itemize}
    \item \textbf{Learning by teaching (LBT)}: a teaching method in which learners not only articulate and restructure their existing knowledge but also engage in reflective knowledge-building, wherein they extend beyond provided materials to craft deeper explanations, analogies, and inferential connections.
    
    \item \textbf{Teachable agent}: virtual agents that can learn declarative and procedural knowledge from learners’ explanations and demonstrations, taking the role of peer learners in LBT.
    
    \item \textbf{Knowledge-telling}: activities that summarize knowledge with little monitoring or elaboration), which should lead to stronger or weaker learning, respectively.
    
    \item \textbf{Knowledge-building}: activities that include self-monitoring of comprehension, integration of new and prior knowledge, and elaboration and construction of knowledge.
    
    \item \textbf{Reflect flow}: a data processing flow in the Reflect-Respond pipeline that incorporates the learned knowledge from the conversation into the knowledge state of AlgoBo.
    
    \item \textbf{Response flow}: another data processing flow that generates a response to a conversation based on the current knowledge AlgoBo holds.
    
    \item \textbf{Reconfigurability}: how precisely we can set AlgoBo’s performance in question-answering and problem-solving.
    
    \item \textbf{Persistence}: how AlgoBo’s knowledge level is maintained consistently throughout a conversation until it is taught new information.
    
    \item \textbf{Adaptability}: how well AlgoBo updates its knowledge as it acquires new information from tutors in conversations.

    \item \textbf{Conversational antipatterns}: sequences of messages of certain message types that inhibit learning in LBT. Antipatterns include Commanding, Spoon-feeding, and Under-teaching.

    \item \textbf{Metacognitive feedback}: feedback throughout the conversation to help learners reflect on the overall teaching session and offer overarching guidance on steering the discussion.
\end{itemize}

\end{document}
\endinput